 \documentclass[iop,revtex4,apj]{emulateapj}
 \usepackage{lscape}
 \slugcomment{{\sc Accepted to ApJ:} September 19, 2014}
\usepackage{epsfig,amsmath}
\usepackage[dvips]{color}
\usepackage{verbatim}
\usepackage{graphicx}
\usepackage{txfonts}
\usepackage{natbib}
\usepackage{hyperref}

\newcommand{\Lsun}{\ifmmode {\rm L}_{\odot} \else L$_{\odot}$\fi}
\newcommand{\qo}{\ifmmode q_{\rm o} \else $q_{\rm o}$\fi}
\newcommand{\Ho}{\ifmmode H_{\rm o} \else $H_{\rm o}$\fi}
\newcommand{\ho}{\ifmmode h_{\rm o} \else $h_{\rm o}$\fi}
\newcommand{\Halpha}{\ifmmode {\rm H}\alpha \else H$\alpha$\fi}
\newcommand{\Hbeta}{\ifmmode {\rm H}\beta \else H$\beta$\fi}
\newcommand{\Hgamma}{\ifmmode {\rm H}\gamma \else H$\gamma$\fi}
\newcommand{\Hdelta}{\ifmmode {\rm H}\delta \else H$\delta$\fi}
\newcommand{\Lya}{\ifmmode {\rm Ly}\alpha \else Ly$\alpha$\fi}
\newcommand{\Lyb}{\ifmmode {\rm Ly}\beta \else Ly$\beta$\fi}
\newcommand{\HeI}{\ifmmode {\rm He}\,{\sc i}\,\lambda5876 \else 
	          He\,{\sc i}\,$\lambda5876$\fi}
\newcommand{\HeII}{\ifmmode {\rm He}\,{\sc ii}\,\lambda4686 \else 
	           He\,{\sc ii}\,$\lambda4686$\fi}

\newcommand{\ciii}{\ifmmode {\rm C}\,{\sc iii} \else C\,{\sc iii}\fi}
\newcommand{\civ}{\ifmmode {\rm C}\,{\sc iv} \else C\,{\sc iv}\fi}

\newcommand{\Flamunit}{\ifmmode {\rm erg\ s}^{-1} cm^{-2}\ \AA^{-1} \else erg s$^{-1}$\ cm$^{-2}$\,\AA$^{-1}$\fi}

\begin{document}

\title{ULTRALUMINOUS INFRARED GALAXIES IN THE AKARI ALL SKY SURVEY}

\author{ E.~Kilerci Eser\altaffilmark{1},
            T.~Goto\altaffilmark{2},
            Y.~Doi\altaffilmark{3}
          }                        
\altaffiltext{1}{Dark Cosmology Centre, Niels Bohr Institute, University of Copenhagen,
Juliane Maries Vej 30, DK-2100 Copenhagen \O, Denmark; ecekilerci@dark-cosmology.dk}
\altaffiltext{2}{National Tsing Hua University, No. 101, Section 2, Kuang-Fu Road, Hsinchu, Taiwan 30013; tomo@phys.nthu.edu.tw}
\altaffiltext{3}{The University of Tokyo, Komaba 3-8-1, Meguro, Tokyo, 153-8902, Japan; doi@ea.c.u-tokyo.ac.jp}

\begin{abstract}\label{S:abstract}
We present a new catalog of 118 Ultraluminous Infrared Galaxies (ULIRGs) and one Hyperluminous Infrared Galaxy (HLIRG) 
by crossmatching {\it{AKARI}} all-sky survey with the Sloan Digital Sky Survey Data Release 10 (SDSS DR10) and the Final Data Release of the Two-Degree Field Galaxy Redshift Survey (2dFGRS).  
40 of the ULIRGs and one HLIRG are new identifications. 
We find that ULIRGs are interacting pair galaxies or ongoing/post mergers. 
This is consistent with the widely accepted view: ULIRGs are major mergers of disk galaxies. 
We confirm the previously known positive trend between the AGN fraction and IR luminosity. 
We show that ULIRGs have a large off-set from the `main sequence' up to $z\sim$ 1; their off-set from the $z\sim$ 2 `main sequence' is relatively smaller. 
We find a consistent result with the previous studies showing that compared to local star forming SDSS galaxies of similar mass, local ULIRGs have lower oxygen abundances.
 We for the first time demonstrate that ULIRGs follow the fundamental metallicity relation (FMR). 
 The scatter of ULIRGs around the FMR (0.09 dex$-$0.5 dex) is comparable with the scatter of $z\sim$ 2-3 galaxies. 
 Their optical colors show that ULIRGs are mostly blue galaxies and this agrees with previous findings. 
 We provide the largest local (0.050 $< z <$ 0.487) ULIRG catalog with stellar masses, SFRs, gas metallicities and optical colors. 
 Our catalog provides us active galaxies analogous to high-$z$ galaxies in the local Universe where they can be rigorously scrutinized.
\end{abstract}


%***********MAIN BODY STARTS HERE********************************************

\section{INTRODUCTION\label{introduction}}

Luminous infrared galaxies (LIRGs), ultraluminous infrared galaxies (ULIRGs) and hyperluminous infrared galaxies (HLIRGs) are 
defined by their high IR luminosities that are 
in the $10^{11}L_{\odot} \le L_{IR} < 10^{12}L_{\odot}$, $10^{12}L_{\odot} \le L_{IR} < 10^{13}L_{\odot}$ and $10^{13}L_{\odot} \le L_{IR}$ ranges, respectively \citep[see the reviews by][]{Sanders1996,Lonsdale2006}. 
The observed enormous IR luminosity is driven by the optical and UV radiation generated by intense star formation and active galactic nuclei (AGN), that is 
absorbed by dust and re-emitted in the IR. 
ULIRGs have been considered as a transition phase from mergers to dusty quasars \citep{Sanders1988a,Veilleux2002} such that 
when gas-rich spiral galaxies merge, the molecular gas clouds channeling towards the merger nucleus trigger nuclear starbursts and 
AGN activity via the accretion of the available fuel on to the central super massive black hole (SMBH). 
According to this scenario, the starburst phase evolves to a dust-enshrouded AGN phase, and once the gas and dust are consumed the system evolves to a bright QSO phase. 

Tidal interactions and merger processes between galaxies play a major role in the formation of elliptical galaxies \citep{ToomreToomre1972}. 
Especially, the proposed scenario by \citet{Sanders1988a} motivated further investigation of the link between mergers and quasars in numerical simulations. 
Hydrodynamical simulations of mergers show that merger processes lead gas inflows towards the center that trigger starbursts and AGN activity \citep[e.g][]{Springel2005}. 
In the merger-driven galaxy evolution simulations ULIRGs represent a contemporary starburst- and AGN-phase 
at the beginning of a rapid self-regulated SMBH growth \citep[e.g.][]{DiMatteo2005,Hopkins2007b}.
ULIRGs evolve to red/elliptical-type remnants by the `negative feedback' mechanisms (e.g. in the form of powerful winds and outflows) that inhibit  star formation and AGN activity \citep[e.g.][]{Hopkins2006,Hopkins2008a,Hopkins2008b,Hopkins2009a}.

The emerged link between ULIRGs and QSOs is supported by several observational evidences. 
Morphological properties of ULIRGs indicate that they are interacting galaxies in pre/ongoing/late merger stages \citep{Farrah2001,Kim2002,Veilleux2002,Veilleux2006}. 
Compared to LIRGs that are disk galaxies (if $\log(L_{IR}/L_{\odot})<$ 11.5) or interacting systems (if 11.5$\le \log(L_{IR}/L_{\odot})<$ 12.0), ULIRGs are mostly advanced mergers \citep{Veilleux2002,Ishida2004}.
Their dynamical masses obtained from near-infrared (NIR) spectroscopy show that they are major mergers of nearly equal mass galaxies \citep{Veilleux2002,Dasyra2006a,Dasyra2006b}. 
CO observations proved that ULIRGs contain the required cold molecular gas for central starbursts \citep{Downes1998}. 
Additionally, their mid-infrared (MIR) images show that MIR emission generated in a region of diameter $\sim$ 1 kpc \citep{Soifer2000}. 
At least $\sim$70\% of 164 local ($z\le$0.35) ULIRGs harbor an AGN \citep{Nardini2010}. 
The coexistence of a starburst and an AGN show that both energy sources contribute to the total IR luminosity. 
The AGN fraction and the strength of the AGN emission increases with IR luminosity; high-luminosity ULIRGs ($\log(L_{IR}/L_{\odot})>$ 12.5) and HLIRGs have a larger 
AGN contribution compared to lower luminosity IR galaxies \citep{Veilleux1995,Veilleux1999b,Veilleux2002,Veilleux2009,Genzel1998,Goto2005,Imanishi2009,Nardini2010}. 
ULIRGs show starburst-and AGN-driven powerful outflows \citep[e.g][ and references therein]{Heckman2000,Rupke2002,Rupke2005c,Rupke2011,Rupke2013a,Spoon2013,Veilleux2013b} that are consistent with the expected negative feedback mechanisms for their evolution. 

The significance of ULIRGs in the galaxy evolution is not limited to the local ($z<$ 0.3) Universe because, at high redshift ($z>$1) they are more numerous and have a substantial contribution to the total IR luminosity 
density \citep{LeFloch2005,Caputi2007} compared to local ULIRGs \citep{Soifer1991,Kim1998a}. 
There is a significant population of ULIRGs beyond $z\sim$1 \citep[e.g][]{Goto2011}. 
An important question is the powering mechanism of these sources: are they powered by interaction-induced nuclear starbursts/AGN or are they normal/undisturbed star forming galaxies? 
The key properties that would answer this question are: morphologies, spectral energy distributions (SEDs) and the extend of star forming regions. 
Observations have shown that ULIRGs at high redshift (1.5 $< z <$ 3.0) are mostly ($\sim$ 47\%\ ) mergers or interacting galaxies, but this sample also includes non-interacting disks, spheroids and irregular galaxies  \citep{Kartaltepe2012}. 
Beyond $z>$ 2 morphological properties of sub-millimeter galaxies (SMGs) are consistent with mergers and interacting systems \citep[e.g][]{Tacconi2008}. 
Morphologies of high $z$ samples show that mergers/interactions taking place in these systems and even  
comparison of $z\sim$ 2 and  $z\sim$ 1 samples indicates a hint for a morphological evolution such that $z\sim$ 1 sample have slightly more mergers and interacting galaxies \citep{Kartaltepe2012}. 
The SEDs of high redshift ULIRGs are different from those of local ones: for example they exhibit prominent polycyclic aromatic hydrocarbon (PAH) features similar 
to those of local lower IR luminosity (10.0$\le \log(L_{IR}/L_{\odot})<$ 11.0) star forming galaxies rather those of local 
ULIRGs  \citep[e.g][]{Farrah2008,Takagi2010}. 
Since  PAH emission indicates ongoing star formation, observations support that high $z$ ULIRGs are starburst dominated. 
A similar conclusion is also achieved by the X-ray studies of high $z$ ULIRGs \citep[e.g][]{Johnson2013}.
The size of the star forming regions of high $z$ ULIRGs are larger than those of local ULIRGs with similar $L_{IR}$ \citep{Rujopakarn2011}. 
This suggest that in these galaxies star formation do not occur in merger nuclei but, it is distributed galaxy wide. 
The similarities of star forming regions of high $z$ ULIRGs and local quiescent star forming galaxies point out a different origin than merger-induced star formation \citep{Rujopakarn2011}. 
Although the evolution of ULIRGs is not fully understood yet, observations provide evidence for changing properties with redshift.

Understanding the role of ULIRGs in galaxy evolution through cosmic time require extensive studies/comparison of local and high $z$ samples. 
Local ULIRGs establish a ground to understand the nature of ULIRGs, the origin of their extreme luminosities and the interplay between star formation and AGN activity in the nearby mergers.
Therefore, it is important to have a large local sample and to master its overall properties. 
The great majority of local ULIRGs are discovered with InfraRed Astronomy Satellite \textit{(IRAS)}. 
\textit{IRAS} performed an all sky scan in four IR bands centered at 12\micron, 25\micron, 60\micron\ and 100\micron. 
\textit{IRAS}  Bright Galaxy Survey (BGS) catalog \citep{Soifer1987} includes 10 ULIRGs selected on the basis of 60\micron\ flux, F(60\micron).  
This catalog replaced with the \textit{IRAS} Revised Bright Galaxy Sample (RBGS) \citep{Sanders2003} that provided more accurate infrared luminosities and increased the number of ULIRGs to 21. 
The \textit{IRAS} 2-Jy \citep{Strauss1992} and 1.2-Jy \citep{Fisher1995} redshift surveys identified new ULIRGs. 
\citet{Sanders1988} showed that ULIRGs with `warm' colors (F(25\micron)/F(60\micron) $>$ 0.2) have Seyfert like spectra and thereon ULIRGs separated as `warm' AGN 
hosting and `cold' star formation dominated systems. 
Analysis of BGS sample showed that the F(60\micron)/F(100\micron) color increases with higher $L_{IR}$ \citep{Soifer1991}. 
A widely studied large sample of local ULIRGs is the  \textit{IRAS} 1-Jy sample \citep{Kim1998}. 
This is a complete flux-limited sample at 60\micron\ that is composed of 118 ULIRGs identified from \textit{IRAS} Faint Source Catalog (FSC) \citep{Moshir1992} and 
a dedicated redshift survey \citep{Kim1998}. 
Since previous studies  \citep{Soifer1991,Strauss1992} showed that F(60\micron)/F(100\micron) color increases with higher $L_{IR} $ and ULIRGs 
colors are in the range of -0.2$<$F(60\micron)/F(100\micron)$<$0.13, 
\textit{IRAS} 1-Jy sample ULIRGs were selected based on their warm colors (F(60\micron)/F(100\micron)$>$ 0.3) \citep{Kim1998}. 
With other redshift surveys such as the QDOT all sky \textit{IRAS} galaxy redshift survey \citep{Lawrence1999}, the \textit{IRAS} Point Source Catalog Redshift (PSCz) 
survey \citep{Saunders2000} and the FIRST/IRAS radioÐfar-IR sample of \citep{Stanford2000} number of IRAS ULIRGs increased. 
Large galaxy redshift surveys like the Sloan Digital Sky Survey (SDSS) \citep{York2000} and Two-Degree Field Galaxy Redshift Survey (2dFGRS) \citep{Colless2001} 
provide redshifts of millions of galaxies. 
Especially SDSS made it possible to study optical properties of large sample of IR galaxies. 
\citet{Goto2005} cross-correlated the \textit{IRAS} FSC with the SDSS Data Release 3 (DR3) \citep{Abazajian2005} spectroscopic catalog and identified 178 ULIRGs. 
\citet{Pasquali2005} cross-correlated SDSS DR2 \citep{Abazajian2004} with the \textit{IRAS} FSC and investigated IR properties of local AGNs and star forming galaxies. 
\citet{Cao2006} cross-correlated \textit{IRAS} FSC and the Point Source Catalog (PSC) with SDSS DR2 and identified a small sample of ULIRGs. 
\citet{Hwang2007} identified 324 ULIRGs, including 190 new discoveries, by cross correlating \textit{IRAS} FSC with SDSS DR4 \citep{Adelman-McCarthy2006}, 2dFGRS and the 
Second Data Release of the 6dF Galaxy Survey (6dFGS) \citep{Jones2004}. 
\citet{Hou2009} cross-correlated  \textit{IRAS} FSC with the SDSS DR6 \citep{Adelman-McCarthy2008} and identified 308 ULIRGs. 

The largest all sky IR survey after \textit{IRAS} completed by the Japanese IR satellite launched in 2006, \textit{AKARI} \citep{Murakami2007} that 
scanned almost all-sky in 9\micron, 18\micron, 65\micron, 90\micron, 140\micron\ and 160\micron\ bands.
The resolution and sensitivity of \textit{AKARI} is better than those of \textit{IRAS}: the Point Spread Function (PSF) of \textit{AKARI} is $\sim$39\arcsec\ (for 90\micron\ band) while the PSF of \textit{IRAS} is 
$\sim$4\arcmin\ (for 100\micron\ band); at 18\micron\ \textit{AKARI} is 10 times more sensitive. 
An other advantage of \textit{AKARI} is that, it has a wider and longer wavelength coverage compared to \textit{IRAS}.
 In particular the 140\micron\ and 160\micron\ bands are very important to measure the 
peak of the dust emission near 100\micron\ and therefore obtain more accurate IR luminosity. 
\citet{Goto2011} matched \textit{IRAS} IR sources with SDSS DR7 \citep{Abazajian2009} galaxies and measured the local IR luminosity function. 
In this study, \citet{Goto2011} identified ULIRGs among \textit{AKARI} sources, but did not provide a detailed catalog of these sources. 
In this work we search for ULIRGs and HLIRGs in the \textit{AKARI} all-sky survey. 
We cross-correlate \textit{AKARI} all-sky survey with 2dFGRS and the largest SDSS spectroscopic redshift catalogue DR 10 \citep{Ahn2013}. 
Beside the redshift information, SDSS has a rich view of optical properties of the sources in this database. 
Optical images, spectra, colors, value added catalogs with emission line properties provided by SDSS D10 give us an opportunity to investigate the morphologies, colors, stellar mass 
and metallicities of local ULIRGs identified in the \textit{AKARI} all-sky survey. 
We provide the first catalog of ULIRGs identified in \textit{AKARI} all-sky survey. 

This paper has the following structure. 
We introduce the data to identify ULIRGs/HLIRGs and our final sample in \S\ref{S:iden}. 
Our results are presented in \S \ref{sec:Results}. 
In \S\ \ref{S:dis} we discuss our results. 
The summarized conclusion of this work is given in \S\ \ref{S:conc}.
Throughout this work, we adopt a cosmology with $H_0=70$\,km\,s$^{-1}$\,Mpc$^{-1}$, $\Omega_\Lambda = 0.7$ and $\Omega_{\rm m}=0.3$. 

\section{IDENTIFICATION OF ULTRALUMINOUS AND HYPERLUMINOUS INFRARED GALAXIES  IN THE \textit{ \textbf{AKARI}} ALL-SKY SURVEY} \label{S:iden}

\subsection{The Samples}\label{S:samples}

\subsubsection{The AKARI All-Sky Survey Catalogs}\label{S:AKARIcatalogs}
The \textit{AKARI} all-sky survey provides two catalogs of the IR sources across more than $\sim$97\% of the whole sky with fluxes centered on two mid-IR and four FIR bands. 
The \textit{AKARI}/IRC all-sky survey point source catalog version 1\footnote[1]{http://www.ir.isas.jaxa.jp/AKARI/Observation/PSC/Public/RN/AKARI-IRC\_PSC\_V1\_RN.pdf} 
includes 870973 IR sources with fluxes at 9\micron\ and 18\micron\ mid-IR bands. 
The \textit{AKARI}/FIS all-sky survey bright source catalog version 1\footnote[2]{http://www.ir.isas.jaxa.jp/AKARI/Observation/PSC/Public/RN/AKARI-FIS\_BSC\_V1\_RN.pdf} \citep{Yamamura2010} 
contains 427071 sources detected at 90\micron\ with flux measurements at 65\micron\, 90\micron\, 140\micron\ and 160\micron\ FIR bands. 
Especially 140\micron\ and 160\micron\ fluxes are very important to constrain the FIR SED peak and to measure $L_{IR}$.

In order to have a single  \textit{AKARI}/FIS/IRC catalog with both FIR and mid-IR fluxes, 
we cross-match the IR sources in the \textit{AKARI}/FIS all-sky survey bright source catalog 
with the \textit{AKARI}/IRC all-sky survey point source catalog within a radii of 20\arcsec. 
The resulting \textit{AKARI}/FIS/IRC catalog contains 24701 sources based on  90\micron\ detections.

To measure the IR luminosity we obtain the spectroscopic redshifts of the IR galaxies from their optical counterparts. 
We cross-correlate the \textit{AKARI}/FIS/IRC catalog with large optical redshift catalogs as described in the following. 

\subsubsection{The AKARI$-$SDSS DR10 Sample}\label{S:samples1}

The Sloan Digital Sky Survey is the largest ground-based survey providing a unique photometric and spectroscopic database of stars, galaxies and quasars. 
SDSS is a red magnitude limited $r<17.7$ survey over 14555 deg$^{2}$ of the sky. 
We have downloaded the SDSS DR 10 \citep{Ahn2013} catalogs \textit{photoObj}\footnote[3]{http://www.sdss3.org/dr10/spectro/spectro\_access.php} 
and \textit{specObj}$^{3}$ to extract both photometric and spectroscopic information. 
The \textit{photoObj} catalog includes all photometric information from previous data releases and the \textit{specObj} catalog includes new spectra from the  Baryon Oscillation Spectroscopic
Survey\footnote[4]{http://www.sdss3.org/survey/boss.php} (BOSS). We combined the two catalogs by matching `OBJID' in \textit{photoObj} to `BESTOBJID' in \textit{specObj} 
and obtain a full SDSS catalog of 2,745,602 sources with spectroscopic and photometric information. 

\textit{AKARI}/FIS/IRC catalog is cross-matched with the full SDSS catalog. 
The astrometric precision of SDSS \citep[$\sim$ 0.1\arcsec\ at $r$=19 mag][]{Pier2003} is much better than that of \textit{AKARI} \citep[$\sim$4.8\arcsec][]{Yamamura2010}).
We follow \citet{Goto2011} and select matching radii as 20\arcsec\ 
because this radius is large enough to contain different emission regions (e.g IR and optical) in a single galaxy; 
also it is small enough not to allow too many optical chance identifications that are not physically related to the IR source. 
Although we pick 20\arcsec\ to be inclusive not to miss any real association, in order to eliminate any miss-association later, we check the positional overlap of the IR and the optical emission from each ULIRG candidate by eye. 
We avoid any duplicated matches, i.e each IR galaxy is allowed to match only with one SDSS counterpart. 
We obtain 6468 matches of  \textit{AKARI}$-$SDSS sources. Among those we removed the sources that were classified as star in the \textit{specObj} catalog. 
This resulted into a \textit{AKARI}$-$SDSS sample of 6373 galaxies. For the IR sources in this sample we adopt the SDSS spectroscopic redshifts.

\subsubsection{The AKARI$-$2dFGRS Sample}\label{S:samples2}

The Two-Degree Field Galaxy Redshift Survey \citep{Colless2001} measured redshifts of 245951 galaxies within $b_{j} < 19.45$ limit. 
The median redshift of this survey is $z\sim0.1$ \citep{Colless2004}. 
We use the Final Data Release of the 2dFGRS , the catalog of \textit{best spectroscopic observations}\footnote[5]{http://www2.aao.gov.au/$\sim$TDFgg/}. 
We cross-match the \textit{AKARI}/FIS/IRC catalog with the 2dFGRS catalog with a matching radii of 20\arcsec. 
We obtain a \textit{AKARI}$-$2dFGRS sample of 954 galaxies with spectroscopic redshifts from 2dFGRS. 

\subsection{Infrared Luminosity Measurements}\label{S:SEDs}

\begin{figure*} 
\begin{center}$
\begin{array}{lcr}
\includegraphics[scale=0.35]{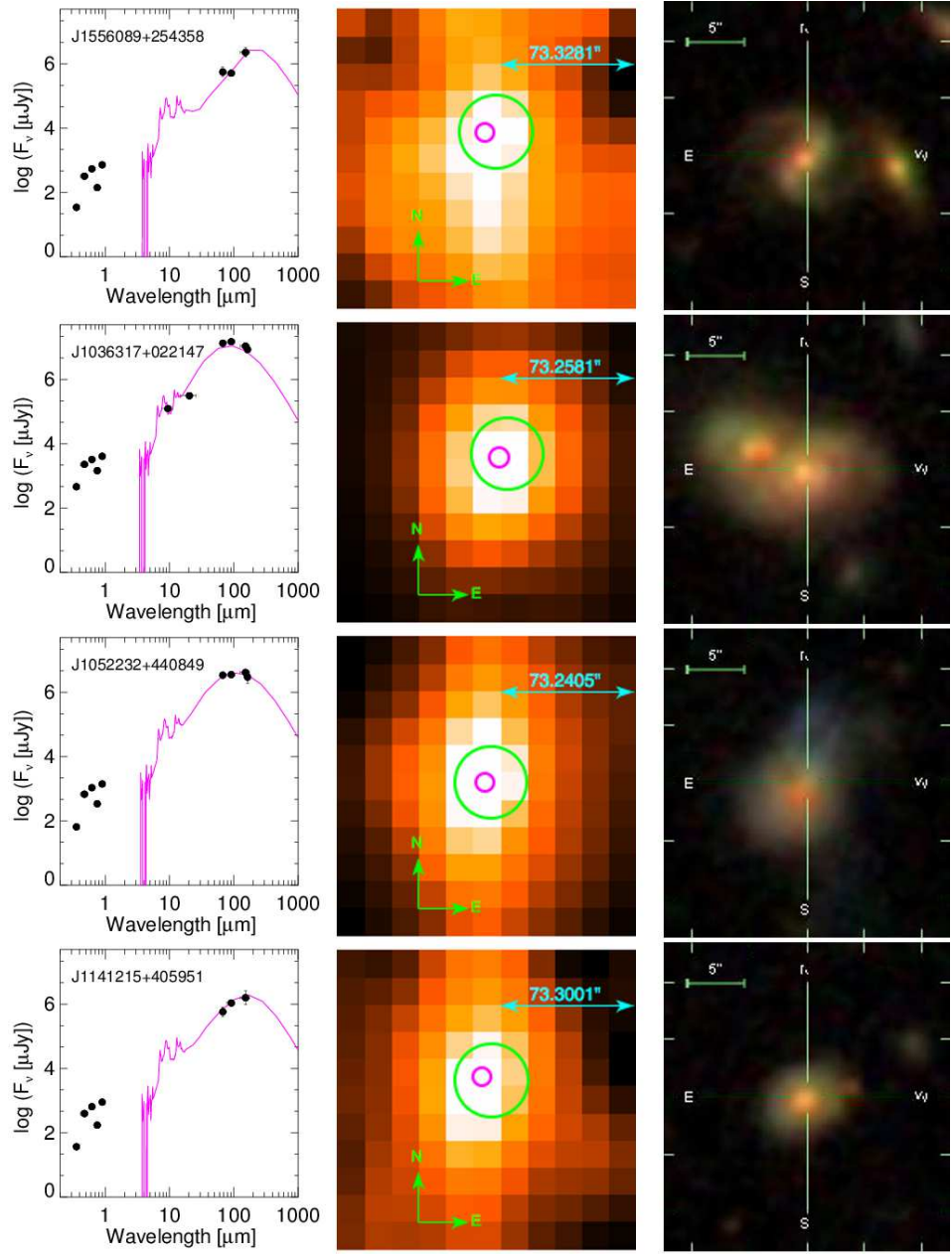} &
\end{array}$
\end{center}
\caption{SEDs (left), \textit{AKARI} (middle) and SDSS $g$-$r$-$i$ colors combined images  (right) of four nearest ULIRGs classified as `IIIa' (first row), `IIIb' (second row), `IV' (third row), `IV' (fourth row). 
The scale of the \textit{AKARI} 90\micron\ images are 165\arcsec$\times$165\arcsec.  
The small 5\arcsec\ radius (colored magenta) and the large 20\arcsec\ radius (colored green) 
circles, mark the optical and IR source, respectively.}
\label{fig:sedsimages1}
\end{figure*}

To estimate the total IR luminosity for the galaxies in the  \textit{AKARI}$-$SDSS and \textit{AKARI}$-$2dFGRS samples 
we perform a SED-fitting using the \textit{LePhare}\footnote[6]{http://www.cfht.hawaii.edu/$\sim$arnouts/lephare.html} 
(Photometric Analysis for Redshift Estimations) code \citep{Arnouts1999,Ilbert2006}. 
The main function of the \textit{LePhare} is to compute photometric redshifts, 
but it can also find the best fitting galaxy template by a $\chi^{2}$ fit for the given photometric magnitudes among the input template libraries. 
For the \textit{AKARI}$-$SDSS and \textit{AKARI}$-$2dFGRS samples we use the six \textit{AKARI} bands with their associated uncertainties adopted from 
the \textit{AKARI} catalogs; if the flux uncertainty is not given we adopt 25\% of the measured flux as the uncertainty. 
We use the FIR SED templates of \citet{Dale2002} as the input library.  
\citet{Dale2002} provide 64 SED templates generated semi-empirically to represent the IR SEDs of star forming galaxies. 
Compared to other SED models, such as the models of \citet{Chary2001}, these templates include FIR improvements based on $ISO$/$IRAS$/$SCUBA$ observations. 
However, they do not include more sophisticated dust emission modeling as provided by \citet{Siebenmorgen2007}. 
Since our main focus is to measure $L_{IR}$ we avoid more sophisticated models and prefer the templates of \citet{Dale2002} for the SED fitting. 
We fix the redshift of each galaxy and fit the FIR region of the SED with the \textit{AKARI} broadband photometry. 
In the fitting procedure $k$-corrections are applied to \textit{AKARI} fluxes. 
In order to obtain the $k$-correction our model flux is computed integrating redshifted SED model flux through \textit{AKARI}'s filter response function. 
The best-fit dust templates of \citet{Dale2002} are shown in Figure \ref{fig:sedsimages1} (left column) for representative cases.
The \textit{AKARI}/FIS name is given in the top left corner. 
The best-fit templates are shown as solid magenta line. 
The black filled circles are the optical (shown only for illustration purposes) and \textit{AKARI} photometric fluxes; the x-axis error bars represent the wavelength range of each photometric band. 

As a result of the SED fitting we obtain the total infrared luminosity integrated between 8\micron\ $-$ 1000\micron, $L_{8-1000}$ with the 
maximum and minimum possible $L_{8-1000}$ value based on the flux errors. 
These are used to determine the the upper and lower uncertainties of $L_{8-1000}$. 

Based on the obtained $L_{8-1000}$ our initial sample includes 170 ULIRG and 10 HLIRG candidates: 
the \textit{AKARI}$-$SDSS sample has 135 ULIRG and eight HLIRG candidates; the \textit{AKARI}$-$2dFGRS has 35 ULIRG and two HLIRG candidates. 
In order to have a reliable sample of ULIRGs and HLIRGs, we check each case to avoid any wrong identification as described in the following.

%FIG 1

\subsection{Elimination of the Mismatches}\label{S:eliminate}

The (H)/ULIRG candidates in our initial sample are selected based on the optical spectroscopic redshifts; by matching the closest optical galaxy to the \textit{AKARI} source. 
If there are more than one source satisfying the cross-match condition, then the one with the smallest positional difference is considered as a match. 
Even the positions of the optical and IR sources are close in the sky, it does not necessarily mean that the IR and optical emissions are counterparts of the same galaxy; further care is required to make this decision. 

Although the optical and IR galaxies are matched within a 20\arcsec\ radius, we visually check the positional overlap of the IR and optical emission in \textit{AKARI} images for each galaxy. 
For the optical counterpart we use SDSS (if available) or Digitized Sky Survey images. 
Examples of \textit{AKARI} (middle panel) and optical (right panel) images are represented in Figure \ref{fig:sedsimages1}.
The SDSS images are $gri$ combined color images downloaded from the SDSS DR10 Finding Chart Tool\footnote[8]{http://skyserver.sdss3.org/dr10/en/tools/chart/chartinfo.aspx}. 
In the \textit{AKARI} images \citep{Doi2012} the green circle represents the 20\arcsec\ radius limit, whereas the optical source is marked with a 5\arcsec\ radius magenta circle. 
Once we make sure of the positional overlap of the matched IR and optical sources, next we check if there are any other sources overlapping with the IR source and possibly contaminate the IR emission. 
Such contaminating sources can be stars or other galaxies. 
Especially nearby bright galaxies lying over the IR source have a contribution to the observed IR emission, therefore such cases are eliminated from the initial sample. 
If there are more than one overlapping optical galaxies with similar separation values within the 20\arcsec\ radius region, the closest one does not necessarily mean the true match. 
Since it is difficult to select the true optical counterpart for these four cases, these are eliminated. 
It is a worry if we are automatically removing compact groups of galaxies in these cases, but before we eliminate these we consider the redshifts of these galaxies and check if they are in groups.

Although SDSS provides a large redshift database, not all galaxies have the spectroscopic information. 
Related to this, in some cases the optical source with the smallest positional difference is not included in the cross-match procedure. 
Therefore, the images show that instead of the `true' optical counterpart, some other optical galaxy with a large separation  (8.12\arcsec\ $-$ 18.87\arcsec) is matched with the IR emission. 
For these cases we look at the literature \citep[e.g.][]{2009MNRAS.398..109W} and check if the `true' optical counterpart has a spectroscopic redshift. 
If the redshift is not known for the `true' optical counterpart we eliminate these cases. 
If the redshift is known, we adopt it for the `true' optical counterpart and re-obtain $L_{8-1000}$. 
Three cases (marked with an asterisk in column (5) of Tables \ref{tab:newULIRGs}, \ref{tab:knownULIRGs}) for which $10^{12}L_{\odot} \le L_{8-1000}$ are kept in the sample. 

After we secure the optical and IR galaxy match by visual inspection, as an additional control we check the adopted spectroscopic redshifts. 
For the \textit{AKARI}$-$2dFGRS sample we require a redshift quality of $\ge 3$. 
This requirement let to eliminate two ULIRG and two HLIRG candidates from the \textit{AKARI}$-$2dFGRS sample. 
For the \textit{AKARI}$-$SDSS sample, we go through the SDSS spectra\footnote[9]{http://dr10.sdss3.org/basicSpectra}. 
The SDSS spectra are reduced through the spectroscopic pipeline \citep{Bolton2012}. 
The SDSS pipeline determines the classification and the redshift of the spectra by applying a $\chi^{2}$ fit with rest-frame templates of stars, galaxies and quasars. 
By looking at the SDSS spectra we eliminate the following cases from the sample: (1) The sources that are classified as a galaxy but show a spectrum of a star; 
(2) The spectra showing an unreliable template fit and therefore indicating a wrong redshift. 

\subsection{The Final Sample}\label{S:finalsample}

Our final sample of (H)/ULIRGs consist of 119 galaxies, 97 are identified in the \textit{AKARI}$-$SDSS sample and 22 are identified in the \textit{AKARI}$-$2dFGRS sample. 
In order to specify the newly identified (H)/ULIRGs in this work we check our final sample against previously studied samples: \citet{Clements1996}, \citet{Kim1998}, \citet{RowanRobinson2000}, \citet{Hwang2007}, \citet{Hou2009}, \citet{Nardini2010}.
40 ULIRGs and one HLIRG are newly identified in this work. 
The IR images of the newly identified ULIRGs and one HLIRG are available in the appendix. 
We divide the final sample into three subsamples: (1) New ULIRGs identified in this work; (2) Known ULIRGs; (3) New HLIRG identified in this work. 
The properties of these subsamples are listed in Tables \ref{tab:newULIRGs}, \ref{tab:knownULIRGs} and \ref{tab:newHLIRG}, respectively. 
These tables contain the \textit{AKARI} name (column 1), \textit{AKARI} coordinates: RA and DEC (columns 2 and 3, respectively), 
other name (column 4), redshift (column 5), total IR luminosity, $L_{IR}$, (column 6), 
{\it{AKARI}} photometric fluxes of the 65\micron\ (F(65\micron)), 90\micron\ (F(90\micron)), 140\micron\ (F(140\micron)) and 160\micron\ (F(160\micron)) bands (columns 7, 8, 9 and 10, respectively), 
SDSS Petrosian $r$ magnitude (column 11), Interaction Class (IC; column 12), reference for IC (column 13), 
note related to optical images indicating if there is a star or other galaxies in the field (column 14), spectral classification (column 15). 
Since we have only a few sources detected in the 9\micron\ and 18\micron\ bands we do not list the photometric fluxes at these bands. 

We note that 3 of the new ULIRGs listed in Table \ref{tab:newULIRGs} have 65\micron\ fluxes above 1 Jy, the flux threshold of the 1-Jy sample of \citet{Kim1998}, 
and have declination $\delta > -40\deg$ and galactic latitude $b > 30\deg$, and therefore should have made it in the 1-Jy sample. 
However, two of these sources (J1036317+022147, J1125319+290316,) are not observed with \textit{IRAS}. 
The 60\micron\ flux of J0857505+512037 is below 1 Jy \citep[$\sim$0.6 Jy][]{Moshir1992}  and therefore it is not in the 1-Jy sample of \citet{Kim1998}.

In Table \ref{tab:susULIRGs} we list five additional ULIRG candidates that are considered as unconfirmed cases either because their 
IR detection is not significant (almost at 5 $\sigma$), or the separation between the matched optical and IR coordinates are large ($\sim20\arcsec$). 
We do not include those five sources in the final sample.

\section{Analysis and Results}\label{sec:Results}

\subsection{Basic Properties Of the \textit{ \textbf{AKARI}} ULIRGs and HLIRG Samples }\label{S:colprop}

\subsubsection{Redshift and $L_{IR}$ Distributions} \label{S:dist}

\begin{figure} 
\begin{center}$
\begin{array}{c}
\includegraphics[scale=0.34,angle=270]{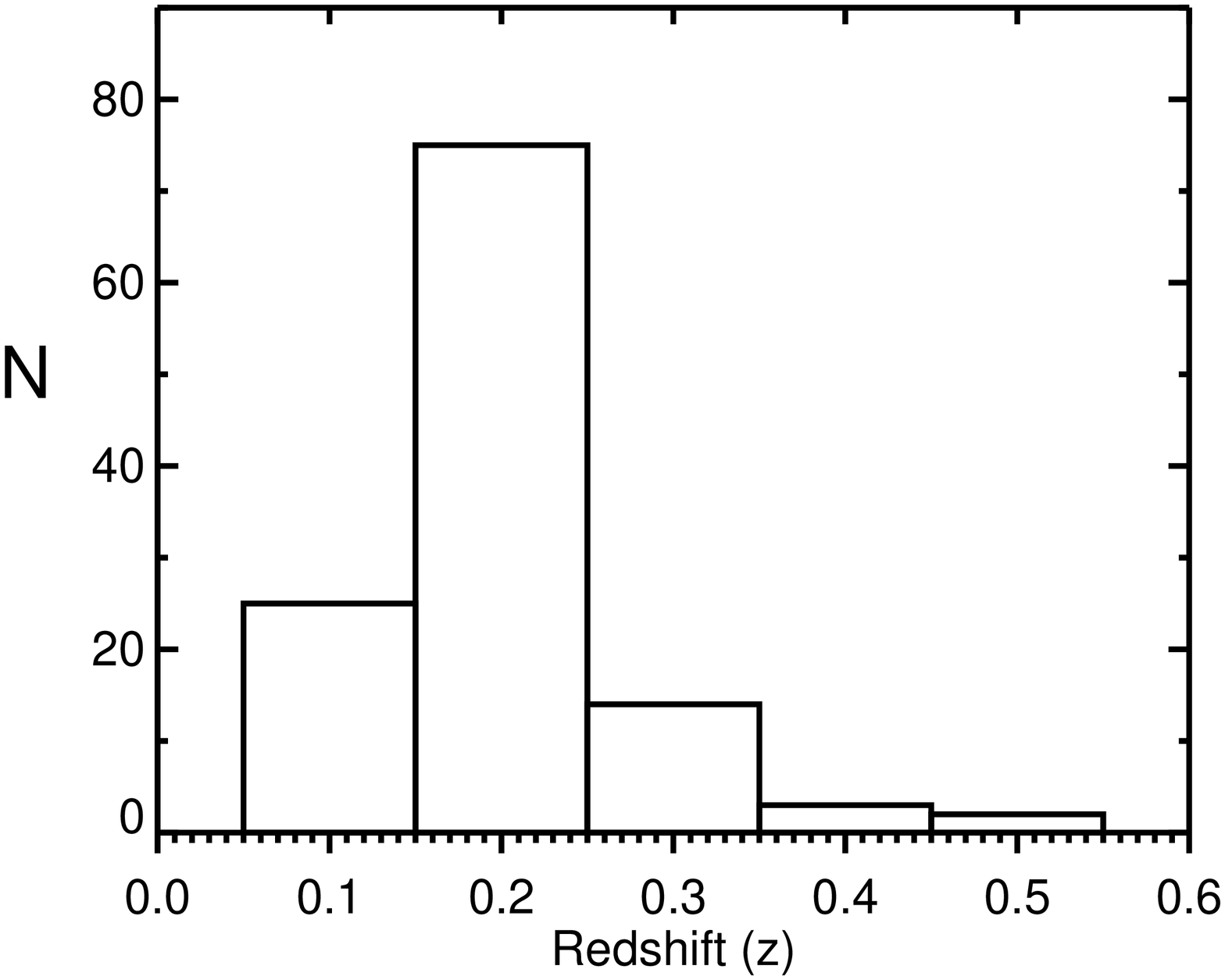}\\
\includegraphics[scale=0.34,angle=270]{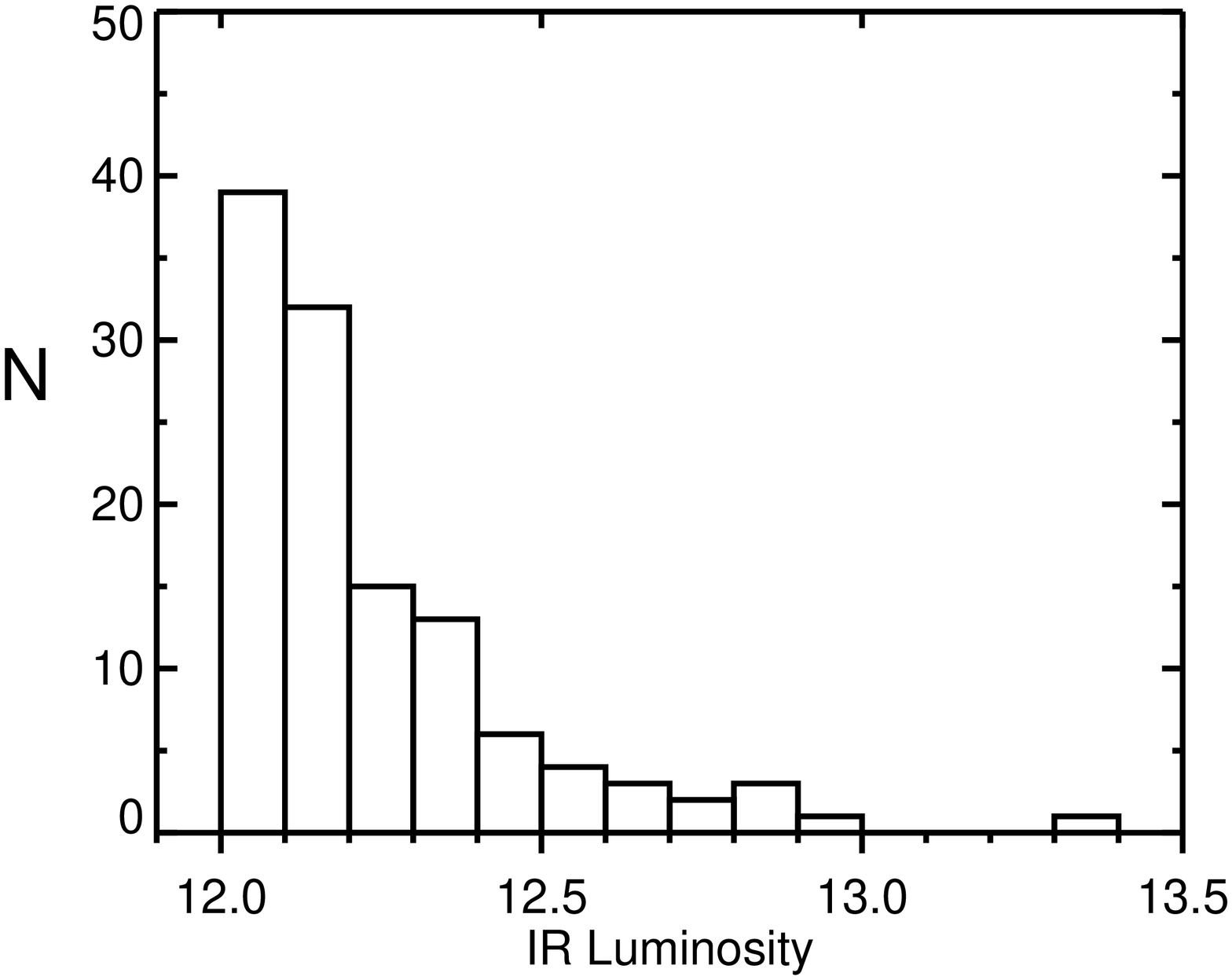}\\
\end{array}$
\end{center} 
\caption{Distributions of redshift (top) and IR luminosity, $\log(L_{IR}/L_{\odot})$, (bottom) for the final (H)/ULIRG sample.} 
\label{fig:zlhist}
\end{figure}

The redshift and IR luminosity distributions of our final sample are presented in the top and bottom panels of Figure \ref{fig:zlhist}, respectively. 
The redshift distribution covers 0.050 $ < z <$ 0.487, with a median redshift of $\bar{z}=0.181$. 
We have 104 ULIRGs distributed over 0.050 $<z\le0.270$ range and 14 ULIRGs are within the 0.270 $< z <$ 0.487 range. 
The IR luminosity distribution of 80 ULIRGs covers $12.0\le L_{IR}\le12.25$ range. 
The higher luminosity range of 12.25 $<L_{IR}\le$ 12.91 includes 38 ULIRGs. 

%FIG 2

\begin{figure} 
\begin{center}$
\begin{array}{c}
\includegraphics[scale=0.3]{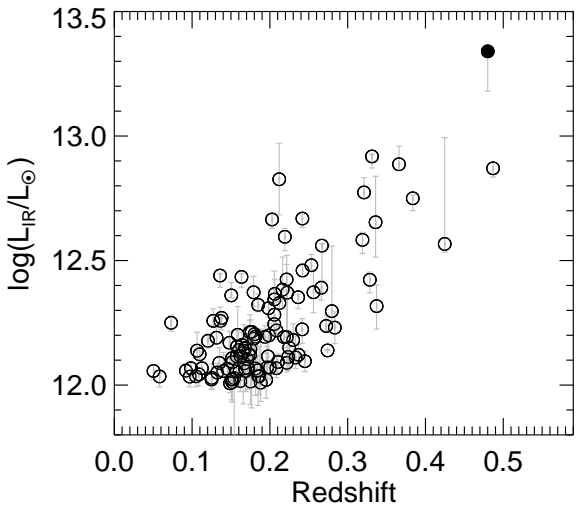}\\
\end{array}$
\end{center}
\caption{IR luminosity vs. redshift for 118 ULIRGs (open circles) and one HLIRG (filled circle) in the final sample.}
\label{fig:zvsl}
\end{figure}

Figure \ref{fig:zvsl} shows the IR luminosity of our sample as a function of redshift. 
As it is expected from the \textit{AKARI} PSC detection limit (0.55 Jy at 90\micron), $L_{IR}$ increases with redshift and only the bright sources can be detected towards the higher redshifts. 

%FIG 3

\subsubsection{FIR Color Properties of Our Sample}\label{S:colprop}

The IR emission of the so called `normal' star forming galaxies (that are not dominated by AGN activity) is mostly due to the thermal radiation from dust grains heated by star formation. 
The `normal' star forming galaxies detected by \textit{IRAS} showed a clear trend of decreasing 60- to 100-\micron\ flux ratios,
 F(60\micron)/F(100\micron), with increasing 12- to 25-\micron\ flux ratios, F(12\micron)/F(25\micron), \citep{Helou1986}. 
This trend associated with  the intensity dependence of IR colors,  such that `warm' colors 
(greater F(60\micron)/F(100\micron) values) are related to active star formation with high IR luminosities \citep{Helou1986}. 

\citet{Dale2001} construct single parameter dust models of normal star forming galaxies  based on the F(60\micron)/F(100\micron) color and the 
intensity of the interstellar radiation field, $U$. 
They  characterize the overall IR SED as a power law distribution of dust mass over $U$ such that: $dM(U)\sim U^{-\alpha} dU$, where $\alpha$ is 
the exponent of the power-law distribution. 
\citet{Dale2002} provide 64 SED models for a wide range of $U$ or equivalently \textit{IRAS} F(60\micron)/F(100\micron) color
(between -0.54 and 0.21) and $\alpha$ values (0.0625$\le \alpha \le$ 4.0). 
In \S \ref{S:SEDs} we measured $L_{IR}$ based on these models. 
In the following, we investigate the \textit{AKARI} color properties of our ULIRG sample and compare the observed colors with the SED models of \citet{Dale2002}. 

\begin{figure*} 
\begin{center}$
\begin{array}{cc}
\includegraphics[scale=0.3]{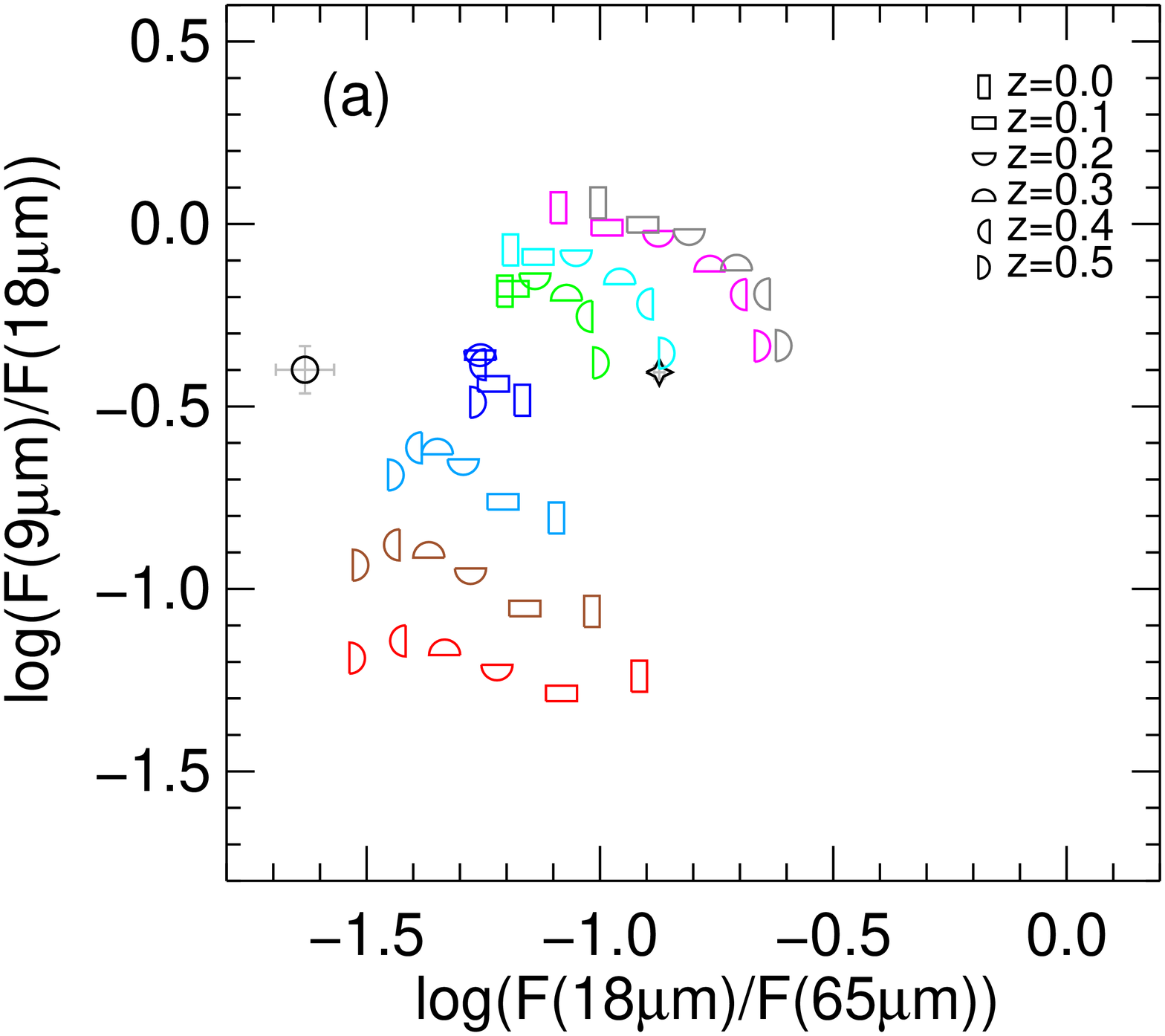}&
\includegraphics[scale=0.3]{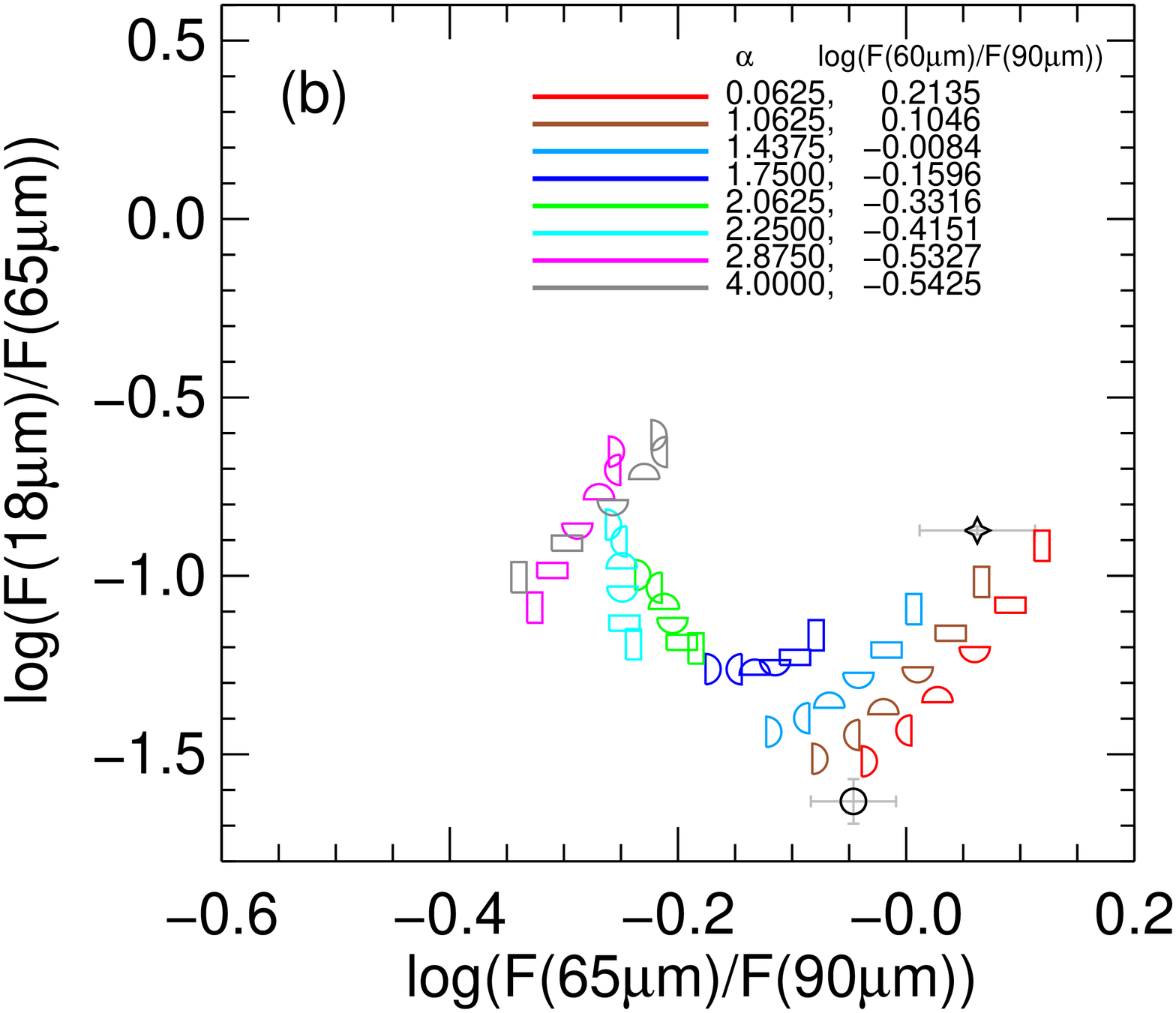}\\
\includegraphics[scale=0.3]{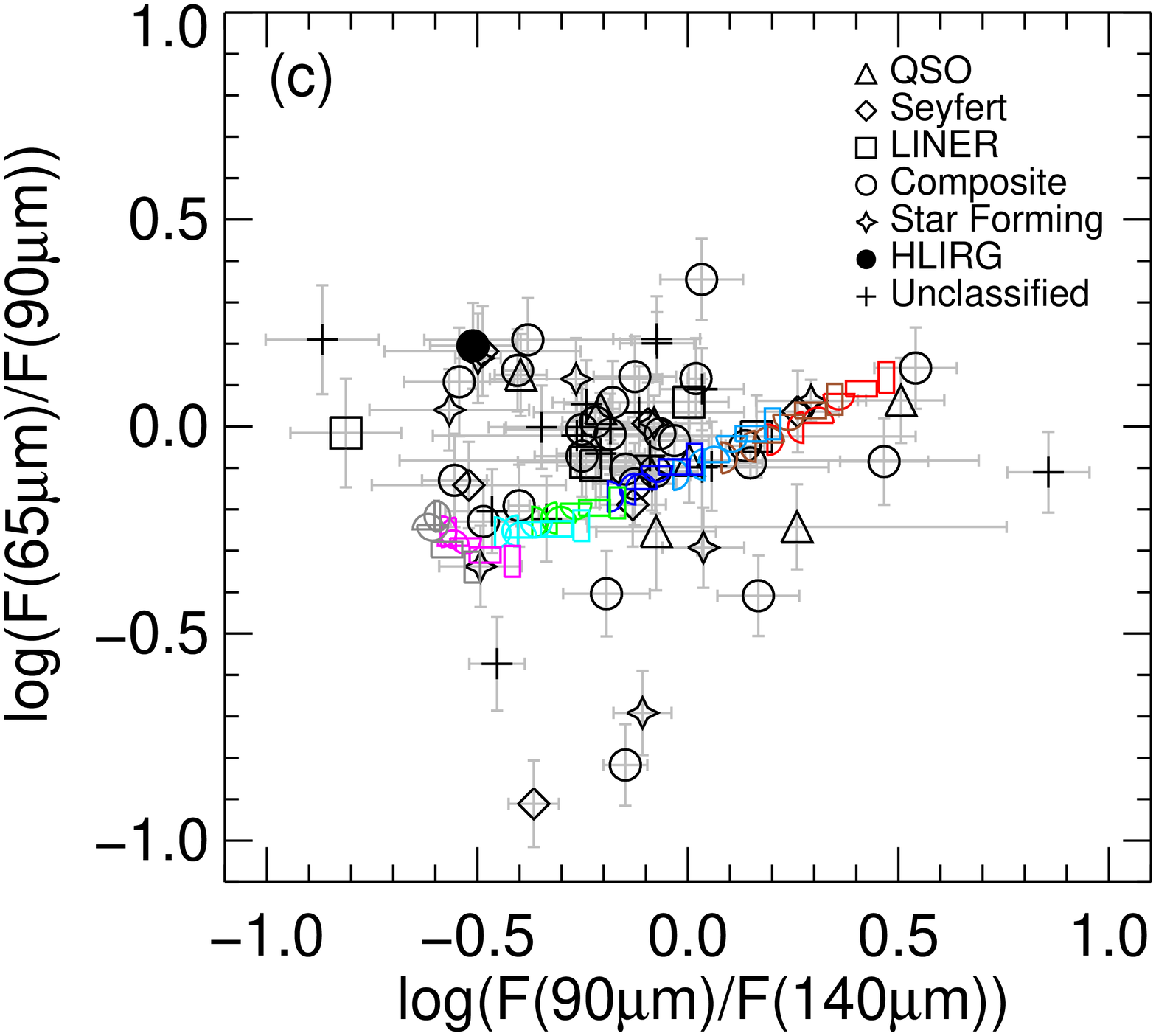} & 
\includegraphics[scale=0.3]{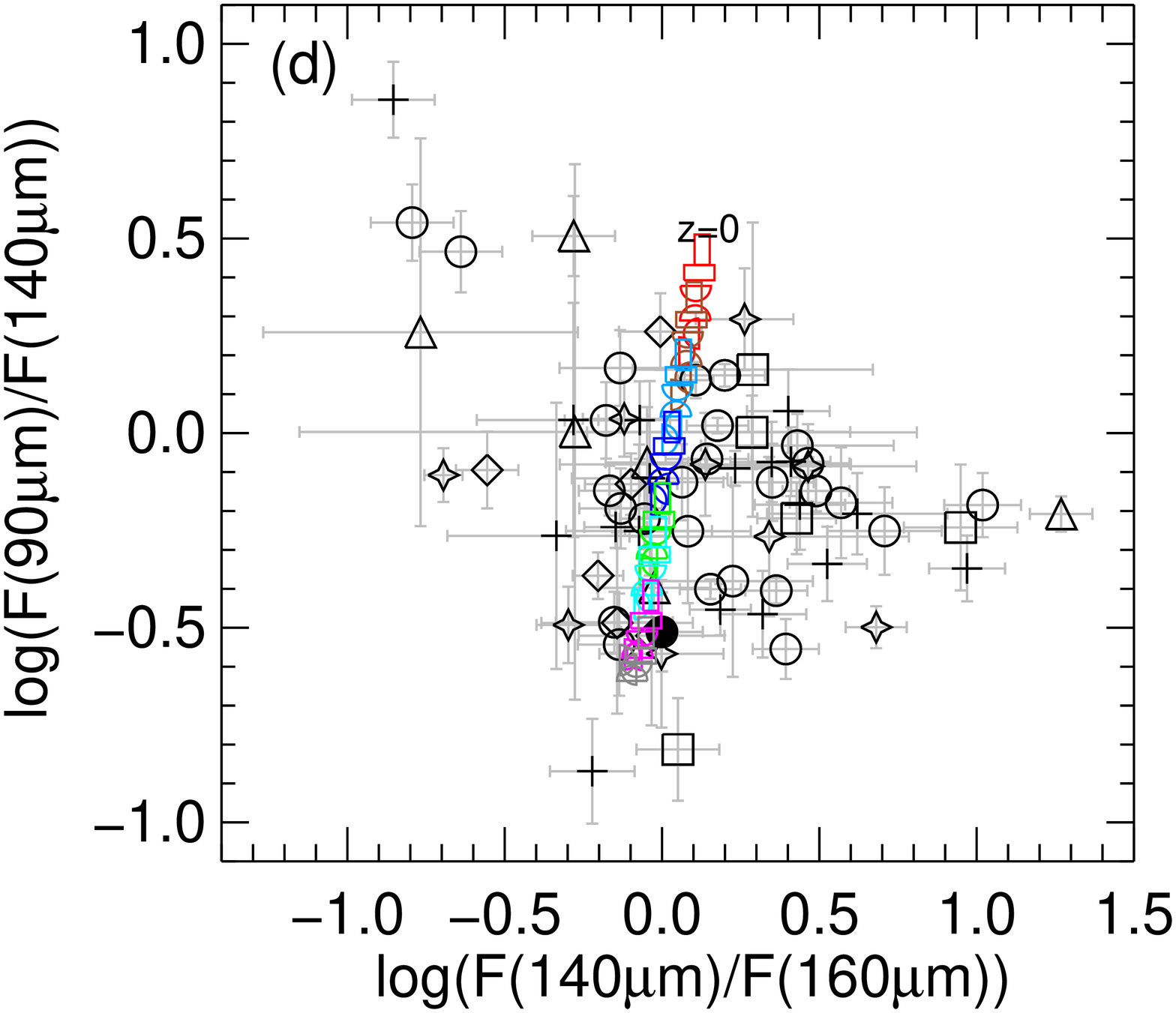}
\end{array}$
\end{center}
\caption{The \textit{AKARI} color-color diagrams. 
The top left (a) and right (b) panels show $\log(F(9\micron)/F(18\micron))$ versus $\log(F(18\micron)/F(65\micron))$ 
and $\log(F(18\micron)/F(65\micron))$ vs.  $\log(F(65\micron)/F(90\micron))$ for the two ULIRGs detected in all \textit{AKARI} bands. 
The bottom panels show $\log(F(65\micron)/F(90\micron))$ versus $\log(F(90\micron)/F(140\micron))$ (c)  and 
$\log(F(90\micron)/F(140\micron))$  versus $\log(F(140\micron)/F(160\micron))$ (d) colors for 71 sources that are detected in \textit{AKARI}  65\micron, 90\micron, 140\micron, and 160\micron\ bands. 
Symbol code is given in the legend of panel (c). 
The colored symbols in each panel indicate the expected colors from the SED templates of \citet{Dale2002} at redshifts $z=$0.0, 0.1, 0.2, 0.3, 0.4, 0.5. 
Symbol code of the redshifts are given in the legend of panel (a). 
We only show the expected colors for eight SED templates, 
different colors represent different models; parameters of the models are given in the legend of panel (b).}
\label{fig:col-col}
\end{figure*}

For this investigation we use the \textit{AKARI} F(9\micron) and F(18\micron) fluxes from the \textit{AKARI}/IRC all-sky survey point source catalog. 
The F(65\micron), F(90\micron), F(140\micron) and F(160\micron) fluxes are listed in Tables \ref{tab:newULIRGs}, \ref{tab:knownULIRGs} and \ref{tab:newHLIRG}. 
Figure \ref{fig:col-col} presents the observed \textit{AKARI} color-color diagrams: 
(a) F(9\micron)/F(18\micron) vs. F(18micron)/F(65\micron); 
(b) F(18\micron)/F(65\micron) vs. F(65\micron)/F(90\micron); 
(c) F(65\micron)/F(90\micron) vs. F(90\micron)/F(140\micron); 
(d) F(90\micron)/F(140\micron) vs. F(140\micron)/F(160\micron). 
Panels (a) and (b) show only the two sources that are detected in all \textit{AKARI} bands. 
Panels (c) and (d) include 71 sources that are detected in all \textit{AKARI} FIS bands.  
Different symbols represent spectral class as listed in Tables \ref{tab:newULIRGs}, \ref{tab:knownULIRGs} and \ref{tab:newHLIRG} (see \S \ref{S:classprop}): 
circle (composite), star (star forming), square (LINER), diamond (Seyfert\footnote[10]{Seyfert galaxies are low-redshift ($z \le$0.1), less luminous cousins of quasars \citep[e.g.,][]{Richards2002}}), triangle (QSOs), plus (unclassified).
In panels (c) and (d) the FIR colors of different class of galaxies distribute over the entire color range. 
Therefore, AGNs or star forming galaxies can not be distinguished by their FIR colors. 
However, this is expected because FIR is tracing star formation activity at low temperature dust and is not sensitive to AGN activity. 
It is known that \textit{IRAS}  mid-IR F(25\micron)/F(60\micron) color is an indicator of `warm' dust and AGN component (F(25\micron)/F(60\micron) $\ge$ 0.2) \citep[e.g.][]{Sanders1988}. 
Unfortunately, the majority of our ULIRG sample are not detected in the mid-IR colors and therefore we do not have enough data to explore the mid-IR color properties of our sample.  

%FIG 4 

The color-color diagrams in panels (c) and (d)  do not show a clear correlation.  
It is important to keep in mind that detection limits of  \textit{AKARI}  bands affect the shape of the color-color diagrams. 
Detection limits of \textit{AKARI} FIS bands are 3.2 Jy, 0.55 Jy, 3.8 Jy and 7.5 Jy for 65\micron, 90\micron, 140\micron\ and 160\micron\ bands \citep{Yamamura2010}, respectively. 
The WIDE-S filter centered at 90\micron\ is the broadest and therefore it has the deepest detection limit compared to other bands. 
In panel (d) the distribution of the colors is shaped by the observational detection limits. 
This is mainly because 140\micron\ is common in both axes. 
In the x-axis as 140\micron\ flux gets brighter the $\log(F(140\micron)/F(160\micron))$ color moves towards right, but at the same time 
in the y-axis the $\log(F(90\micron)/F(140\micron))$ color moves downwards. 
This behavior creates a boundary on the top right corner of this diagram.   
Even if there was an intrinsic color-color correlation in panel (d) it would be truncated on the upper right corner due to observational limits. 
We expect to have a similar detection limit effect in panel (c) because 65\micron\ and 140\micron\ detection limits are brighter than 90\micron, and this may cause 
colors to hit the boundaries of 65\micron\ and 140\micron\ before the limit of 90\micron. 

In Figure \ref{fig:col-col} the colored symbols in each panel show the expected colors by the IR SED models of \citet{Dale2002}. 
We choose eight SEDs with different $\alpha$ and F(60\micron)/F(100\micron) values among 64. 
The selected models have a sequence in terms of $\alpha$ and $\log$(F(60\micron)/F(100\micron)): 0.06 $\le \alpha \le$ 4.0 and  -0.54 $\le \log$(F(60\micron)/F(100\micron)) $\le$ 0.21. 
The expected colors from the selected SEDs are shown with different colors; the  $\alpha$  and $\log$(F(60\micron)/F(100\micron)) parameters of each model is given in top corner of panel (b). 
We show colors expected from each model as a function of redshift from 0 to 0.5 in order to illustrate the redshift dependence of the colors. 
The symbol code for $z=$0.0, 0.1, 0.2, 0.3, 0.4, 0.5 is given the legend of panel (a).  
In panels (c) and (d) the data show a large spread around the model colors. 
In panel (c), the large vertical and horizontal spreads of  $\log$(F(65\micron)/F(90\micron)) and  $\log$(F(90\micron)/F(140\micron)) colors around the models are mainly due to the limited parameter coverage of the SED models. 
The models cover the ranges between -0.54$-$0.21 and -0.62$-$0.47 in the y- and x-axis, respectively. 
Therefore, the models do not overlap with the colors exceeding this range. 
In panel (d), especially the $\log$(F(140\micron)/F(160\micron)) colors have a large scatter around the models, the models covers only the $-0.11 - 0.13$ range while the observed colors can exceed up to 1.27. 
We do not see a clear trend in $\log$(F(140\micron)/F(160\micron)) colors with redshift. 
Since we show the expected colors for $z=$0.0$-$0.5, the observed scatter does not seem to be due to the range in redshift of our sample. 
We further discuss such outliers in color-color diagrams in \S\ \ref{S:dis}. 

Since the \textit{AKARI}  $\log$(F(65\micron)/F(90\micron)) color is equivalent to \textit{IRAS} $\log$(F(60\micron)/F(100\micron)) color, 
it is possible to make a comparison of \textit{AKARI} and \textit{IRAS} color distributions of ULIRGs. 
\citet{Hwang2007} investigate \textit{IRAS} colors of 324 ULIRGs and report $\log$(F(60\micron)/F(100\micron)) within -0.80$-$ 0.22 range with a mean of -0.19. 
The ULIRGs in our sample have a slightly larger range between -0.91$-$0.36, but still the \textit{AKARI}  $\log$(F(65\micron)/F(90\micron)) colors overlap with the \textit{IRAS} $\log$(F(60\micron)/F(100\micron)) colors. 

\subsubsection{FIR Colors versus IR Luminosity}\label{S:colprop}

IR bright galaxies ($10^{9.5}L_{\odot} \le L_{IR} < 10^{13}L_{\odot}$) studied with \textit{IRAS} show a correlation between the IR colors and the IR luminosity: 
$\log$(F(12\micron)/F(25\micron)) color decreases, and $\log$(F(60\micron)/F(100\micron)) color increases with increasing IR luminosity \citep{Soifer1991}. 
As stated in \S \ref{S:colprop} $\log$(F(60\micron)/F(100\micron)) color is related to the intensity of the radiation field. 
The SED models of \citet{Dale2002} cover a wide range of \textit{IRAS} $\log$(F(60\micron)/F(100\micron)) colors that correlate with $L_{IR}$; 
higher $\log$(F(60\micron)/F(100\micron)) colors indicate higher $L_{IR}$ \citep{Dale2002}. 
In the following we investigate the color dependence of the IR luminosities for our (H)/ULIRG sample. 

\begin{figure*} 
\begin{center}$
\begin{array}{cc}
\includegraphics[scale=0.26]{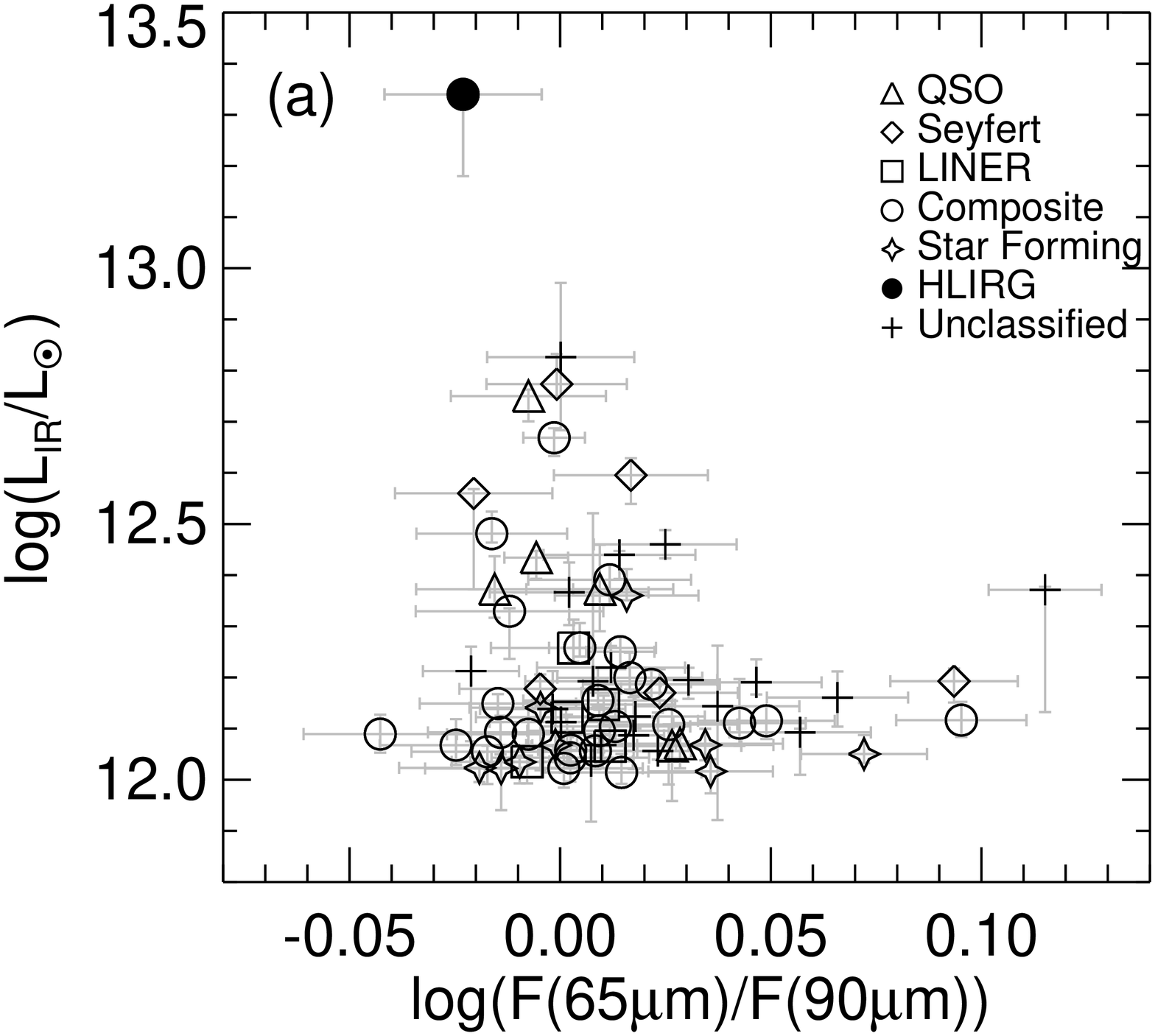} & 
\includegraphics[scale=0.26]{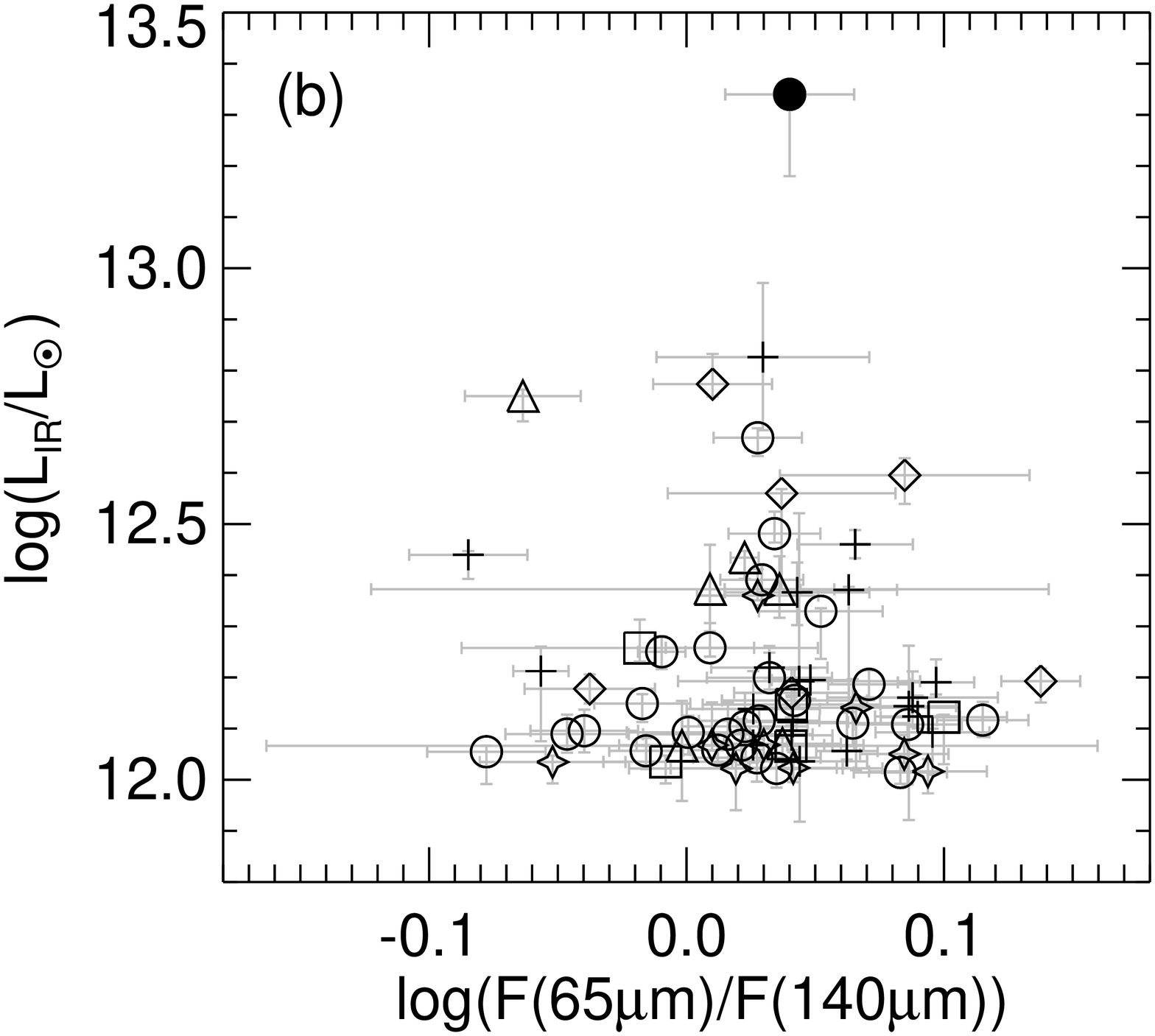}\\
\includegraphics[scale=0.26]{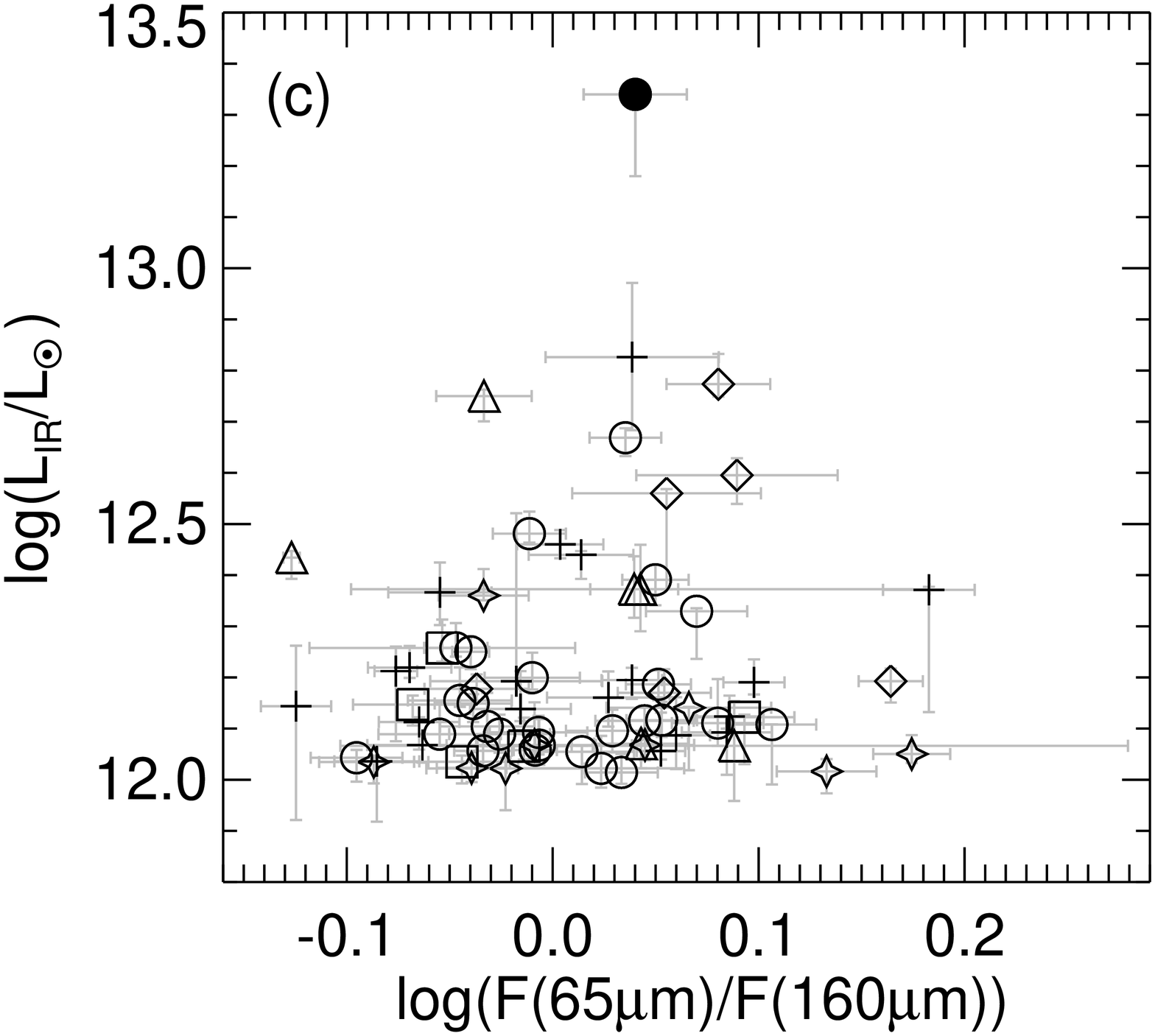} & 
\includegraphics[scale=0.26]{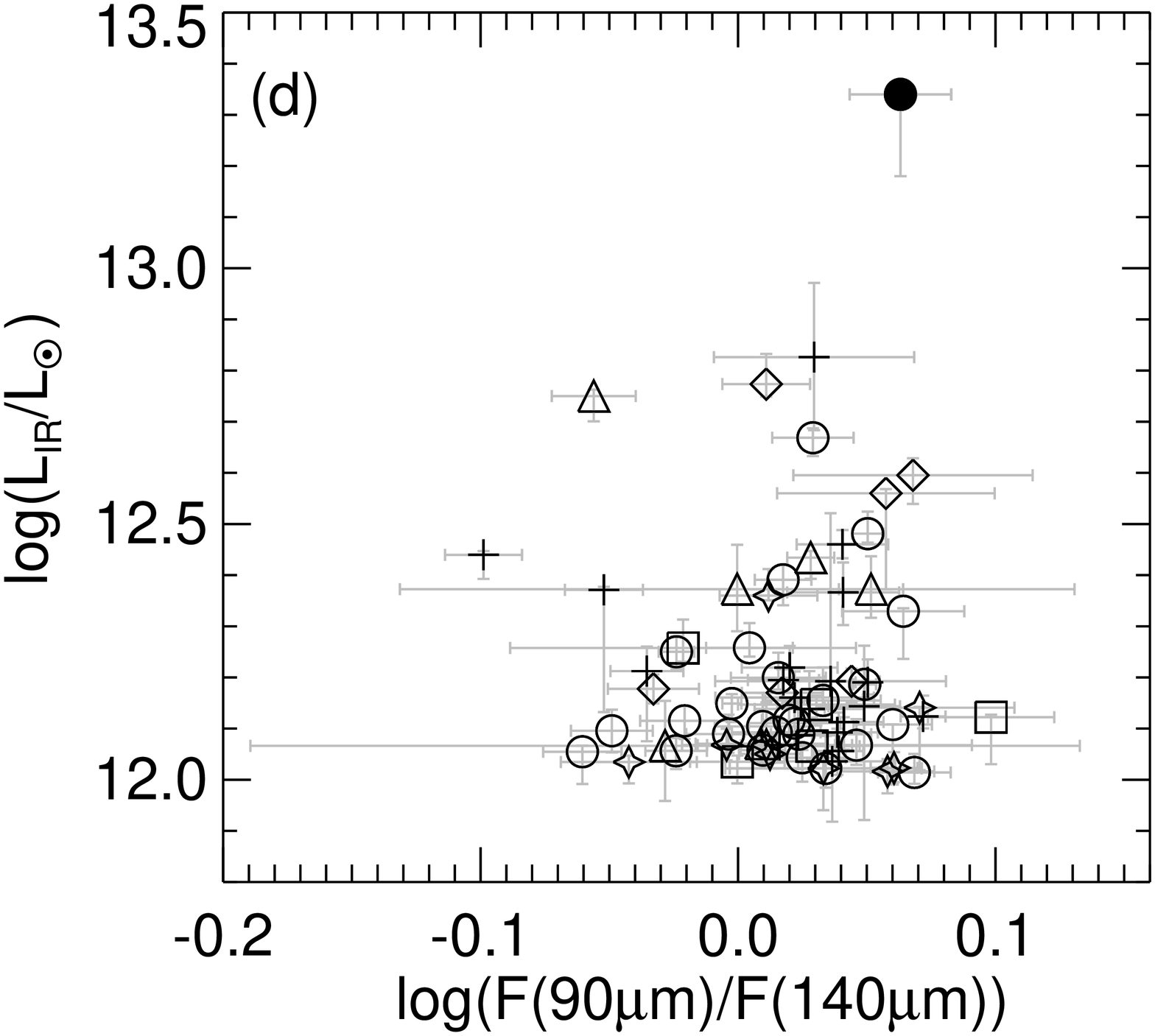}\\
\includegraphics[scale=0.26]{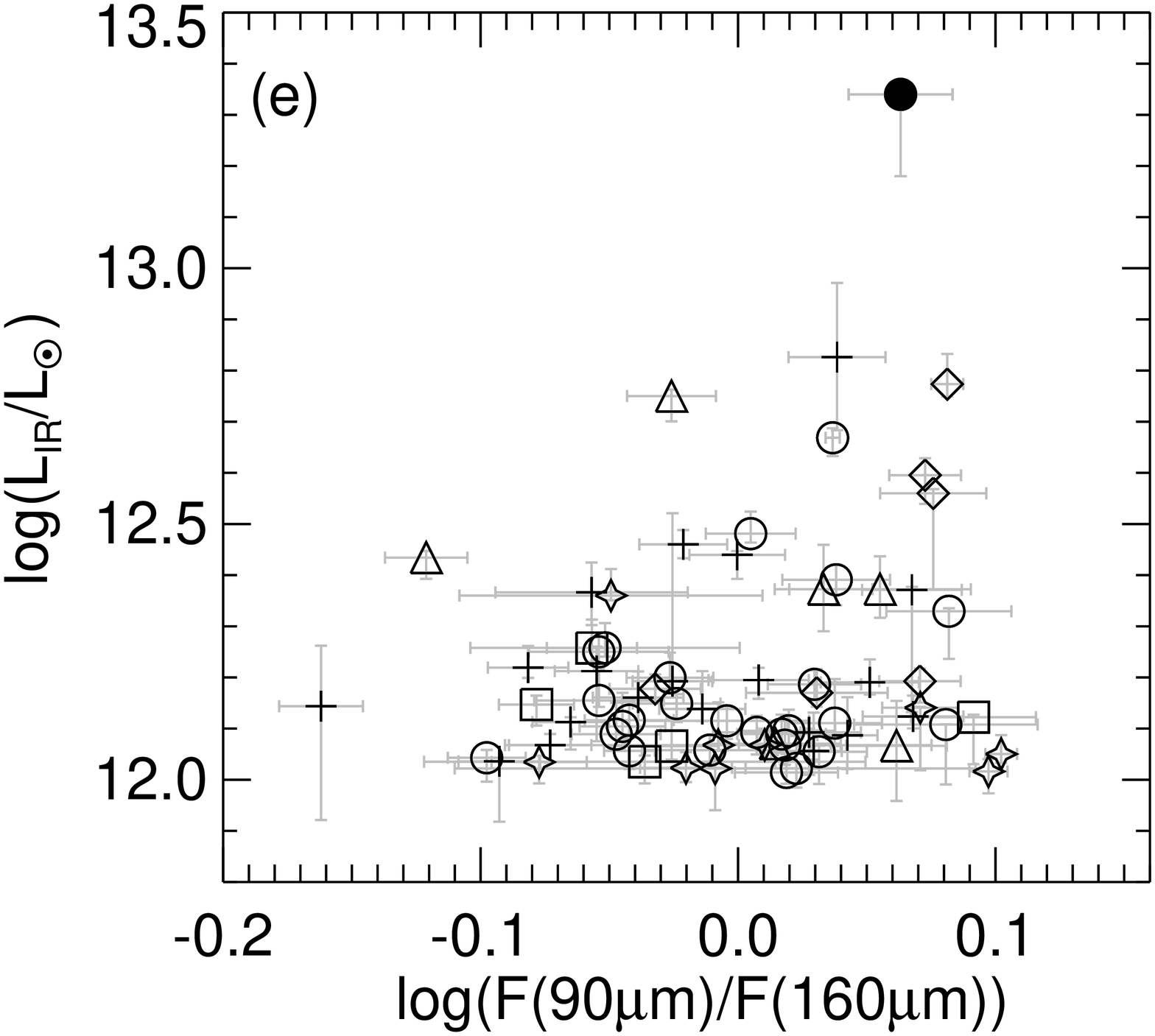} & 
\includegraphics[scale=0.26]{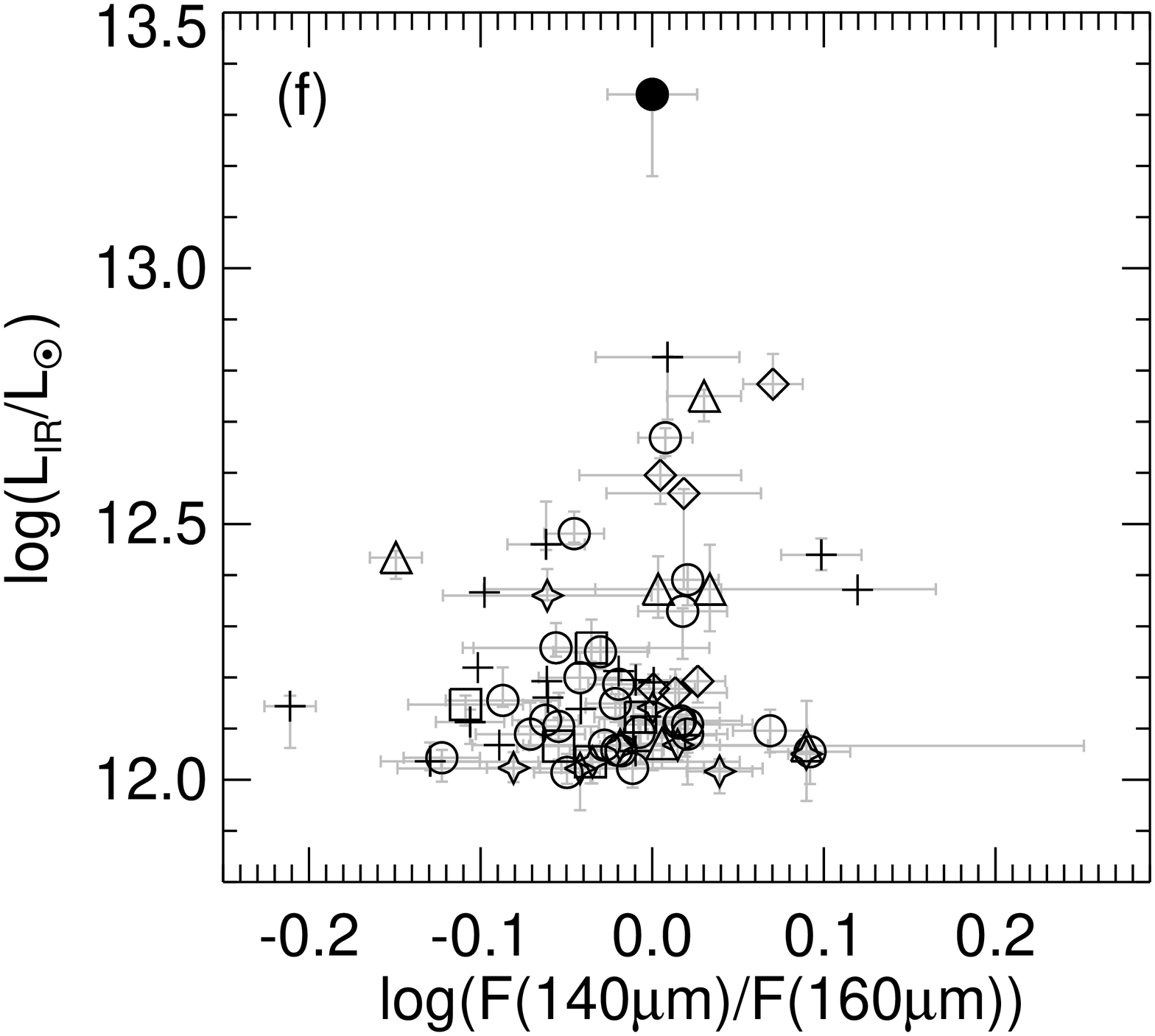}\\
\end{array}$
\end{center}
\caption{The color $-$ luminosity diagrams of 71 (H)/ULIRGs that are detected in all {\it{AKARI}} FIS bands. Symbol code is given in panel (a).}
\label{fig:col-lir}
\end{figure*}

Figure \ref{fig:col-lir} presents IR luminosity versus: (a) $\log$(F(65\micron)/F(90\micron)), (b) $\log$(F(65\micron)/F(140\micron)), (c) 
$\log$(F(65\micron)/F(160\micron)), (d) $\log$(F(90\micron)/F(140\micron)), 
(e) $\log$(F(90\micron)/F(160\micron)), \\*
(f) $\log$(F(140\micron)/F(160\micron)),  
for 71 sources that are detected in all \textit{AKARI} FIS bands. 
As noted before, since we have only very few sources detected in 9- and 18-\micron\ bands we do not include those in this investigation. 
Since the observed colors change as a function of redshift and luminosity depends on redshift, we apply k-correction to the \textit{AKARI} FIS colors shown in Figure \ref{fig:col-lir}. 

%FIG 5

As noted before \textit{AKARI} $\log$(F(65\micron)/F(90\micron)) color is similar to \textit{IRAS} $\log$(F(60\micron)/F(100\micron)) color and therefore, we would expect a strong correlation in panel (a). 
However, none of the \textit{AKARI} colors in Figure \ref{fig:col-lir} show a clear dependence in $L_{IR}$ between 12.0 $\le \log(L_{IR}/L_{\odot}) <$ 13.3. 
Since the interested $L_{IR}$ range in this study is very narrow compared to the $L_{IR}$ range probed in previous studies \citep[e.g][]{Soifer1991} it is natural not to see the previously discovered significant color$-L_{IR}$correlations. 
The representative SED models shown in Figure \ref{fig:col-col} show a luminosity dependence with color, but the observed colors show a large scatter around the models (discussed in \S \ref{S:colprop} and \S \ref{S:dis}). 
The large differences between the SED models and the observed colors weakens the color$-L_{IR}$correlation expectation.  

In Figure \ref{fig:col-lir}, apart from the color dependence of IR luminosity, different galaxy types do not show a significant dependence on color.

\subsection{The Visual Morphologies and Interaction Classes}\label{S:morphprop}

Morphological studies of local ULIRGs showed that they are mostly interacting galaxies showing tidal features or disturbed morphology \citep[e.g][]{Farrah2001,Veilleux2002,Veilleux2006}. 
\citet{Surace1998} introduced an interaction classification scheme based on the evolution sequence that merging galaxies follow in simulations \citep[e.g][]{Mihos1996}. 
Such an interaction classification scheme is important to interpret the morphological properties of ULIRGs in the context of galaxy evolution triggered by mergers.
\citet{Veilleux2002} classified 117 local ULIRGs based on this scheme and showed that ULIRGs are interacting or advanced merger systems. 

Here we investigate the morphological properties of our sample with the aim of identifying interaction classes. 
We use the following widely preferred classification scheme that is described by \citet{Veilleux2002} : 
\begin{itemize}
\item I: First approach. Separated galaxies with no signs of interaction or merging. 
\item II: First contact. Overlapped discs without interaction signs. 
\item III: Pre-mergers. Two nuclei separated by more than 10 kpc (a; wide binary) or less than 10 kpc (b; close binary), with interaction signs.
\item Tp1: Interacting triplet system.
\item IV: Merger. One nucleus with prominent tidal features.
\item V: Old merger.  Disturbed central morphology without clear tidal tail signs. 
\item NI: Non interacting. Isolated single galaxy, no signs of disturbed morphology.
\end{itemize}
Note that we added class NI to represent isolated single galaxies showing  no signs of disturbed morphology. 
Also note that we do not subdivide class IV into two as done by \citet{Veilleux2002} because, we do not have $K$ band luminosities. 

We (two classifiers: EKE and TG) examined SDSS $g$-$r$-$i$ colors combined images and classified only the galaxies for which SDSS images are available. 
For the known ULIRGs, we adopt the interaction classifications from the literature. 
We prefer to adopt the classifications mainly from \citet{Veilleux2002} and \citet{Hwang2007}. 
The interaction classifications of the galaxies in our sample are given in column (12) of Tables \ref{tab:newULIRGs}, \ref{tab:knownULIRGs} and \ref{tab:newHLIRG}. 
The references for the interaction classes are given in the column (13) of Tables \ref{tab:newULIRGs}, \ref{tab:knownULIRGs} and \ref{tab:newHLIRG}.
In additional to the above interaction classification we also note if the galaxies are in a group with (G). 
We define groups as galaxy systems with more than two members with similar colors. We note that our group definition is subjective and the group classification given in this work is only for guidance. 
SDSS images showing examples of different interaction classes are represented in Figures \ref{fig:sedsimages1}. 

\begin{figure} 
\begin{center}$
\begin{array}{c}
\includegraphics[scale=0.3]{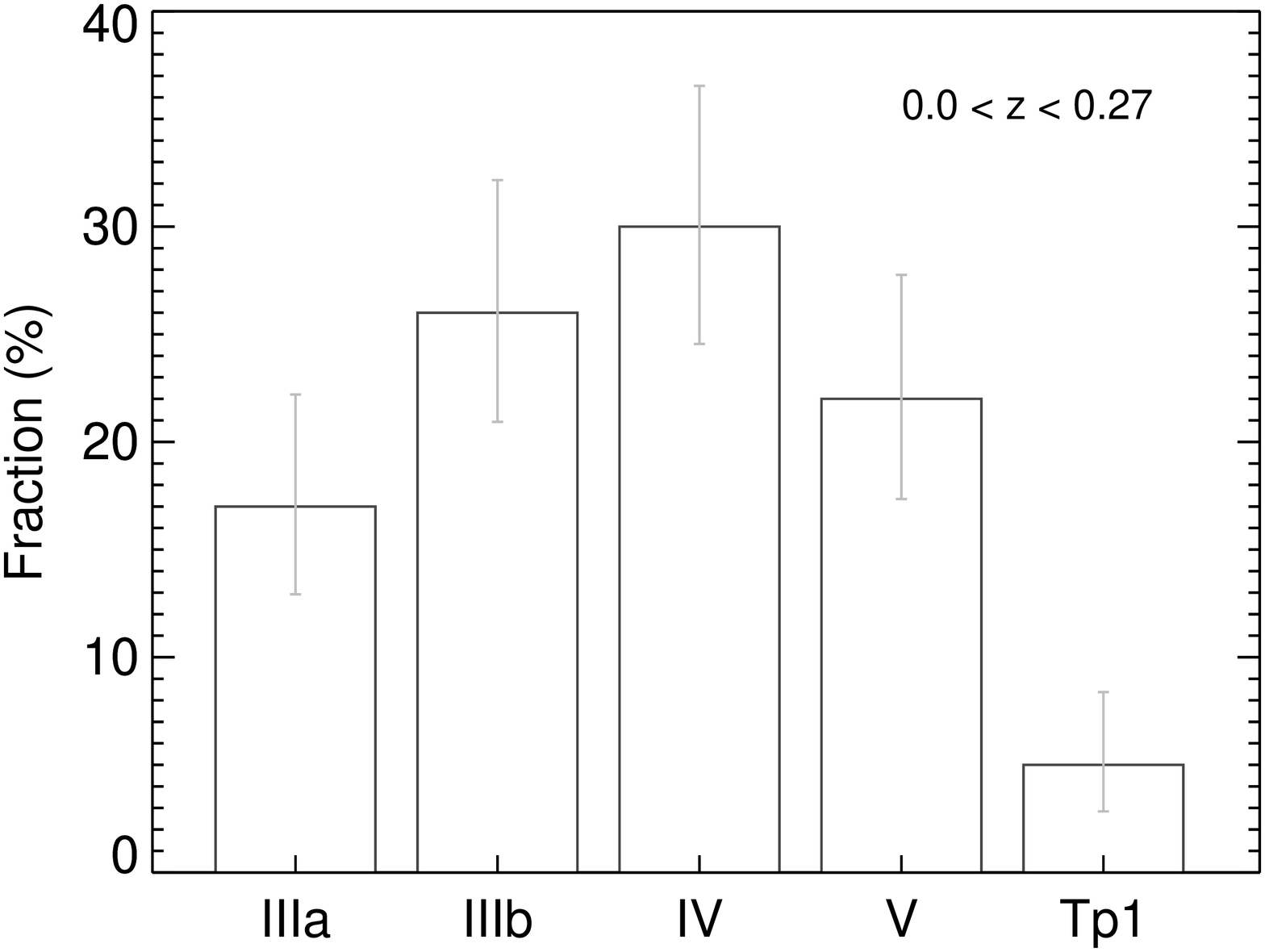}\\
\end{array}$
\end{center}
\caption{The distribution of the interaction classes for 100 ULIRGs within 0.0$<$z$<$0.27 limit. 
Interaction classes are described in \S\ \ref{S:morphprop}. The fraction of the late/old mergers that are classified as IV or V is 52\%. 
Error bars represent the 1$\sigma$ Poisson errors \citep{Gehrels1986}.}
\label{fig:morphst}
\end{figure}

As shown in Figure \ref{fig:zvsl} luminosity is correlated with distance and it becomes more difficult to identify the morphological details for more distant sources. 
To avoid uncertainties in interaction classifications due to the distances, in the following analysis we focus on a redshift limited sample of 100 ULIRGs. 
For comparison reasons we apply a redshift cut as $z$=0.27; this is the limit of the \citet{Veilleux2002} ULIRG sample. 
The distribution of interaction classes of 100 ULIRGs is shown in Figure \ref{fig:morphst}. 
This Figure presents the percentage of different interaction classes. 
There are no ULIRGs classified as I and II, so they are not in an early interaction phase. 
The fraction of triplets (Tp1) in our sample is very small (5\%).
The faction of binary systems showing strong interaction features (IIIa and IIIb) is 43\%. 
Most of the ULIRGs (52\%) are single nucleus galaxies classified as IV and V indicating a late/post merger phase.   
\citet{Veilleux2002} study 117 ULIRGs from \textit{IRAS} 1 Jy sample \citep{Kim1998} and report 56\% of the sample as single nucleus systems at a late merger stage. 
The fraction of such systems (IV or V) in our sample is 52\% and this is a consistent result with \citet{Veilleux2002}. 
35 of the 100 ULIRGs in our morphology subsample are also part of the sample of 117 ULIRGs studied by \citet{Veilleux2002}, therefore our results can be considered as independent from those derived by \citet{Veilleux2002}. 
We also find 11\% of the ULIRGs to be in a group environment. 

%FIG 6 

\begin{figure} 
\begin{center}$
\begin{array}{c}
\includegraphics[scale=0.3]{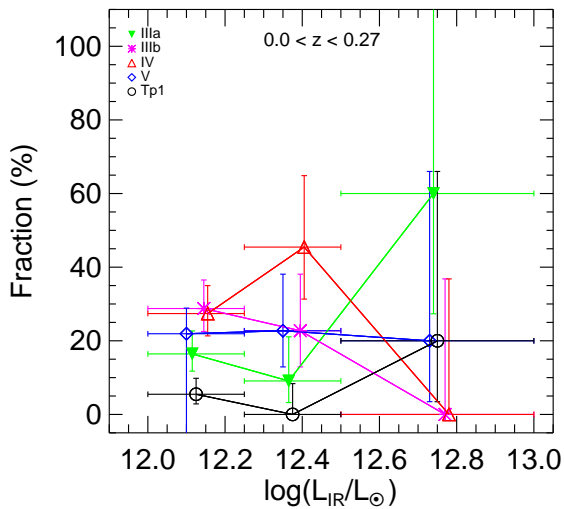}\\
\end{array}$
\end{center}
\caption{Fraction of interaction classes per IR luminosity for 100 ULIRGs within $z<0.27$ limit. 
X-axis error bars represent the range of the IR luminosity bins (12.0$\le L_{IR}<$12.25, 12.25$\le L_{IR}<$12.5, 12.5$\le L_{IR}<$13.0). 
The y-axis error bars represent the 1$\sigma$ confidence limits of the Poisson errors on the counts given by  \citet{Gehrels1986}.}
\label{fig:morLir}
\end{figure}

In Figure \ref{fig:morLir} we show the fraction of ULIRGs in different interaction classes as a function of IR luminosity. 
We divide IR luminosities into three bins (12.0$\le \log(L_{IR}) <$ 12.25, 12.25$\le \log(L_{IR}) <$ 12.5, 12.5$\le \log(L_{IR})$); the number of sources in each bin is 73, 22 and 5, respectively.  
This Figure shows a hint for a negative trend for pre-mergers (IIIa and IIIb). 
The fraction of galaxies classified as IIIa and IIIb decreases from the first bin to the second, but IIIa galaxies increases in the highest $L_{IR}$ bin. 
The fraction of mergers (IV) increases from the first bin to the second, but decreases in third bin. 
The fraction of old mergers (V) appear to be almost constant with luminosity. 
The fraction of triplets is constant in the first two bins, but increases in the third bin. 
The fractions in the highest luminosity bin is highly uncertain due to the very small number of sources. 
Therefore, we consider the trends including the the highest luminosity bin as unreliable. 
If we only take into account the first two luminosity bins then, 
it is clear that the fraction of pre-mergers have a negative trend while the mergers have a positive trend with increasing luminosity. 
This is a consistent result with \citet{Veilleux2002} who find a positive trend with the fraction of advanced mergers and IR luminosity. 
 
 %FIG 7 

The morphological properties of our sample confirms that ULIRGs are mostly either in pre-merger two galaxy systems or single galaxies in late/post merger phase. 
This is a consistent picture with the general idea that ULIRGs are triggered via strong interactions between galaxies. 
 
\subsection{The Spectral Classification of Our Sample}\label{S:classprop}

The power sources of ULIRGs are high rates of star formation and AGN activity \citep[e.g][]{Nardini2010}. 
The traces of the dominant power source can be detected in optical spectra. 
Properties of the emission lines provide a practical tool to uncover the source of the ionization producing those lines. 
To identify the spectral classes of the ULIRGs in our sample we make use of the available SDSS catalogs providing such a classification. 
SDSS spectroscopic pipeline classify objects as `broad-line AGNs'/`quasars', `galaxies' or 'stars'. 
We adopt this classification to identify the quasars in our sample. 

\citet{Thomas2013} investigate emission line properties of SDSS sources that are already classified as `galaxies' through the pipeline. 
They apply Baldwin-Phillips-Terlevich (BPT) diagnostics \citep{Baldwin1981} 
based on [O\,{\sc iii}]/\Hbeta\ and [N\,{\sc ii}]/\Halpha\ emission line ratios to classify sources into: Seyfert, Low-Ionization Nuclear Emission Region (LINER), Star Forming Galaxy and 
Star Forming/AGN composite. 
\citet{Thomas2013} use the empirical separation between AGN and star forming galaxies according to \citet{Kauffmann2003b}, and they use the separation line defined by \citet{Schawinski2007} to select LINERs. 
We adopt the spectral classification given by \citet{Thomas2013} for the ULIRGs included in their galaxy sample. 
Some of the ULIRGs in our sample are not included in the sample of \citet{Thomas2013}. 
These are mostly AGNs, but a few sources classified as `broad-line AGNs starbursts' by the spectroscopic pipeline.
To classify such sources we adopt the  available line flux measurements in the SDSS database (see the footnotes of Table \ref{tab:newULIRGs} for the SDSS references) and use 
a similar line diagnostic diagram as described by \citet{Thomas2013}. 
The spectral classes listed in column (15) of Tables \ref{tab:newULIRGs}, \ref{tab:knownULIRGs} and \ref{tab:newHLIRG}. 
The spectral classes marked with a star are obtained in this work. 

\begin{figure} 
\begin{center}$
\begin{array}{c}
\includegraphics[scale=0.3]{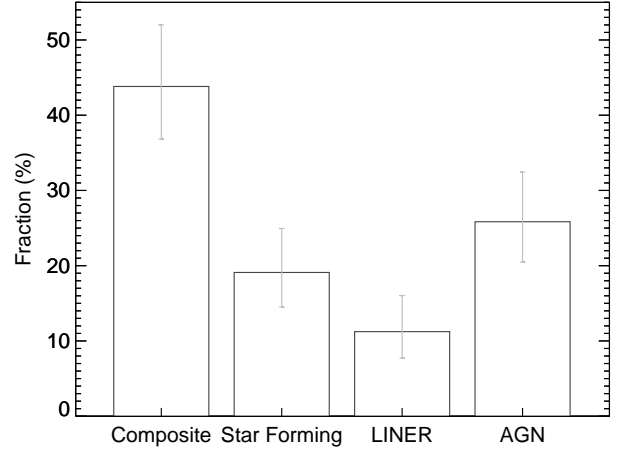}\\
\end{array}$
\end{center}
\caption{The distribution of spectral classes of 89 ULIRGs for which the SDSS spectra are available.
Error bars represent the same quantity as in Figure \ref{fig:morphst}.}
\label{fig:classhst}
\end{figure}

The distribution of the spectral classes is shown in Figure \ref{fig:classhst}; it represents only 89 sources ULIRGs for which SDSS spectra are available. 
The fraction of purely star forming galaxies is 19\%.  
The fraction of composite galaxies in our sample is 44\%. 
The fraction of LINERs in our sample is 11\%. 
LINERs are thought to be powered by AGNs \citep[e.g][]{Nagar2005}, however other power sources can also produce LINER-like emission \citep[e.g][]{Maoz1998,Sarzi2010}. 
Since there is a debate whether LINERs are low-luminosity AGNs or a separate class of objects, to be conservative in this work we separate LINERs from AGNs. 
The faction of AGN (QSOs and Seyferts) ULIRGs in our sample is 26\%.

%FIG 8 

Most of the ULIRGs in our sample are classified as composite galaxies. 
It is important to note that these are star forming galaxies possibly with a hidden AGN component. 
To be conservative we do not include composites to AGNs. 
Since both LINERs and composites may harbor an AGN, the given AGN fraction is only a lower limit. 

\begin{figure} 
\begin{center}$
\begin{array}{c}
\includegraphics[scale=0.3]{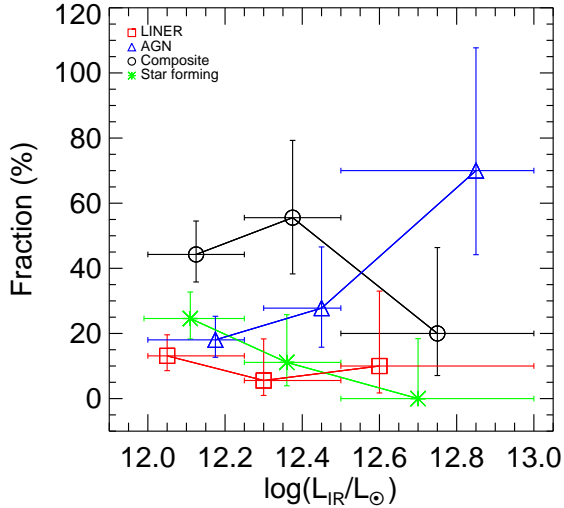}\\
\end{array}$
\end{center}
\caption{Spectral class fraction per IR luminosity bins for 89 ULIRGs. 
See the caption of Figure \ref{fig:morLir} for the error bars.}
\label{fig:classLir}
\end{figure}

Figure \ref{fig:classLir} shows the fraction of ULIRGs in different spectral classes as a function of IR luminosity. 
We use the same $L_{IR}$ bins as in Figure \ref{fig:morLir}. Each bin includes: 61 (12.0$\le \log(L_{IR}) <$ 12.25), 18 (12.25$\le \log(L_{IR}) <$ 12.5) and 10 (12.5$\le \log(L_{IR})$) sources.   
The fraction of AGNs grows with increasing $L_{IR}$. 
Star forming galaxies show an opposite trend, their fraction decreases with increasing $L_{IR}$. 
This is consistent with the results of previous studies showed that the fraction of AGNs in IR galaxies increases with higher IR luminosity \citep{Veilleux1995,Kim1998, Goto2005}. 
LINERs tend to be constant in each luminosity bin. 
Composites also tend to be almost constant in the first two bins, but they show a dramatic decrease in the highest luminosity bin. 
Again, since the LINERs and composites may have AGN contribution that is hidden in optical wavelengths, the AGN factions in each luminosity bin represent the lower limit. 
However, the trends seen in Figure \ref{fig:classLir} still agree with known correlation between the AGN fraction and IR luminosity. 
28 of the 89 ULIRGs in our optical spectral type subsample are also part of the 1-Jy sample of \citet{Veilleux1995,Veilleux1999}, since $\sim$ 31\% is a small fraction our results are mostly independent from those derived for the 1-Jy sample. 

%FIG 9

\subsection{Stellar Masses - Star Formation Rates - Metallicities and Optical Colors of ULIRGs }\label{S:comparison}

ULIRGs are very special galaxies that are selected according to their enormous IR luminosity, which means a rich dust content. 
Dust has an important role in galaxy growth and evolution because it is directly linked to the star formation and metals in the interstellar medium (ISM).  
The interplay between the dust and stellar content with Star Formation Rate (SFR) controls the galaxy evolution. 
For normal star forming galaxies this is evident from the observed correlations between these parameters. 
Stellar mass (M$_{star}$) and SFR tightly correlates within $0 < z < 3$ range and normal star forming galaxies lie on the so called `main sequence' \citep[e.g.][]{Noeske2007a,Elbaz2007,Santini2009,Rodighiero2011,Tadaki2013}. 
Stellar mass also strongly correlates with the metallicity ($Z$), massive galaxies show a higher metallicity than the less massive systems. 
The M$_{star} - Z $ relationship is confirmed for normal star forming galaxies in the local universe (z$\sim$0.15) \citep{Tremonti2004}. 
Although there is not a strong relation between SFR and $Z$,  
metallicity is a function of SFR and M$_{star}$ in the M$_{star} - Z - $SFR plane \citep{Lara-Lopez2010,Mannucci2010}. 
Recently, \citet[][]{Santini2013} showed that there is a tight correlation between the dust mass and SFR and they introduce a fundamental relation between gas fraction, M$_{star}$ and SFR. 
These relationships provide a basis for understanding the evolution of normal star forming galaxies. 
ULIRGs do not belong to this galaxy category and in order to explore their place in galaxy evolution we need to compare them with normal star forming galaxies. 
In the following, we investigate the position of ULIRGs in M$_{star} -$ SFR, M$_{star} - Z $ relationships and in the color$-$magnitude diagram. 
Below we briefly outline the SDSS data used in this investigation. 

Available SDSS photometric and spectral data allow us to obtain stellar masses, metallicities and optical colors of ULIRGs in our sample. 
SDSS DR10 \citep[see][]{Ahn2013} provides stellar masses, emission-line fluxes, stellar and gas kinematics, 
and velocity dispersions derived spectra of galaxies observed by Baryon Oscillation Spectroscopic Survey (BOSS). 
Following the spectroscopic pipeline \citep{Bolton2012}, the objects classified as a galaxy with a reliable redshifts are studied by several groups. 
`Portsmouth' group derives photometric stellar-mass estimates \citep{Maraston2012} and measure emission-line fluxes \citep{Thomas2013}. 

\citet{Maraston2012} estimate stellar masses through SED fitting of stellar population models to $u$, $g$, $r$, $i$, $z$ magnitudes. 
They use both passive \citep{Maraston2009} and star-forming templates \citep{Maraston2005} with \citet{Salpeter1955} and \citet{Kroupa2001} Initial Mass Functions (IMF). 
\citet{Maraston2012} use the fixed BOSS spectroscopic redshift values and do not include internal galaxy reddening in the SED fitting procedure. 
Wisconsin' group also derives stellar masses via full spectral fitting \citep{Chen2012}. 
They use models based on stellar population models of \citet{Bruzual2003} with a \citet{Kroupa2001} initial mass function. 
\citet{Chen2012} and \citet{Maraston2012} use different stellar population models based on different galaxy star formation histories, reddening, and initial mass function assumptions. 
Stellar masses given by \citet{Maraston2012} are $\sim$0.2 dex smaller than the masses estimated by \citet{Chen2012} and for high signal to noise spectra the results from the both methods agree well \citep{Chen2012}. 
Since spectral data quality is an issue, in this work we prefer to adopt the stellar masses given by \citet{Maraston2012}; however this preference does not change the results of this work. 
\citet{Maraston2012} obtain stellar masses for active and passive stellar population models.   
Since ULIRGs are actively star forming galaxies we adopt the stellar masses from `stellarMassPortStarforming'\footnote[3]{http://www.sdss3.org/dr10/spectro/galaxy\_portsmouth.php} catalog. 
These are listed in Table \ref{tab:sparameters}. 
The magnitudes used in the stellar mass estimates include contributions from star formation but also possible AGN contamination. The errors associated with the stellar masses are discussed in \S \ref{S:dis_SFRMass}. 

\citet{Thomas2013} fit stellar population synthesis models of \citet{Maraston2011} and Gaussian emission-line templates to the spectra by using the 
Gas AND Absorption Line Fitting (GANDALF) code of \citet{Sarzi2006}. 
This code accounts for the diffuse dust on the spectral shape according to \citet{Calzetti2001} obscuration curve. 
\citet{Thomas2013} correct for the diffuse dust extinction and provide de-reddened emission-line fluxes (this includes Galactic extinction). 
In the following analysis we adopt the emission-line fluxes from SDSS `emissionlinesPort'$^{3}$ catalog.

\subsubsection{Star Formation Rate and Stellar Mass}\label{S:sfrmass}

IR luminosity measured from the SEDs (\S \ref{S:SEDs}) between 8\micron\ $-$ 1000\micron\ is the obscured emission from young stars that is re-emitted by dust, hence it can be converted to SFR. 
We use Eq. (4) given by \citet{Kennicutt1998} to estimate the SFR based on $L_{IR}$, SFR(IR). 
This conversion assumes a \citet{Salpeter1955} IMF and that $L_{IR}$ is generated by recent star formation and re-emitted by dust. 
Even in the case of AGN we expect this assumption to be still valid to infer SFR(IR) because, 
ULIRGs on average have an AGN contribution from 5.0\% to 40.0\% AGN \citep[e.g.,][]{Genzel1998,Veilleux2009} but they are mostly powered by star formation. 
Therefore we note that the SFR(IR) values of the AGN, LINERS and composites may have on average $\sim$ 40.0\% AGN contamination and may be overestimated up to 80$-$100\% \citep{Veilleux2009}. 
Derived SFR(IR) values are tabulated in Table \ref{tab:sparameters}.

\begin{figure} 
\begin{center}$
\begin{array}{c}
\includegraphics[scale=0.3]{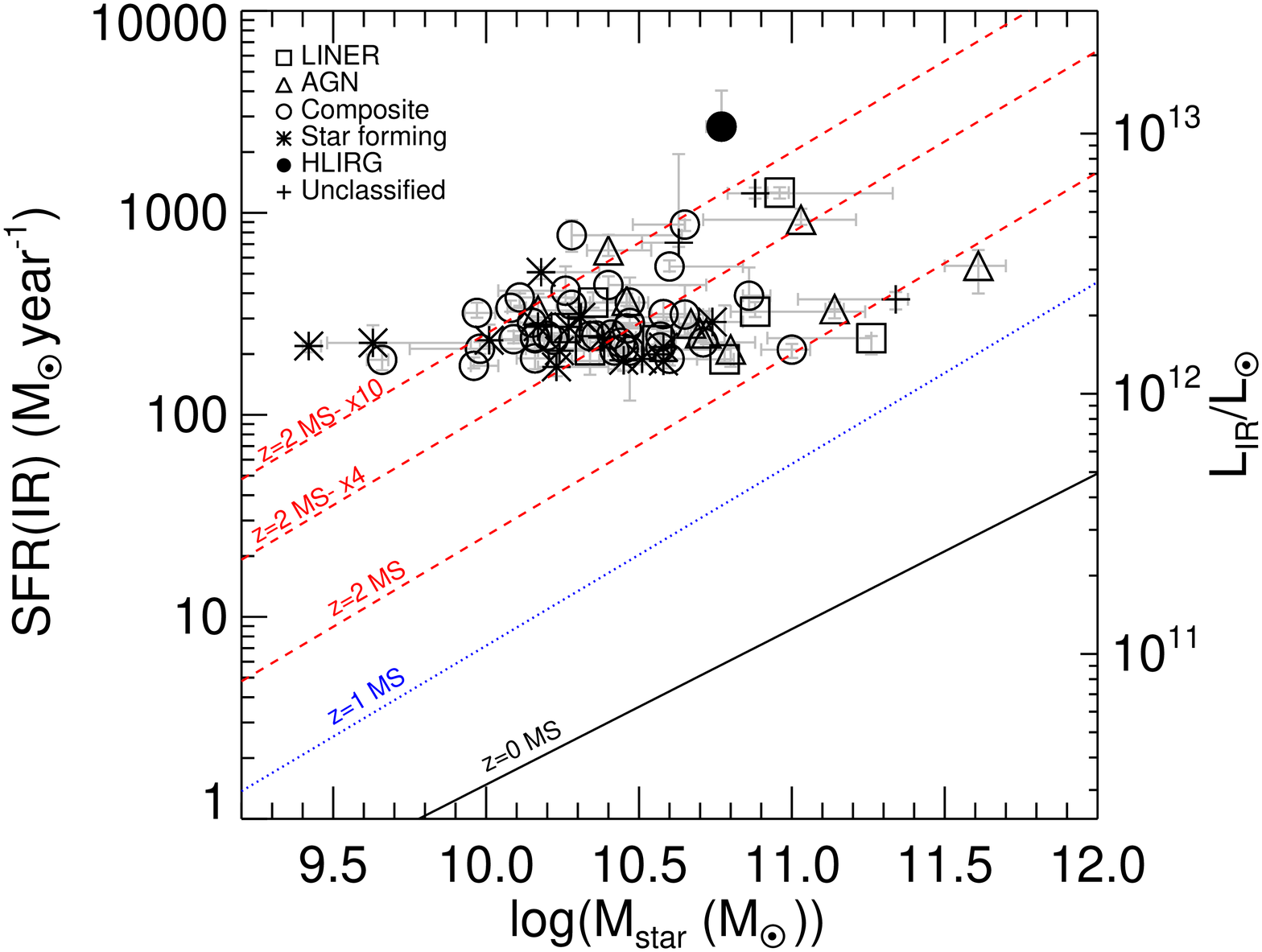}\\
\end{array}$
\end{center}
\caption{SFR(IR) versus stellar mass for 75 ULIRGs and one HLIRG.  
SFR(IR) values are derived from Eq. (4) of \citet{Kennicutt1998}. 
Error bars of SFR(IR) represent the uncertainties propagated through $L_{IR}$ uncertainties.  
The solid line is the z=0 `Main Sequence' (MS) of normal star forming galaxies; Eq. (5) of \citet{Elbaz2007}. 
The dotted (colored blue) line is the SFR$-$M$_{star}$ relationship of z=1 star forming galaxies in the GOODS fields; Eq. (4) of \citet{Elbaz2007}. 
The dashed (colored red) lines represent the SFR$-$M$_{star}$ relationship of z=2 star forming galaxies in the GOODS fields \citep{Daddi2007} and 4 and 10 times above this relationship. 
} 
\label{fig:sfrir}
\end{figure}

Figure \ref{fig:sfrir} shows SFR versus M$_{star}$ for 75 ULIRGs and one HLIRG for which M$_{star}$ estimates are given by \citet{Maraston2012}. 
The solid (black), dotted (blue)  and dashed (red) lines represent the `main sequence' of normal star-forming galaxies (SFGs) at $z\sim$0 \citep{Elbaz2007}, $z\sim$1 \citep{Elbaz2007} and $z\sim$2 \citep{Daddi2007}, respectively. 
For comparison we also show the 4 and 10 times above the $z\sim$2 `main sequence' (MS) relationship (top dashed lines). 
Local ULIRGs exhibit extremely high SFRs compared to normal SFGs with the same masses. 
It is evident from Figure \ref{fig:sfrir} that local ULIRGs lie above the 'main sequence' up to $z\sim$2. 
We note that the `main sequence' relationships represent the total SFR obtained from $L_{IR}$ and UV continuum, SFR(IR+UV). 
Since we do not include SFR from UV continuum, SFR(UV), our SFR(IR) estimates are lower compared to SFR(IR+UV).  
However, the total SFR is dominated by SFR(IR) and therefore the difference between SFR(IR) and SFR(IR+UV) should be small. 

%FIG 10 

Previously, \citet{Elbaz2007} showed that (their Fig. 17) Arp220 (a well studied nearby ULIRG) exhibits a large off-set both from the `main sequence' and $z\sim$1 relationship. 
In the same Figure they also show that M82 (a starburst galaxy) lies above the local `main sequence', but it is located in the 1$\sigma$ confidence level of the $z\sim$1 SFR$-$M$_{star}$ relationship. 
\citet{Cunha2010} also compared local ULIRGs with local star forming SDSS galaxies and showed that ULIRGs have higher SFRs. 
In Figure \ref{fig:sfrir} we show a large local ULIRG sample, 75 ULIRGs and one HLIRG. 
We find that  local ULIRGs do not exhibit typical SFR for their masses even at $z\sim$2. 
Compared to the `main sequence' at $z\sim$0, $z\sim$1 and $z\sim$2, on average ULIRGs have 92, 17 and 5 times higher SFRs, respectively. 
Local ULIRGs seem to be equally distributed around the dashed line representing 4 times above the  $z\sim$2 MS. 
Compared to the `main sequence' at $z\sim$2 on average AGN, LINERs, composite and star forming ULIRGs have 3, 4, 5 and 5 times higher SFRs, respectively.
We do not see a significant systematic offset with optical spectral type. 
However, we note that AGN, LINERs and composites tend to have the highest SFR(IR) which might be a sign of the AGN contamination to $L_{IR}$.
We discuss the impact of star formation history on stellar mass estimates in \S \ref{S:dis_SFRMass}. 

%FIG 11 
\begin{figure*} [ht]
\begin{center}$
\begin{array}{ccc}
\includegraphics[scale=0.18]{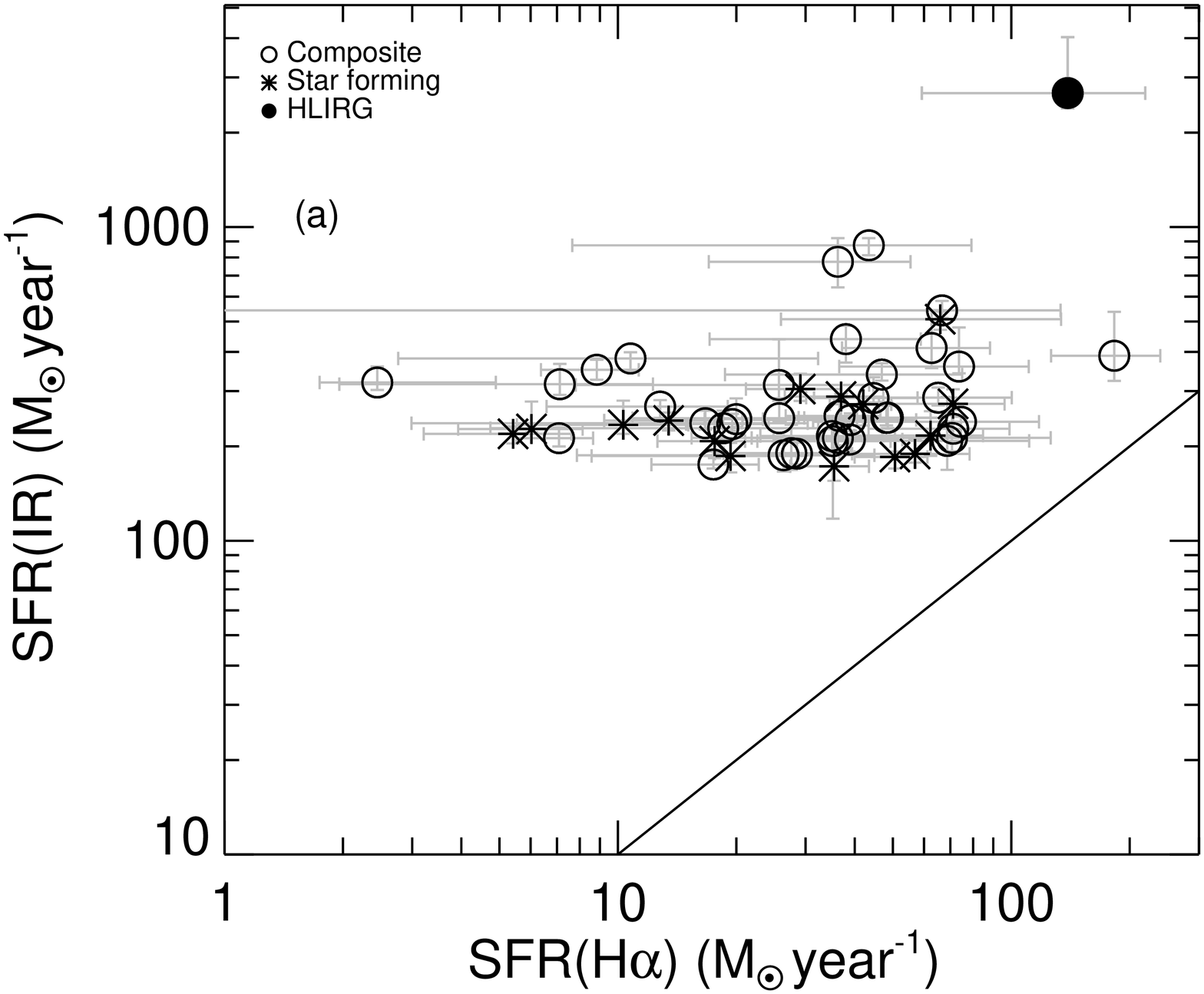} &
\includegraphics[scale=0.18]{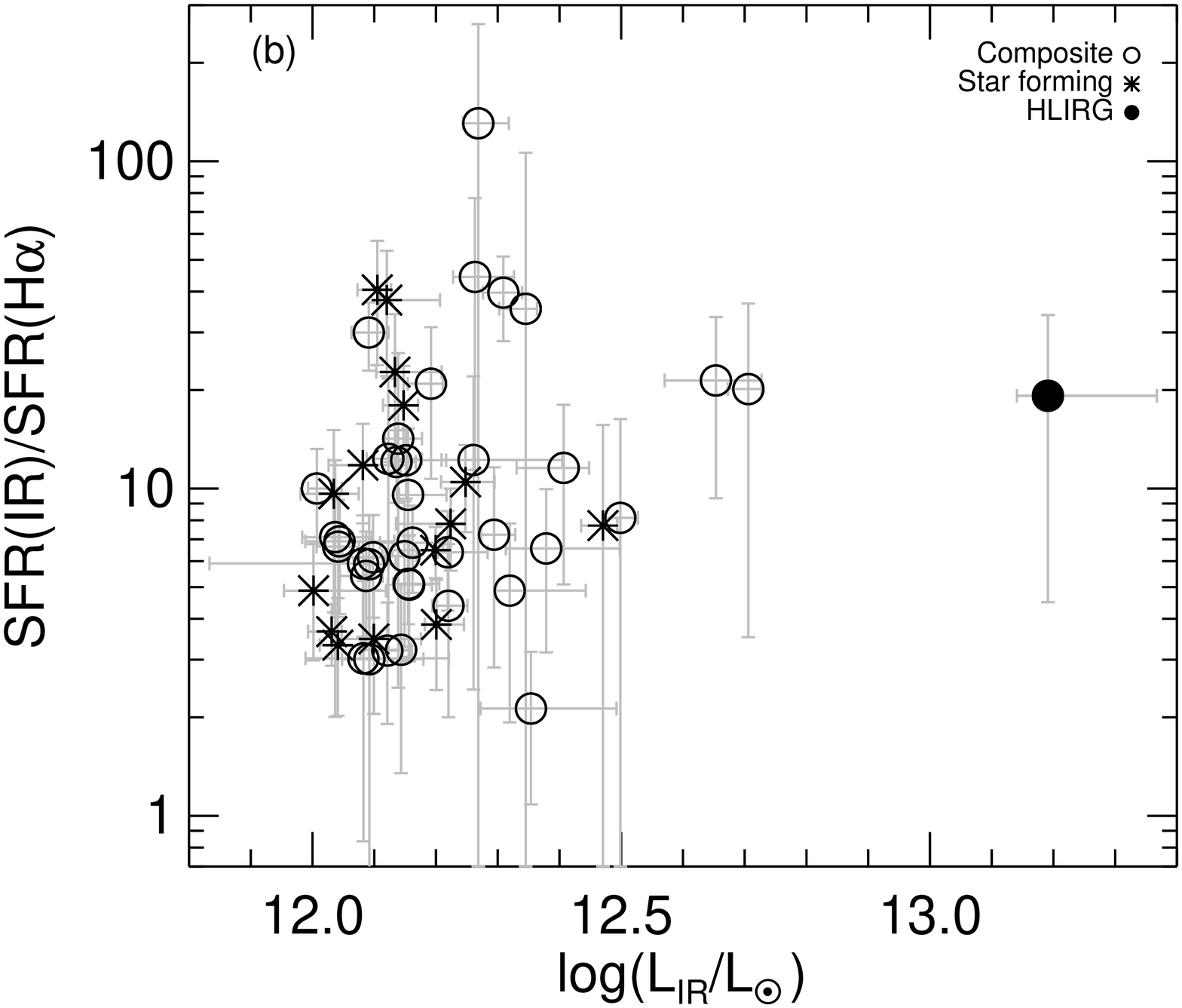} &
\includegraphics[scale=0.18]{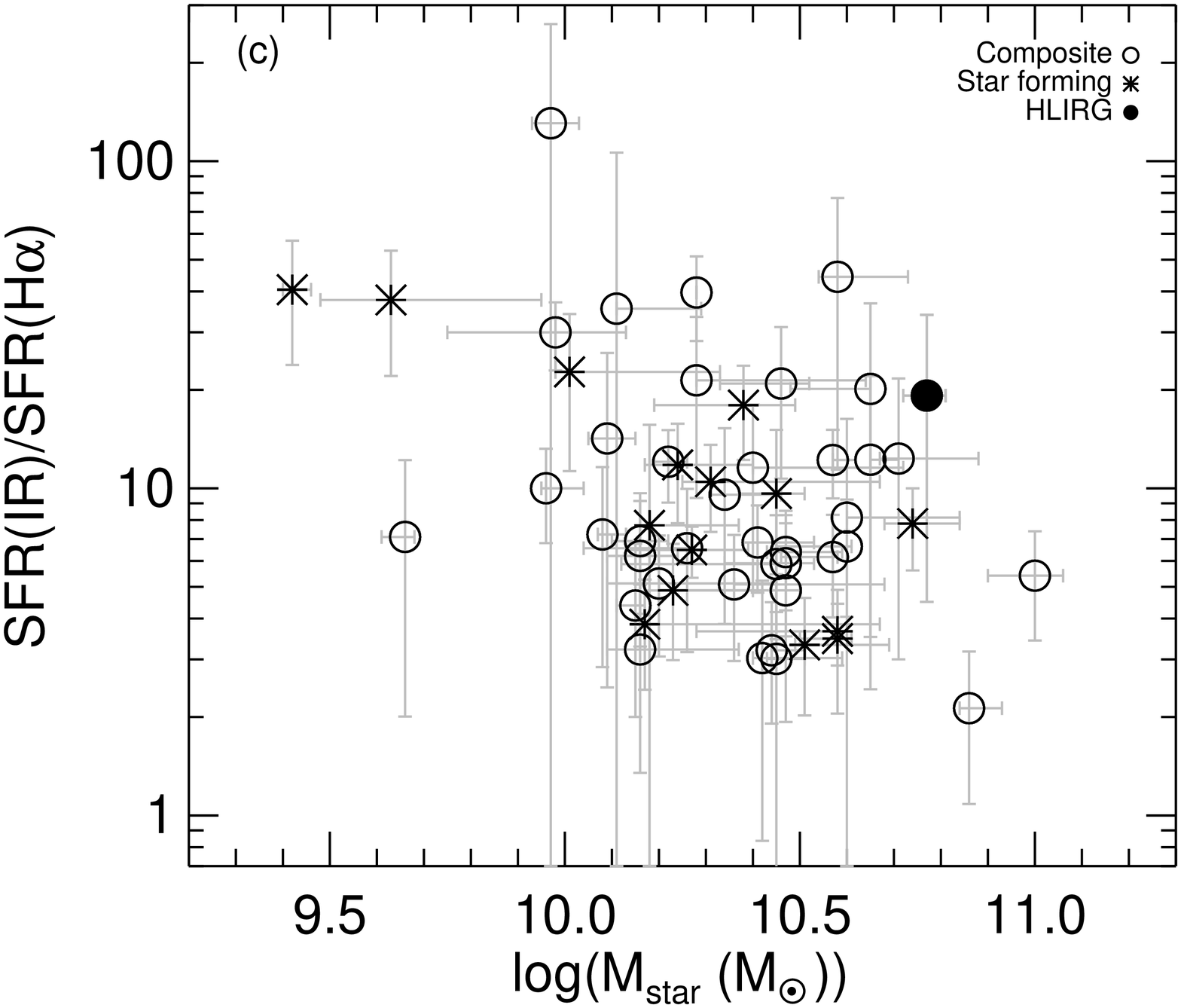}\\
\end{array}$
\end{center}
\caption{ SFR(IR) versus SFR(\Halpha) (a), SFR(IR)/SFR(\Halpha) versus $L_{IR}$ (b) and stellar mass (c) for 55 ULIRG and one HLIRG. 
In panel (a) the solid line represents the one-to-one relation. }
\label{fig:sfirvsuv}
\end{figure*}

\begin{figure} 
\begin{center}$
\begin{array}{c}
\includegraphics[scale=0.3]{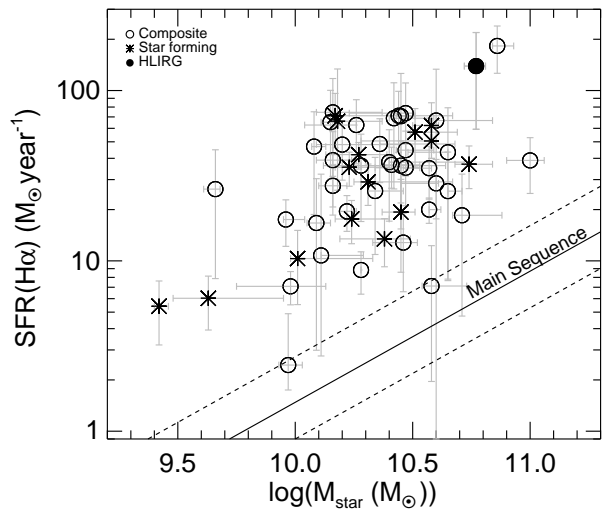}\\
\end{array}$
\end{center}
\caption{ SFR(\Halpha) versus stellar mass. 
 SFR(\Halpha) values are derived from Eq. (2) of \citet{Kennicutt1998}. 
Error bars of SFR(\Halpha) are dominated by the emission-line fluxes. Black solid line is the `Main Sequence' from \citet{Elbaz2007}, dashed lines represent the 1$\sigma$ width of the `Main Sequence'.}
\label{fig:sfruv}
\end{figure}

The ionizing radiation of recently formed young stars produce nebular lines such as \Halpha, hence it traces the unobscured radiation generated by star formation. 
Therefore, it can be used to derive SFR. 
As noted above \citet{Thomas2013} only correct for the diffuse dust extinction widely spread throughout the whole galaxy that affects the emission lines and stellar continuum, but they do not consider embedded dust component local to star forming regions that affects the emission lines.  
This is mainly to avoid highly uncertain dust extinction values measured from Balmer decrement due to low S/N spectra. 
However, if we avoid this additional dust extinction we may underestimate the SFR based on \Halpha\ luminosity, SFR(\Halpha).  
Therefore we use the already de-reddened emission-line fluxes (only for diffuse dust component) given in `emissionlinesPort'$^{3}$ catalog and obtain Balmer decrement. 
Predicted \Halpha/\Hbeta\ ratio is 2.86 for $10^{4}$ K \citep{Osterbrock2006}, we adopt this value to estimate the local dust extinction around nebular regions and correct \Halpha\ emission-line flux for the estimated extinction. 
Applying this additional extinction correction typically result in factor of 3.4 higher \Halpha\ emission-line flux with large uncertainties. 
We apply Eq. (2) given in \citet{Kennicutt1998} to obtain SFR(\Halpha); listed in Table \ref{tab:sparameters}.
Since in the presence of an AGN  \Halpha\ emission represents the photoionization from the AGN, we do not obtain SFR(\Halpha) for AGNs and LINERs. 
Figure \ref{fig:sfirvsuv} shows a comparison between the SFR(IR) and SFR(\Halpha) values (a). 
Note that error bars of SFR(\Halpha) are dominated by the \Halpha\ and \Hbeta\ emission-line flux uncertainties. 
The SFR(IR) values are systematically larger than SFR(\Halpha) values, this difference is between factor of 2 and factor of 130 and the median difference is factor of 8. 
This indicates that even the highest possible dust extinction correction applied \Halpha\ luminosity underestimates SFR at least by factor of 2. 
Therefore, it is evident that \Halpha\ is not sufficient enough to trace SFR for ULIRGs and IR observations are crucial to infer SFR of these galaxies. 
This is consistent with the fact that hydrogen recombination line SFR indicators are under luminous relative to the IR indicators in 
ULIRGs \citep[e.g.,][]{Goldader1995,Kim1998}.
In Figure \ref{fig:sfirvsuv} panels (b) and (c) show that the difference between the SFR(IR) and SFR(\Halpha) does not depend on $L_{IR}$ or M$_{star}$. 
In Figure \ref{fig:sfruv} SFR(\Halpha) versus M$_{star}$ is shown. 
Although SFR(\Halpha) underestimates SFR, Figure \ref{fig:sfruv} shows that ULIRGs still lie above the local `main sequence'. 

%FIG12 

\subsubsection{Stellar Mass and Gas Metallicity}\label{S:massmetallicity}

Nuclear metallicities and stellar masses of normal star forming galaxies show a well established M$_{star} - Z $ correlation \citep[][hereafter T04]{Tremonti2004}. 
In the following, we compare the stellar masses and oxygen abundances of ULIRGs with the mass-metallicity relation of local star-forming SDSS galaxies obtained by \citet{Tremonti2004}. 
Reliable metallicity constraints are difficult to obtain from broad-band SED fitting applied by \citet{Maraston2012}. 
Therefore, in order to measure metallicity we adopt the relevant emission-line fluxes from \citet{Thomas2013}. 
First, we apply  the additional extinction correction based on Balmer decrement (see \S \ref{S:sfrmass}) to the adopted emission-line fluxes. 
Then we compute the line ratio $R_{23}$=([O\,{\sc ii}]\,$\lambda \lambda 3726,3729$+[O\,{\sc iii}]\,$\lambda \lambda 4959,5007$)/\Hbeta.
We convert $R_{23}$ values to oxygen abundances, (O/H), by following Eq. (1) of \citet{Tremonti2004}. We list the derived oxygen abundances in Table \ref{tab:sparameters}.
Note that this conversion and the $R_{23}$ line ratio is only applicable to normal star forming galaxies, and they are not relevant for AGNs because the radiation from the AGN contributes to the line emission. 
Therefore we do not calculate metallicities for AGNs and to be conservative we also exclude LINERs from this investigation. 
After excluding AGNs and LINERs we are left with 48 ULIRGs and one HLIRG for which emission-line fluxes given by \citet{Thomas2013}. 

\begin{figure} 
\begin{center}$
\begin{array}{c}
\includegraphics[scale=0.3]{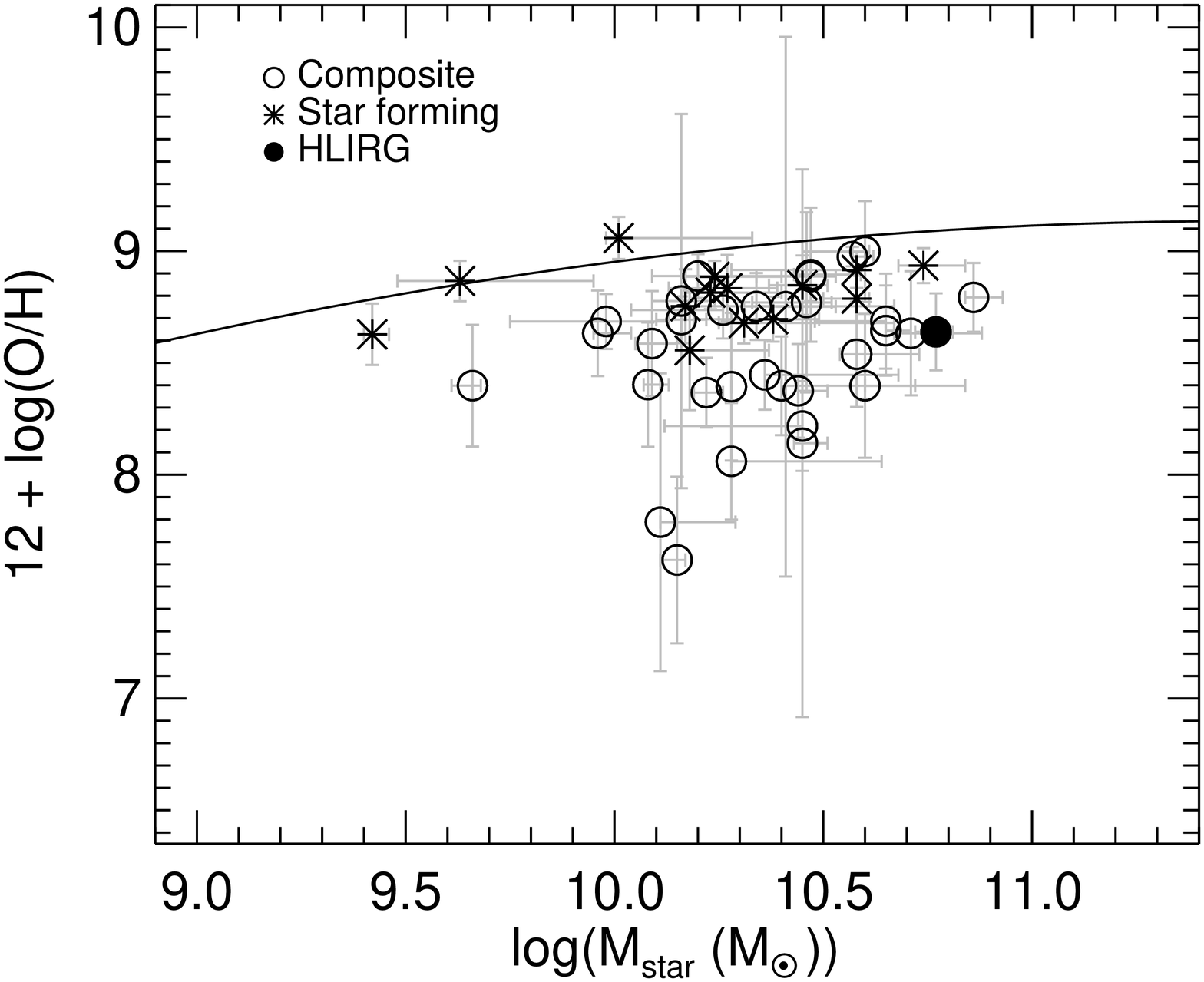}\\
\includegraphics[scale=0.3,angle=270]{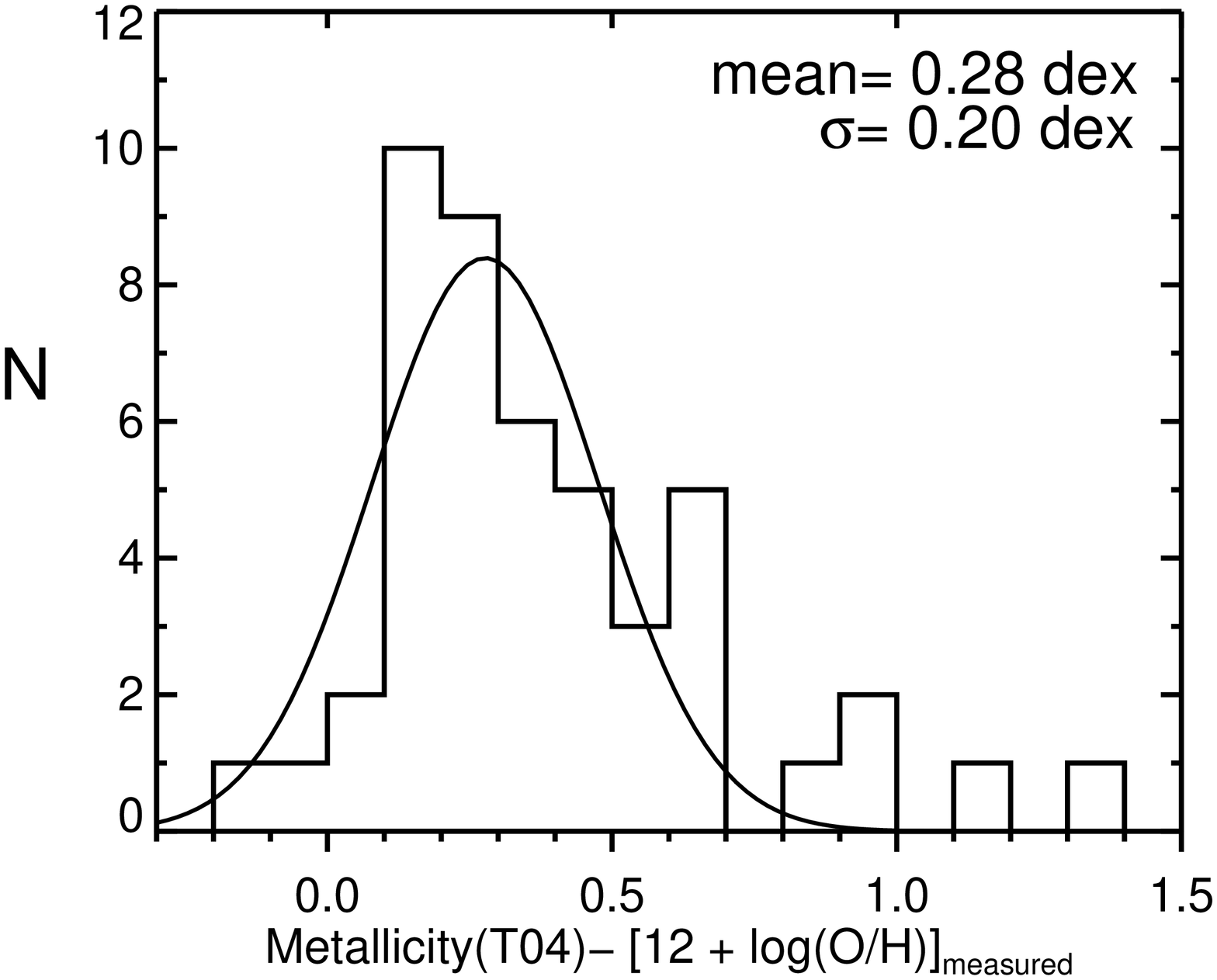}\\
\end{array}$
\end{center}
\caption{ Top: Oxygen abundances versus stellar mass for 48 ULIRGs and one HLIRG. 
The black solid line represents the mass$-$metallicity relation of the local SDSS galaxies given by \citet{Tremonti2004}. 
Error bars of the oxygen abundances represent the uncertainties of emission-line fluxes (including the uncertainties associated additional extinction based on Balmer decrement) propagated through Eq. (1) and (2) of \citet{Tremonti2004}. 
Bottom: The distribution of the residuals between the measured oxygen abundances and the ones expected from T04 relation. The over plotted Gaussian function demonstrates that the residuals have a normal distribution with a few outliers. }
\label{fig:massmetal}
\end{figure}

The M$_{star} - Z $ distribution of our ULIRG sample is shown in Figure \ref{fig:massmetal} (top). 
The error bars of the oxygen abundances represent the uncertainties associated with emission-line fluxes and additional extinction obtained from Balmer decrement. 
The black solid line is the M$_{star} - Z $ relationship, Eq. (3) given by \citet{Tremonti2004}. 
The vast majority of ULIRGs (46 out of 48) have lower metallicities compared to the normal star forming SDSS galaxies at similar masses. 
In the bottom panel of Figure \ref{fig:massmetal}, the distribution of the residuals of the measured oxygen abundances to the expected oxygen abundances from T04 relationship is displayed. 
The distribution of the  residuals are comparable to the over plotted Gaussian distribution with  standard deviation $\sigma$=0.20 dex, therefore we consider the shift of ULIRGs from the T04 relationship as 0.20 dex. 
Normal star forming SDSS galaxies exhibit a scatter between 0.07 dex $-$ 0.2 dex with a mean of 0.1 dex \citep{Tremonti2004} around the stellar mass$-$metallicity relation. 
The scatter of normal star forming galaxies from this relationship is mostly attributed to the observational errors in the mass and metallicity measurements \citep{Tremonti2004}. 
The scatter of ULIRGs (0.20 dex) is equal to the upper limit of the scatter of normal star forming galaxies. 
The median error in metallicity measurements of ULIRGs is large $\sim$0.17 dex, while the median error in mass measurements is smaller $\sim$0.08 dex. 
If we consider the lower error bars, it is very likely that the metallicity distribution of ULIRGs may shift even lower values and this may result in a larger scatter with respect to T04 relationship. 	
However, If we consider the upper error bars only 6 ULIRGs may move above T04 relationship and most of the ULIRGs would still lie below this relationship. 

%FIG 13 

Previously, \citet{Rupke2008} have already shown that ULIRGs have under-abundant compared to the SFGs on the M$_{star} - Z $ relation. 
The position of our ULIRG sample with respect to the M$_{star} - Z $ relationship of normal SFGs is consistent with the results of \citep{Rupke2008}. 
11 of the 48 ULIRGs in our M$_{star} - Z $ subsample are also part of 100 (U)LIRGs of \citet{Rupke2008}. 
Since a small fraction (23\%) of our sample overlaps with the sample of \citet{Rupke2008} our results are highly independent from theirs. 

\subsubsection{Color $-$ Magnitude Distribution of ULIRGs}\label{S:colmag}

The color versus magnitude distribution, the so called `Color-Magnitude Diagram' (CMD), of galaxies out to $z\sim$1 
show two separate distributions \citep[e.g][]{Hogg2003,Blanton2003,Baldry2004,Baldry2006,Cooper2006,Muzzin2012}: 
(1) `red sequence' of early type galaxies, (2) `blue cloud' of late type galaxies. 
The red sequence galaxies are bulge-dominated, more massive, non-star-forming, passive galaxies \citep[e.g.]{Blanton2003,Blanton2005a,Hogg2003,Baldry2006,Driver2006}. 
The blue cloud galaxies are disk-dominated, less massive, actively star forming galaxies \citep[e.g.]{Kauffmann2003,Brinchmann2004,Wyder2007}. 
Observations show that while the number density of blue cloud galaxies has stayed almost constant, the red sequence galaxies has doubled from $z\sim$1 to $z\sim$0 \citep{Bell2004,Faber2007}. 
This suggest that star forming disk galaxies at $z\sim$1 evolve to local passive galaxies. 
Such an evolution involves different physical processes that change galaxy morphology and quench star formation. 
As galaxies go through a transition phase from blue cloud to red sequence they reside in the region in between, the so called `green valley'. 

The transition of a late-type galaxy to an early-type includes physical processes that are not fully understood yet. 
Mergers and AGN feedback are among the proposed star formation quenching mechanisms \citep[e.g][]{BarnesHernquist1996,Hopkins2006,Hopkins2008a}. 
Since ULIRGs are both merging systems and mostly host an AGN they are good candidates for evolving galaxies from blue cloud to red sequence. 
In the following we explore the location of our ULIRG sample in the color-magnitude diagram of local SDSS galaxies. 

For this investigation we have selected a local comparison sample from SDSS DR 10 database. 
We selected sources classified as galaxies that are brighter than $r_{Petrosian}<$17.7 and  have spectroscopic redshifts within 0.018$< z< $0.260 interval ($z_{median}= 0.1$).  
We also select galaxies that have photometric measurements in $u$, $g$ and $r$ bands. 
Our selection criterion leads to 499953 galaxies. 
Throughout this analysis we use the Galactic extinction corrected `modelMag' measurements from SDSS DR10 `PhotoObj' catalog. 
$K-$corrections are calculated using the $kcorrect$ code v4.2 of \citet{Blanton2007}. 
For comparison reasons with previous studies, we derive $K-$corrections for a fixed bandpass shift by $z$=0.1. 
The absolute magnitudes and colors are denoted with $M_{r}^{0.1}$ and u$^{0.1}-$r$^{0.1}$, respectively; these are tabulated in Table \ref{tab:sparameters}.
Figure \ref{fig:colmag} shows CMD , (u$^{0.1}-$r$^{0.1}$) vs $M_{r}^{0.1}$, of the comparison and our ULIRG samples. 
The contours represent the number density of the comparison sample. 
The distribution of local SDSS galaxies shows two separate distributions: the red sequence and the blue cloud. 
We determine the color-magnitude relation of  the red sequence and  the blue cloud by following \citet{Baldry2004}. 
We divide the comparison sample into 16 $M_{r}^{0.1}$ bins from -23.5  to -15.5; bin size is 0.5 mag. 
For each $M_{r}^{0.1}$ bin we fit the color distribution with a double Gaussian and obtain mean and variance for the red and the blue distributions. 
We adopt the color function and the absolute magnitude functions given by \citet{Baldry2004} and obtain the color-magnitude relations as: 

%FIG 14
\begin{figure} 
\begin{center}$
\begin{array}{c}
\includegraphics[scale=0.6]{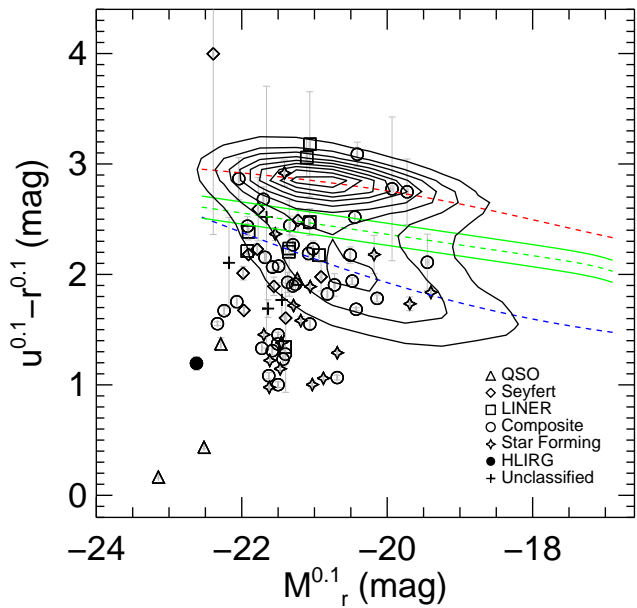}\\
\end{array}$
\end{center}
\caption{ Color-magnitude diagram for local SDSS comparison sample, 82 ULIRGs and one HLIRG. 
Contours represent the number densities for 10 levels. (H)/ULIRGs are shown on top of the contours. 
The color-magnitude relations for the red sequence (the upper dashed, colored red), the blue sequence (the lower dashed, colored blue) 
 and the green valley (the lower dashed, colored green) are shown. The solid (colored green) lines show the $\pm$0.1 mag width of the green valley.  
See the text for the detailed descriptions of these relations. The vertical and horizontal error bars represent the uncertainties in the model magnitude measurements. 
}
\label{fig:colmag}
\end{figure}

\begin{align}\label{E:redseq} 
& (u^{0.1}-r^{0.1})_{red-sequence}= 2.559+(-0.045) \times (M_{r}^{0.1}+20)+ \nonumber \\ 
& (-0.298) \times \tanh(\frac{M_{r}^{0.1}- (-17.757)}{2.833}),
\end{align}

\begin{align}\label{E:blueseq}  
& (u^{0.1}-r^{0.1})_{blue-cloud}= 2.831+0.066 \times (M_{r}^{0.1}+20)+ \nonumber \\ 
& (-2.180) \times \tanh(\frac{M_{r}^{0.1}- (-22.999)}{6.786}).
\end{align}
The upper and lower dashed lines in Figure \ref{fig:colmag} represent the red and blue sequence color-magnitude relations, respectively. 
To derive the color-magnitude relation of the green valley we locate the minimum in the double Gaussian functions. 
We then fit a linear plus a tanh function, the same function used to fit red and blue sequence relations,  to the minimums. 
The resulting relation of the green valley is: 
\begin{align}\label{E:greenvalley}  
& (u^{0.1}-r^{0.1})_{green-valley}= 2.232+(-0.096) \times (M_{r}^{0.1}+20)+ \nonumber \\
& (-0.131) \times \tanh(\frac{M_{r}^{0.1}- (-16.447)}{0.492}).
\end{align}
We choose the width of the green valley to be 0.1. 
The middle dashed line in Figure \ref{fig:colmag} represents Equation \ref{E:greenvalley}, the solid lines represent the 0.1 mag width. 

In Figure \ref{fig:colmag} the color-magnitude distribution of 82 ULIRGs and one HLIRG in our sample is shown on top of the contours of the comparison sample. 
From our ULIRG sample 10 are in the red sequence, 6 are in the green valley and 66 are in the blue cloud. 
The 81\% of the ULIRGs are located in `blue cloud', 12\% are in `red sequence' and only 7\% are in `green valley'. 
Two of the 6 (33\%) ULIRGs are in the `green valley' host an AGN. 
One of the 10 (10\%) ULIRGs are in the `red sequence' and 16\%\ of the ULIRGs are in the `blue cloud' host an AGN. 
The fraction of the AGN hosting ULIRGs is highest in the green valley.
40\% (33 of 82) of the ULIRGs are located out of the 90\% level contour. 
The median absolute magnitude and the u$^{0.1}-$r$^{0.1}$ color of our ULIRG sample are $M_{r}^{0.1}=-21.40\pm0.71$ and $u^{0.1}-r^{0.1}=1.91\pm0.64$. 
The median absolute magnitude of the comparison SDSS sample is $M_{r}^{0.1}=-20.55\pm1.12$. Compared to the local SDSS sample absolute magnitudes of ULIRGs are 0.86 mag brighter. 
 The median $u^{0.1}-r^{0.1}$ of the comparison sample is $u^{0.1}-r^{0.1}=2.56\pm0.55$, so ULIRGs have 0.64 mag brighter colors.
As ULIRGs are selected by their star-formation powered IR luminosity we expect them to be bright optical sources. So in a sense, their bright optical colors are consistent with their identification criteria. 

\citet{Chen2010} study color-magnitude properties of a sample of 54 ULIRGs from \textit{IRAS} 1 Jy sample \citep{Kim1998} and 
show that ULIRGs are mostly in the `blue cloud'. 
They also find that compared to SDSS galaxies local ULIRGs are 0.2 mag bluer in $g - r$. 
Compared to \citet{Chen2010} we study a larger ULIRG sample and find consistent results; we find very similar color$-$magnitude properties. 
The distribution of our ULIRG sample across the color$-$magnitude diagram is also similar to the distribution shown by \citet{Chen2010}. 
We find a smaller fraction for the ULIRGs that lie out of the 90\% level contour. 
While they do not find any AGN hosting ULIRGs in the `green valley' we find two ULIRGs. 
22 of the 82 ULIRGs in our color-magnitude subsample are also part of 54 ULIRGs studied by \citet{Chen2010}. 
Since only a small fraction (27\%) of our color-magnitude subsample overlaps with the sample 
of \citet{Chen2010} our results are independent from theirs. 

We note that the colors of the AGN hosting ULIRGs have a contribution from the central AGN. 
In principle, AGN contamination makes the ULIRG colors bluer and this may result in shifting them from the green valley to the blue cloud. 
However, \citet{Chen2010} show that on average removing the AGN contamination change the color only by a small amount (0.005$-$0.007 mag). 
For their sample only one source moved closer to the green valley and 11 out of 12 remained close to their original positions. \citet{Chen2010} 
show that the lack of AGN in the green valley is not due to AGN contamination. 
3 of 15 AGN ULIRGs in our sample are part of the 12 AGN ULIRGs studied by \citet{Chen2010} and therefore we assume their 
results to be valid for our sample
Since it is beyond the scope of this work we do not  attempt to remove the AGN contribution for the AGN hosting ULIRGs. 
 
\section{Discussion}\label{S:dis}

\subsection{Infrared Luminosities}\label{S:dis_fircolors}
The IR luminosities computed in this work highly depends on the selected SED library of \citet{Dale2002}. 
\citet{Goto2011a} compare the IR luminosities computed from the SED models of \citet{Chary2001}, \citet{Dale2002}, \citet{Lagache2003} and quote the median offsets between the models as $13\%\, - 24\%$. 
The listed IR luminosities in this work might have similar offsets between these models.

On the other hand, the SED models of \citet{Dale2002} represents especially the IR SEDs of normal star forming galaxies and they are not specifically developed for ULIRGs. 
For example, \citet{Rieke2009} compare the observed SEDs of five local purely star forming ULIRGs with a \citet{Dale2002} template with $\alpha=1.5$ and 
point out that at high IR luminosities the FIR bump of their SEDs are more peaked. 
If the intrinsic SEDs of local ULIRGs are more peaked compared to \citet{Dale2002} SED models, then IR luminosities computed in this work might be overestimated. 
However, the comparison of \citet{Rieke2009} is based on a very small sample and a full comparison between possible ULIRG SEDs is beyond the scope of this work. 
But, \citet{Rieke2009} and our sample have one common source, \textit{IRAS}\,12112+0305, for which we compare the IR luminosities based on the \citet{Dale2002} model SED and the observed SED used by \citet{Rieke2009}. 
We find that  IR luminosity computed from \citet{Dale2002} model SED is only $\sim$5\% higher than IR luminosity based on the SED template of \citet{Rieke2009}. 
This shows that while \citet{Rieke2009} state that their SEDs are more peaked compared to \citet{Dale2002} template the difference between the IR luminosities is small. 
These considerations suggest that the IR luminosities presented in this work  do not have a significant systematic offset. 
Even though \citet{Dale2002} models are not special for ULIRGs, the high number of already known ULIRGs in our sample that are identified based on these 
SED templates indicate that the IR luminosity measurements based on the SED templates of \citet{Dale2002} are reliable to identify ULIRGs. 

\subsection{FIR Colors}\label{S:dis_fircolors}
In panel (d) of Figure \ref{fig:col-col} four sources, J1639245+303719, J0159503+002340, J1356100+290538 and J1706529+382010, exhibit extreme $\log$(F(140\micron)/F(160\micron))$>$ 0.9 colors compared to the models. 
For these cases we check the reliability of the flux measurements from the \textit{AKARI} catalogs. 
In all of these cases while the 90\micron\ flux is highly reliable, the 65\micron, 140\micron\ and 160\micron\ flux measurements are low quality and the uncertainty of the 160\micron\ flux is not given. 
In such cases we assumed the uncertainty as the 25\% of the given flux measurement, but it seems these uncertainties could be even larger. 
Since the 90\micron\ flux measurements are secure we still consider the measured IR luminosities as reliable. 
The SEDs of three cases, J1639245+303719, J1356100+290538 and J1706529+382010, show that their flux densities at 140\micron\ is $\sim$0.8 dex larger than that of the models. 
The SED of J0159503+002340 also exhibits a large difference ($>$ 1 dex) between the observed and model flux at 160\micron. 
We also check five more sources with $\log$(F(140\micron)/F(160\micron))$>$ 0.5; J1603043+094717, J0030089-002743, J1102140+380240, J1346511+074720 and J2307212-343838.  
The SEDs of J0030089-002743, J1102140+380240, J1346511+074720 and J2307212-343838  show that their flux densities at 140\micron\ is 0.5-0.8 dex larger than that of the models. 
The SED of JJ1603043+094717 also exhibits a 0.6 dex difference between the observed and model flux at 160\micron. 
These large differences between the observed colors and that are expected from the SED models 
can be attributed to the low quality 140\micron\ and 160\micron\ flux measurements. 

As mentioned in \S \ref{S:colprop} the SED models covers only the $\log$(F(140\micron)/F(160\micron)) color range between -0.5425$-$0.2135 and therefore, 
in panel (c) the three sources, J1202527+195458, J1559301+380843, and J1502320+142132 appear as outliers with $\log$(F(140\micron)/F(160\micron))$<$ -0.58. 
The SEDs of these sources show that their 65\micron\ fluxes $\sim$0.5 dex lower than that of the models. 
Although the quality of 65\micron\ flux measurements are low for these sources, however it is more likely that the limited parameter range of the models is the main reason for their large deviation from the models. 
If the intrinsic SEDs of ULIRGs are more peaked compared to the templates of \citet{Dale2002} as shown by \citet{Rieke2009}, then we might 
expect to have a wider distribution for the FIR colors and this might explain the large scatter seen panels (c) and (d). 

\subsection{Interaction Classes}\label{S:dis_intclass}
 In \S \ref{S:morphprop} the interaction classes of 64 sources are adopted from literature \citep{Veilleux2002,Hwang2007} and 55 sources classified in this work based on visual inspection. 
 Although visual classification is a subjective method, we prefer it due to its practical application. 
 Two classifiers independently classified each source and for most of the cases there was good agreement. 
 There was a disagreement between the classifiers only for a few cases that are single nucleus systems at higher redshifts. 
 In such systems the identification of the disturbed morphologies or weak interaction signs is difficult. 
 However, the number of such systems are only five and most of them are not included in our statistics due to the applied redshift limit. 
 Even if they were included in our statistics, they would be classified as NI  instead of V and this would only decrease the number of sources classified as old mergers. 
 Such a change would not change the high percentage of IV and V systems in the overall population. 

Wide binary (IIIa) systems have the largest uncertainties because, most of the companion galaxies do not have spectroscopic redshifts. 
However, usually wide binary galaxies have similar colors and they show interaction signs. 
Therefore the chance coincidences are low and the assumed physical connection is highly likely. 
Even if most of the IIIa systems were instead IV, the dominancy of the mergers still holds. 
So the overall conclusion of morphology investigation in \S \ref{S:morphprop}, that is the vast majority of ULIRGs in the local universe are single nucleus ongoing/old mergers, is not affected by the 
disagreements of the classifiers or unconfirmed redshifts of the companion galaxies in wide binaries. 

\subsection{AGN Fraction of OUR ULIRG Sample}\label{S:dis_AGNfrac}
In \S \ref{S:classprop} we investigate the optical spectral types of the ULIRGs in our sample. 
The classification of Star Forming galaxies, composites, LINERs and Seyferts are based on empirical emission line diagnostics. 
ULIRGs are dust-rich systems and dust extinction at optical wavelengths is high. 
Therefore, the dusty nature of ULIRGs brings a large uncertainty to their optical emission line diagnostics. 
\citet{Nardini2010} use the rest-frame 5-8\micron\ spectra to disentangle the contribution of star formation and AGN in ULIRGs. 
As shown by \citet{Nardini2010} optical diagnostics do not provide reliable information on the presence of AGN. 
They trace obscured AGN in some LINERs and even some star forming galaxies. 
Therefore, as stated earlier our spectral classification provides only a lower limit on the AGN fraction. 
This brings a large uncertainty to the AGN fraction per $L_{IR}$ bin presented in Figure \ref{fig:classLir}. 
It is very highly likely that most of the composites and LINERs may have AGN component. 
If all of the LINERs and composites had AGN contribution then the correlation between the AGN fraction  and $L_{IR}$ would be still valid. 
                    
Assuming all of the LINERs and composites as AGNs may be an unrealistic overestimation because we would expect at least some fraction of the low luminosity ULIRGs to be star formation dominated. 
To investigate the hidden AGNs among such sources in our sample we look at the result of the mid-IR diagnostic applied by \citet{Nardini2010}. 
In total we have 31 overlapping sources with their sample. 
Our main interest is the AGN component of the star forming galaxies, composites and LINERs in our sample.  
For those sources we adopt the AGN bolometric contribution parameter given by \citet{Nardini2010} ($\alpha_{bol}$ parameter in their Table 1). 
Only one star forming galaxy (J0900252+390400) in our sample seem to have a significant AGN contribution based. 
If we consider this source as an AGN instead of a star forming galaxy, this would not affect the correlation of AGNs and the anti-correlation of star forming galaxies with IR luminosity. 
Also it would have a negligible affect on the fraction of AGNs; fraction of AGNs would increase to 26\% and the fraction star forming galaxies would decrease to 19\%. 

\subsection{The offset of ULIRGs from the Main Sequence of Star Forming Galaxies}\label{S:dis_SFRMass}

Star formation rate and M$_{star}$ tightly correlates from $z\sim$0 to $z\sim$2; the slope is between $\sim$0.6 and $\sim$1.0 (mostly depends on the galaxy sample) but the normalization decreases with redshift. 
This indicates that the overall SFR increases from $z\sim$0 to $z\sim$2 and SFGs were forming stars more actively in the past compared to lower redshift galaxies at the same masses.  
Observations indicate that high redshift SFGs contain larger molecular gas reservoir \citep[e.g][]{Daddi2010a,Tacconi2010}, and therefore the star formation rate per stellar mass is higher at $z\sim$2; 
in time this reservoir is used up and result in lower SFRs at $z\sim$0.  
Figure \ref{fig:sfrir} clearly demonstrates that local ULIRGs are outliers with respect to the `main sequence'  of the normal star-forming galaxies up to $z\sim$2. 
Local ULIRGs are already known to be outliers compared to the local `main sequence' \citep{Elbaz2007}. 
This is not surprising because, in the first place, ULIRGs are defined by their enormous IR luminosities powered by intense star-formation and in order to be defined as ULIRGs they should have 
$172 \le SFR(IR) \le 1721$. 
So their position on the y-axis is a pure selection effect and we expect them to have higher SFRs compared to normal star forming galaxies. 
We note that Figure \ref{fig:sfrir} includes type 2 AGNs, LINERs and composites. 
As mentioned earlier, even the AGN has a contribution to $L_{IR}$, the measured IR luminosities are mainly dominated by the FIR emission. 
As mentioned in \S \ref{S:sfrmass} the average AGN contamination is $\sim$40.0\% \citep{Veilleux2009}. 
But the offset of the local ULIRGs from the `main sequence' relations from $z\sim$0-2 is relatively large and it can not be attributed to 
the AGN contribution in the SFR(IR) estimates alone.

Normal starburst galaxies are also outliers off the `main sequence' at $z\sim$0.7 \citep{Guo2013} and at $z\sim$2 \citep{Rodighiero2011}. 
\citet{Guo2013} show their best fit `main sequence' and the`main sequence' relationships given by \citet{Elbaz2007} and \citet{Daddi2007} in their Fig. 7 where they report starburst galaxies as outliers.    
Since the local ULIRG sample lie above these `main sequence' relationships and their galaxy sample, it can be concluded that compared to normal starburst galaxies at $z\sim$0.7 local ULIRGs exhibit higher SFRs. 
\citet{Rodighiero2011} define off-sequence galaxies (see their Fig. 1) as the ones lying factor of 10 above the $z\sim$2 SFR$-$ M$_{star}$ relation of \citet{Daddi2007}. 
Compared to these extreme outliers at $z\sim$2, as seen in Figure \ref{fig:sfrir}, 90\% of the local ULIRGs have lower SFRs and only 10\% have comparable SFRs.  
SMGs, often referred as high redshift analogues of local ULIRGS, are also known to be outliers compared to $z\sim$2 SFR$-$ M$_{star}$ relation \citep{Tacconi2008,Daddi2007,Daddi2009,Takagi2008}. 
Compared to massive SFGs at the same masses, SFRs of SMGs are 10 times higher \citep{Daddi2007}. 
As noted by \citet{Daddi2007} SMGs at $z\sim$2 and local ULIRGs have similar properties; both are rare sources and outliers in SFR$-$ M$_{star}$ relations. 
However compared to the location of SMGs shown by \citet{Daddi2007} (their Figure 17b) local ULIRGs occupy a wider M$_{star}$ range and they are closer to the $z\sim$2 SFR$-$ M$_{star}$ relation. 

%FIG 15
\begin{figure}[ht]
\begin{center}$
\begin{array}{c}
\includegraphics[scale=0.3]{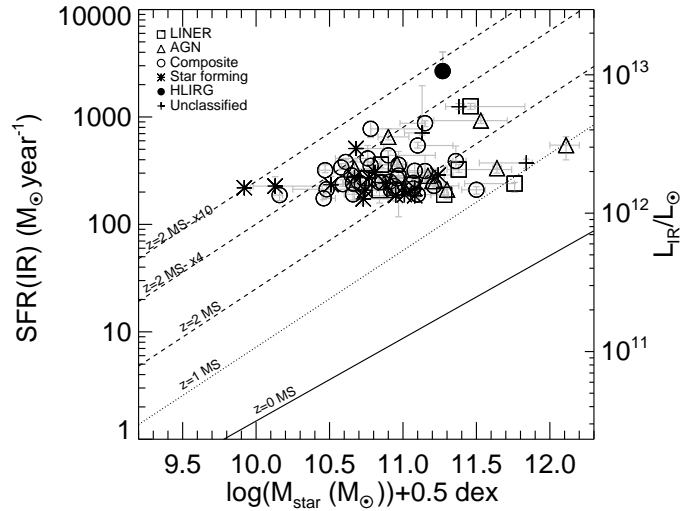}\\
\end{array}$
\end{center}
\caption{ Same as Figure \ref{fig:sfrir} but stellar mass are shifted by 0.5 dex.}
\label{fig:sfrir_shift}
\end{figure}

As expected galaxies with similar IR luminosities should have similar SFRs and positions in the SFR$-$M$_{star}$ diagram. 
In particular we call attention to the role of the stellar mass as the distinguishing parameter. 
At this point it is important to consider the uncertainties of the stellar masses to interpret Figure \ref{fig:sfrir}.   
ULIRGs in our sample have moderate stellar masses within 9.42$<\log(M_{star} (M_{\odot}))<$11.61 range where the median is 10.41. 
We compare the adopted stellar masses from  \citet{Maraston2012} with the M$_{star}$ estimates given by previous ULIRG studies. 
\citet{RodriguezZaurin2010} provide M$_{star}$ estimates for 36 local ULIRGs derived by performing spectral synthesis modeling to high quality optical spectra. 
We have two overlapping sources with their sample: J0900252+390400,  J1052232+440849 and they report 1.0 dex and 0.5 dex higher stellar masses, respectively. 
However, we note that J0900252+390400 is the lowest mass ULIRG in our sample, and as mentioned in \S \ref{S:dis_AGNfrac} it has an AGN. 
Therefore we consider the difference of 1 dex in M$_{star}$ for this particular object as an exceptional case. 
\citet{Cunha2010} also provide M$_{star}$ estimates for a sample of 16 purely star forming ULIRGs based on full SED modeling including UV to FIR wavelengths. 
We have one common source with this sample (J1213460+024844); and for this source the M$_{star}$ estimates agree well, their estimate is just 0.06 dex higher than the adopted value from \citet{Maraston2012}. 
As shown by \citet{RodriguezZaurin2010} ULIRGs contain different stellar populations (very young, young, intermediate-young and old stellar populations) at the same time and their optical light is mainly dominated by the 
less massive young stellar populations. 
Therefore, we expect the stellar mass estimates of ULIRGs to be highly dependent on the followed approach (SED or spectral fitting) together with the used data and the assumed star formation histories (SFHs). 
Especially M$_{star}$ estimates of ULIRGs like complex galaxies from SED fitting can be very sensitive to assumed SFHs. 
As shown by \citet{Michalowski2012} using multicomponent SFHs that fit young and old populations result in systematically higher stellar masses compared to exponentially declining SFH. 
Therefore, it is very likely that the adopted M$_{star}$ values in this work are underestimated. 
Obtaining the most robust stellar mass estimates of ULIRGs is beyond the scope of this paper. 
But, with the available data we are able to assign an uncertainty limit. 
Considering the M$_{star}$ differences of two (since it is an exception case we exclude J0900252+390400) ULIRGs with respect to the values reported 
by \citet{RodriguezZaurin2010} and \citet{Cunha2010} all of the adopted M$_{star}$ in this work values might be underestimated by 0.06 dex$-$0.5 dex. 
A natural consequent question is the affect of this underestimate in Figure \ref{fig:sfrir}. 
To be conservative we may assume that M$_{star}$ are underestimated by 0.5 dex.  
As shown in Figure \ref{fig:sfrir_shift}, if we shift stellar masses by 0.5 dex ULIRGs are still exhibit a large offset from the $z\sim$0 and $z\sim$1 `main sequence', 
but they are consistent with the $z\sim$2 `main sequence'. 
This shows that even the stellar masses adopted in Figure \ref{fig:sfrir} are underestimated, this does not change the main conclusion that ULIRGs are outliers compared to the $z\sim$0 and $z\sim$1 `main sequence'. 
However, it indicates that their offset from the $z\sim$2 `main sequence' is very likely due to their underestimated stellar masses. 
Of course Figure \ref{fig:sfrir_shift} is a simple illustration and might not reflect the M$_{star}$ distribution of ULIRGs at all, thus we caution against its interpretation. 

\subsection{Comparison of SFRs With Observations and Simulations of Mergers}\label{S:dis_compmergers}

\begin{figure} 
\begin{center}$
\begin{array}{c}
\includegraphics[scale=0.35,angle=270]{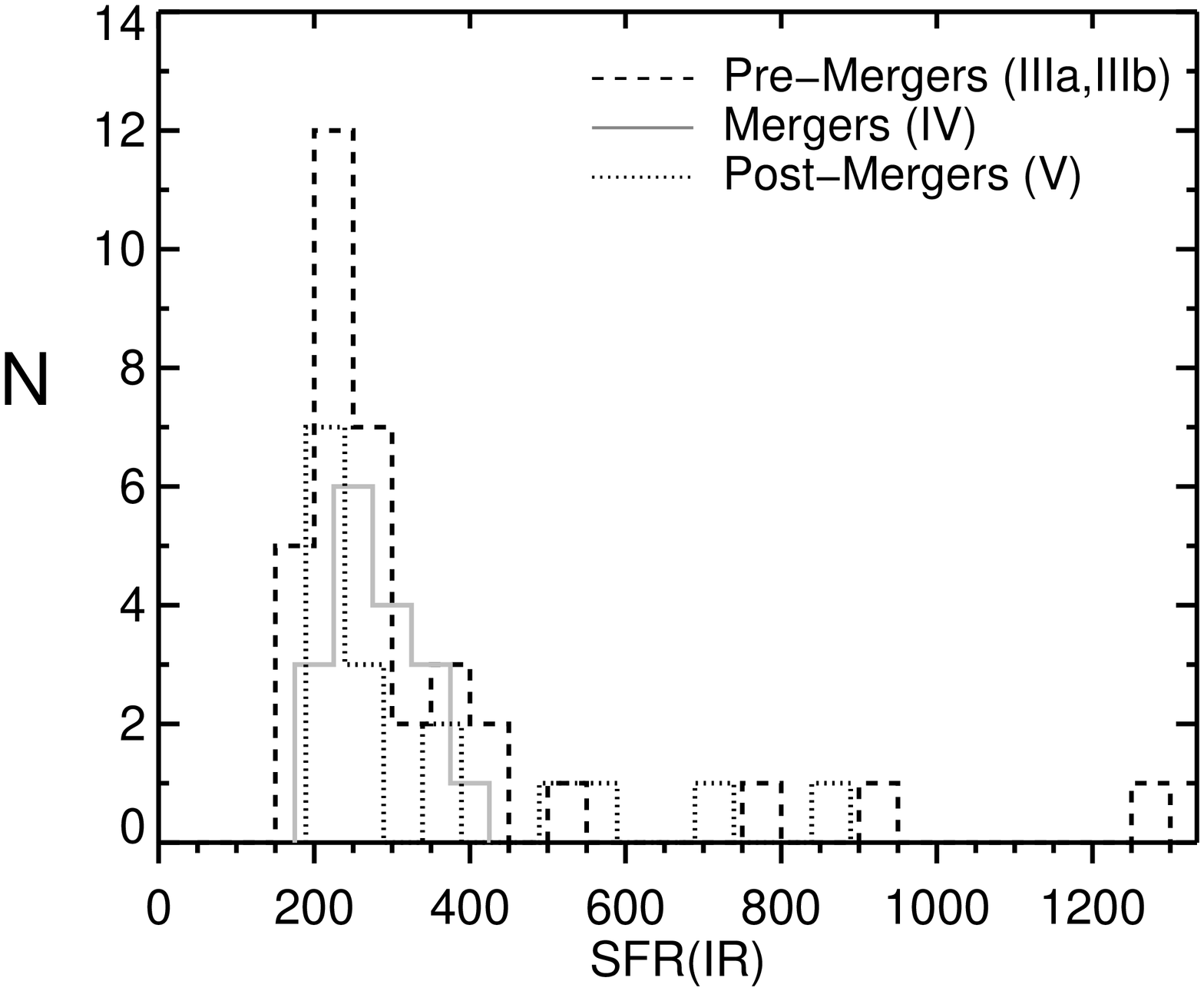}\\
\includegraphics[scale=0.35,angle=270]{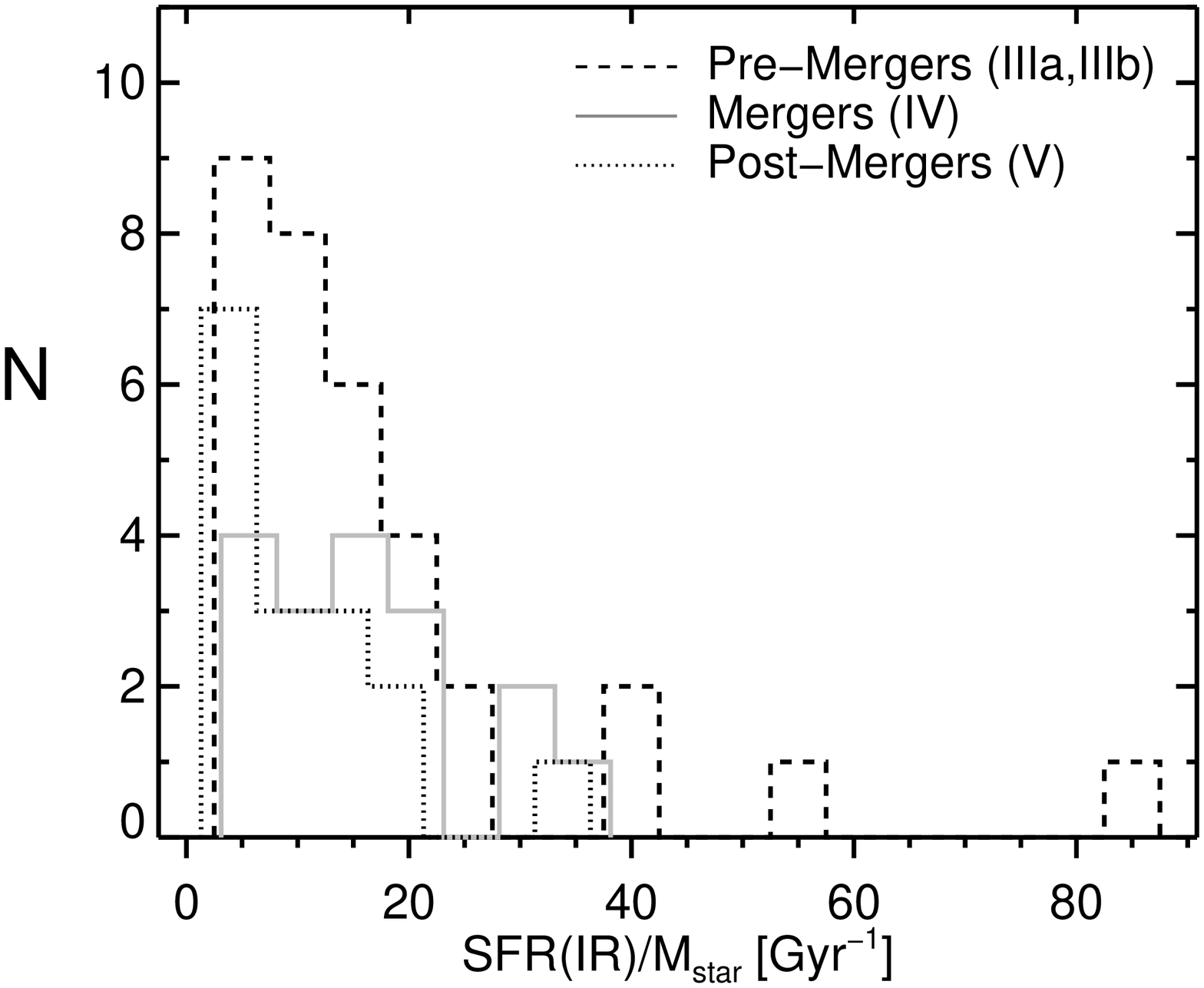}\\
\end{array}$
\end{center}
\caption{ SFR(IR) (top) and specific star formation rate (SFR/M$_{star}$) (bottom) distribution of 68 ULIRGs as a function of interaction stage.}
\label{fig:SFRir_IC}
\end{figure}

ULIRGs are interacting systems and they are mostly ongoing/late mergers and their extreme SFRs are generally attributed to merger events. 
Observations support this link, SFRs of local ULIRGs are consistent with the observed enhanced SFR of mergers \citep[e.g.][]{Ellison2008b,Ellison2013a}. 
Moreover, the role of merger processes on triggering SFR is a general prediction of merger models showing that major mergers cause nuclear gas inflows \citep{BarnesHernquist1991,BarnesHernquist1996} and these 
inflows generate intense SFR that peaks around when merging galaxies coalescence \citep[e.g][]{DiMatteo2005,Dimatteo2007,Springel2005,Montuori2010,Torrey2012}. 
Merger models show that star formation activity increases after the first peri-center passage and reaches its maximum level when two galaxies coalescence. 
In this picture we expect to observe lower SFRs in the pre-merger (widely or closely separated binaries) ULIRGS compared to the ULIRGs in the coalescence phase. 
In order to check if the observed SFRs of our ULIRG sample is consistent with this prediction, in the top panel of Figure \ref{fig:SFRir_IC} SFR(IR) distribution of ULIRGs is shown as a function of interaction class (defined in \S \ref{S:morphprop}). 
We find that ULIRGs do not show a systematic difference in SFR(IR) for different interaction stages. 
We do not find coalescence stage as the peak of the SFR as suggested by general merger simulations \citep[e.g][]{Torrey2012}. 
Since SFR is correlated with stellar mass, in the bottom panel of Figure \ref{fig:SFRir_IC} we show specific star formation rate, sSFR (SFR(IR)/M$_{star}$), as a function of interaction class. 
This panel shows a similar distribution as the top one, sSFR does not depend on the interaction stage. 
Of course this does not mean that ULIRGs are completely inconsistent with the merger models, because we are not tracing single merger events in time as simulations do. 
Instead we are looking at different snapshots of merger events for different sources. 
Therefore, Figure \ref{fig:SFRir_IC} is rather consistent with the observations \citet{RodriguezZaurin2010} showing that ULIRGs have complex multi-stellar populations. 
Probably, in some ULIRGs the SF activity triggered at pre-coalescence epochs are comparable with that of the others coalescence phase, thus we see a similar distribution for different interaction phases. 

%FIG 16

\begin{figure}[ht]
\begin{center}$
\begin{array}{c}
\includegraphics[scale=0.35,angle=270]{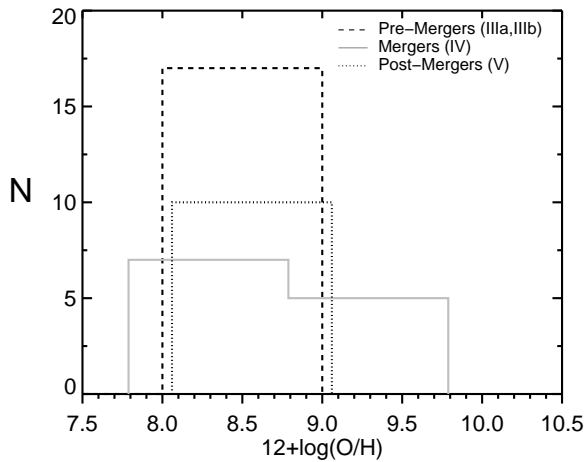}\\
\end{array}$
\end{center}
\caption{ Oxygen abundances distribution of 39 ULIRGs as a function of interaction stage.}
\label{fig:metallicity_IC}
\end{figure}

Merger simulations also predict that nuclear gas inflows in the periods prior to increasing SFR epochs cause nuclear metallicity dilution, but following that high SFRs cause metallicity enhancement \citep[e.g][]{Torrey2012}. 
So, the overall metallicity change has a rather complex fluctuating nature as the merger progress. 
In Figure \ref{fig:metallicity_IC} we show the oxygen abundance distribution of ULIRGs as a function of merger stage. 
Again the three distributions (pre-merger, mergers and post-mergers) overlap and do not show a significant difference. 
As discussed above, we do not probe the evolution of oxygen abundances for individual ULIRGs as simulations do, thus based on Figure \ref{fig:metallicity_IC} we can not conclude any inconsistency with their predictions. 
However, when we compare oxygen abundances of ULIRGs with that of normal SFGs we find that they systematically have lower oxygen abundances and this is consistent with the predictions of the numerical simulations \citep[e.g][]{Torrey2012}. 
Similarly, interacting galaxies such as close pairs \citep[e.g][]{Kewley2006,Ellison2008b} do not lie on the M$_{star} - Z $ relation. 
These interacting, merging galaxies exhibit a lower metallicity compared to the non-interacting normal SFGs.  

%FIG 17

\subsection{ULIRGs in the Fundamental Metallicity$-$Mass$-$SFR Plane}\label{S:dis_MassMetallicity}

Figure \ref{fig:massmetal} shows that ULIRGs have lower metallicities with respect to the M$_{star} - Z $ relation. 
The possible systematic uncertainties discussed in \S \ref{S:dis_SFRMass} are relevant to Figure \ref{fig:massmetal}, too. 
However, since M$_{star} - Z $ relation is rather flat with increasing stellar mass, a shift of 0.5 dex in M$_{star}$ does not change the observed scatter of ULIRGs. 

Figure \ref{fig:sfrir} indicates that local ULIRGs have comparable SFRs with  $z\sim$2.0 galaxies and it is known that $z\sim$2.2 galaxies have lower metallicities compared to local galaxies with the same masses \citep{Erb2006,Tadaki2013}. 
A similar result also found for even higher redshift galaxies $z$=3-4 \citep{Maiolino2008,Mannucci2009}.  

\begin{figure*}[ht]
\begin{center}$
\begin{array}{cc}
\includegraphics[scale=0.3]{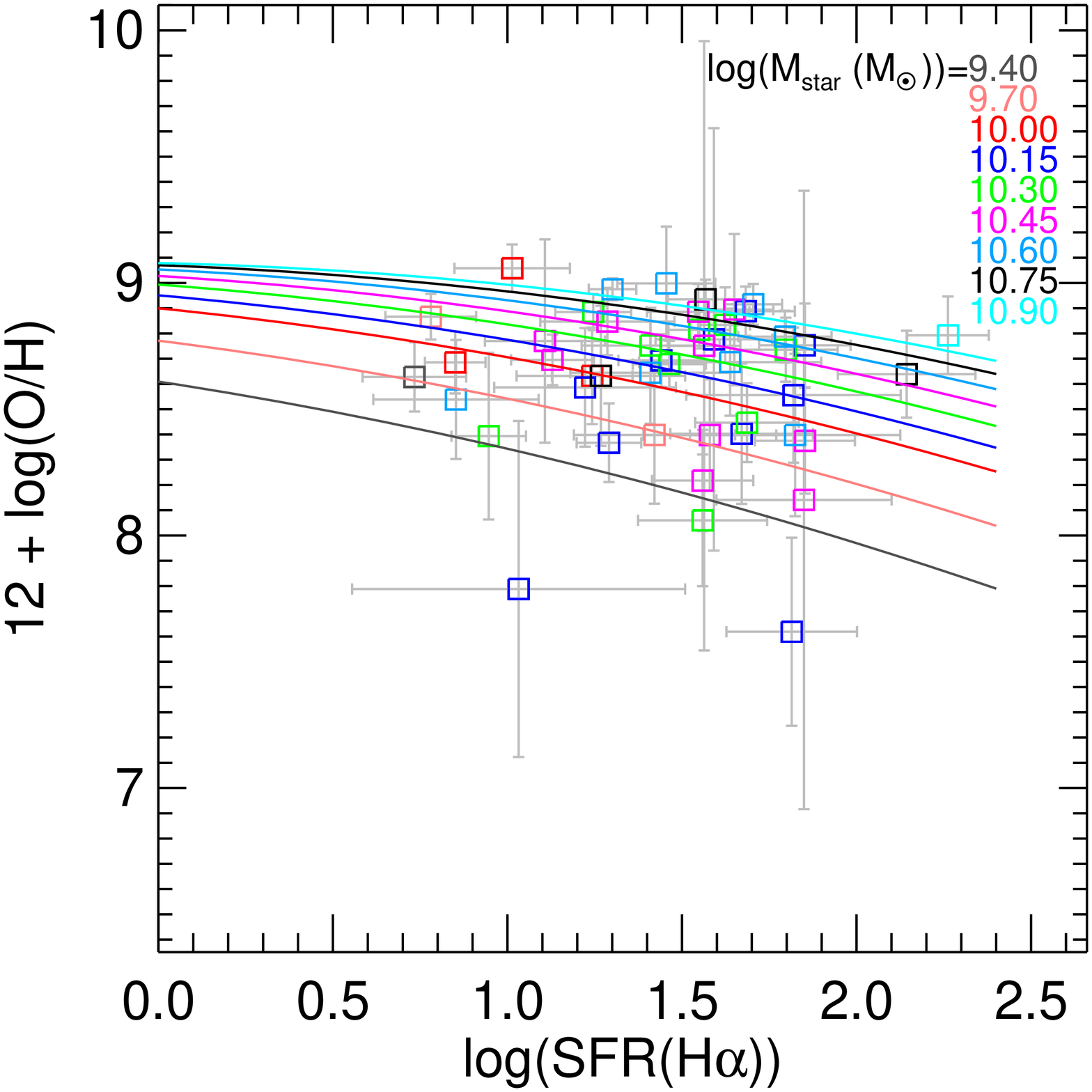}&
\includegraphics[scale=0.3]{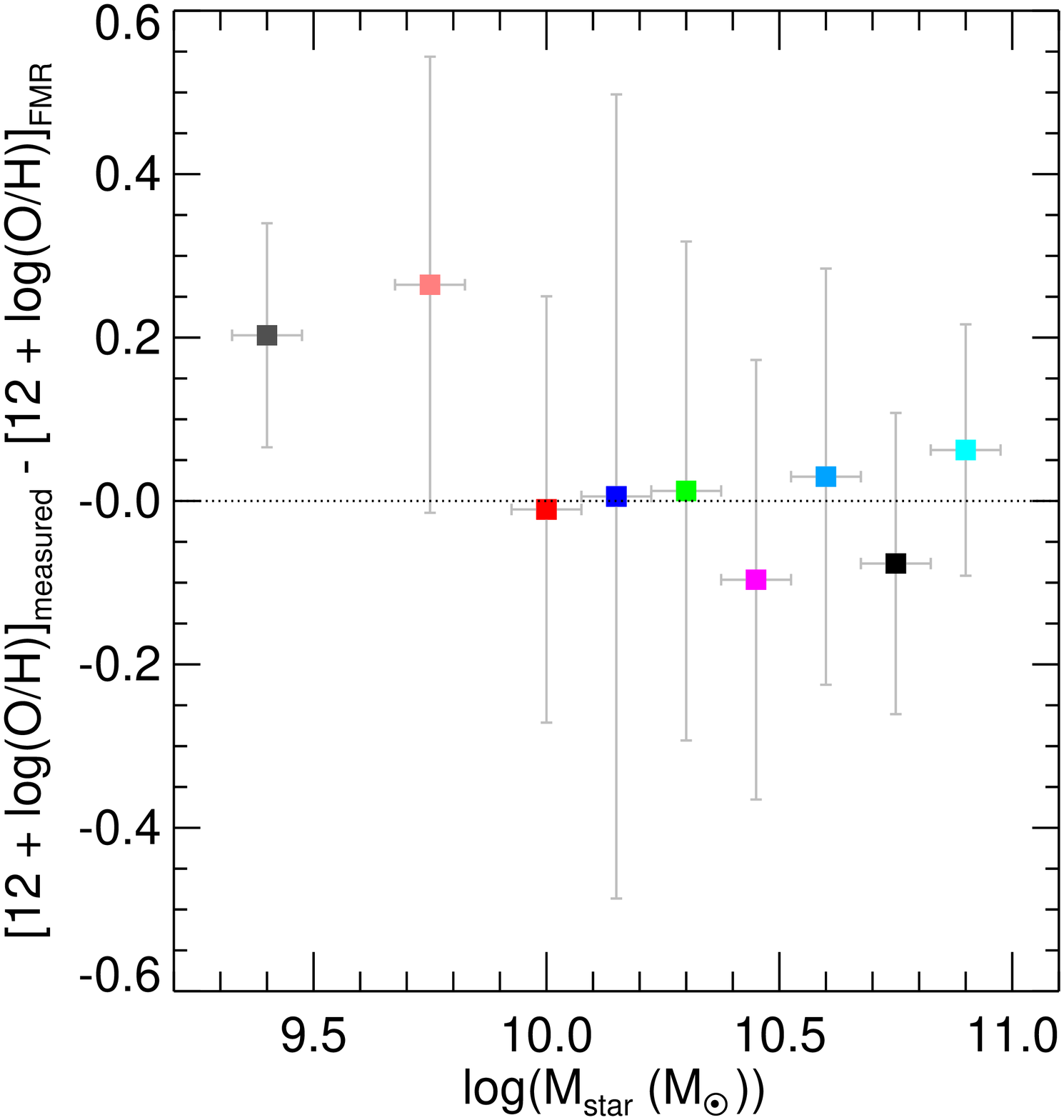}
\end{array}$
\end{center}
\caption{ Left panel: FMR \citep[Eq. 2 of][]{Mannucci2010} for different mass bins as a function of SFR. The colored lines show the mass bin. The colored open squares show the ULIRGs in each mass bin. 
Right panel: Metallicity residuals of ULIRGs from the FMR, colors represent the same mass bins labeled in the left panel. 
The residuals represent the median values in each bin, except the first and the last bins that have single measurements.}
\label{fig:FMR_2D}
\end{figure*}

Star forming galaxies up to $z\sim$2.5 follow the fundamental metallicity relation (FMR), a tight relation between M$_{star}$, gas metallicity and SFR \citep{Mannucci2010}. 
This relation indicates that metallicity decreases with increasing SFR for low M$_{star}$, but for high M$_{star}$ it does not change with SFR.  
So, according to FMR at a fixed mass we expect to have lower metallicities with increasing SFR. 
In order to understand if the lower metallicities of ULIRGs are due to higher SFRs we need to check if they are on the FMR plane. 
We base this investigation on the FMR defined by \citet{Mannucci2010} for local SDSS galaxies. 
Following \citep{Mannucci2010} we divide 47 (H)/ULIRGs into eleven mass bins of 0.15 dex from $\log(M_{star}(M_{\odot}))$=9.70 to 10.90. 
We only consider the bins containing at least 1 galaxy, this selection results in 9 mass bins. 
To be consistent with \citet{Mannucci2010} we use SFR(\Halpha) estimates obtained in \S \ref{S:sfrmass}. 
Since ULIRGs typically have larger SFRs, we extrapolate Eq. (2) of \citet{Mannucci2010} up to $\log$SFR(\Halpha)=2.4. 
Left panel in Figure \ref{fig:FMR_2D} shows the local FMR \citep[Eq. 2 of][]{Mannucci2010} for these mass bins (color coded), open circles show the distribution of ULIRGs in each mass bin (color coded with respect to mass).   
Right panel in Figure \ref{fig:FMR_2D} shows the residuals between the measured metallicities of ULIRGs and FMR. These are the median values in each bin, but the first and the last bin represent residuals of single measurements. 
Without considering the uncertainties, the residuals of ULIRGs from FMR are between 0.09 dex $-$ 0.26 dex.  
This is of course larger than the dispersions of the local SDSS galaxies that is $\sim$0.05 dex, but it indicates that local ULIRGs are consistent with FMR. 
We also note that, the residuals of local ULIRGs are comparable with that of high redshift $z\sim$2 galaxies \citep{Mannucci2010}. 
If we consider the uncertainties, the largest residual is $\sim$0.5 dex, this might indicate an inconsistency with FMR. 
We used the same recipe to infer oxygen abundances and SFRs therefore, the off-set of 0.5 dex can not be due to metallicity or SFR measurements themselves. 
However, the largest contribution to the metallicity uncertainties directly comes from the emission-line flux uncertainties and this point can only be addressed with higher quality data. 
So the large  uncertainties showing $\sim$0.5 dex residuals do not necessarily mean a real off-set from FMR. 
But, also note that ULIRGs are  interacting rare local galaxies with very high SFR, and they are expected to show a large scatter around FMR \citep{Mannucci2010}.

%FIG 18

\subsection{ULIRGs in Color $-$ Magnitude Diagram }\label{S:discolmag}
In \S \ref{S:colmag} we found that local ULIRGs are optically bright and blue galaxies. 
As noted before this is consistent with their starburst nature. 
On the other hand, ULIRGs are dusty galaxies and one might expect them to have redder colors due to dust extinction. 
However, as suggested by \citet{Chen2010} dust distribution in ULIRGs might not be uniform and therefore, their stellar light is not completely obscured.  

The low fraction of ULIRGs in the `green valley', as suggested by \citet{Chen2010}, indicates that ULIRGs are rapidly star forming galaxies and they are not evolved into a transition phase yet. 
The evolution tracks of ULIRGs in the color$-$magnitude diagram is beyond the scope of this paper, therefore for a discussion on this topic we refer \citet{Chen2010}. 

\section{Conclusions}\label{S:conc}
We identified ULIRGs in the \textit{AKARI} all-sky survey by crossmatching \textit{AKARI} catalogs with SDSS DR 10 and 2dFGRS. 
With the advantage of \textit{AKARI} and available SDSS data, we are able to investigate morphologies, stellar masses, SFRs, gas metallicities and optical colors of a large sample of local ULIRGs. 
We have examined the SFR $-$ M$_{star}$, M$_{star} - Z $, SFR $-$M$_{star} - Z $ and color$-$magnitude relations of our local ULIRG sample. 
The following summarizes the main conclusions from this work: 
\begin{itemize}
\item A sample of 118 ULIRGs and one HLIRG with F(90\micron) $\ge$ 0.22Jy have been identified in the \textit{AKARI} all-sky survey. 
40 of the ULIRGs and one HLIRG are newly identified in the \textit{AKARI} all-sky survey based on the spectroscopic redshifts from SDSS DR10 and 2dFGRS. 
The redshift range of our ULIRG sample is 0.050$ < z < $0.487 and the median redshift is 0.181. 
\item In the redshift ($z<$0.27) limited sample of 100 ULIRGs all show interaction features either between two galaxies or in a single system. 
Only 5\%\ are interacting triplets. 
43\%\ of the ULIRGs are two galaxy systems with strong tidal tails, bridges. 
52\%\ of the ULIRGs are ongoing/post mergers showing strong tidal tails or disturbed morphology. 
Our results support the known picture of ULIRGs as mergers.   
\item Based on the adopted optical emission line diagnostics, we confirm the known trend of increasing AGN faction with higher IR luminosity.  
\item Compared to SFR(IR), SFR(\Halpha) strongly underestimates SFR of local ULIRGs, by a factor of $\sim$8. 
This implies that IR observations provide the best estimate of SFR for highly star forming dusty galaxies. 
\item ULIRGs have significantly higher star formation rates compared to the `main sequence' of normal SFGs up to $z\sim$2. 
Local ULIRGs have 92, 17 and 5 times higher SFRs compared to `main sequence' galaxies with similar mass at $z\sim$0, $z\sim$1 and $z\sim$2, respectively. 
Most of the local ULIRGs have lower SFRs compared to the off-main sequence galaxies at $z\sim$2. 
\item We find that ULIRGs have lower gas metallicities compared to the M$_{star} - Z $ relation of normal star forming galaxies; hence we confirm previous studies. 
We also find that local ULIRGs follow the FMR with high dispersions between 0.09 dex$-$0.5 dex, that is similar to that of high redshift ($z\sim$2-3) galaxies. 
\item Compared to previous studies we investigate color properties of a larger ULIRG sample and find that 81\% of the ULIRGs are in `blue cloud', 12\% are in `red sequence' and 7\% are in `green valley'. 
The vast majority of local ULIRGs in our sample are blue galaxies. 
\end{itemize}

We provide the largest local ULIRG comparison sample to further study the M$_{star}$, SFRs, gas metallicities and optical colors of high redshift ULIRGs.  

\subsection*{Acknowledgements}  
We thank the anonymous referee for many insightful comments. 
We thank Marianne Vestergaard and Jens Hjorth for their comments. 
The Dark Cosmology Centre is funded by the Danish National Research Foundation. 
AKARI is a JAXA project with the participation of universities and research institutes in Japan, the European Space Agency (ESA), the IOSG
(Imperial College, UK, Open University, UK, University of Sussex, UK, and University of Groningen, Netherlands) Consortium, and Seoul National University, Korea.
Funding for SDSS-III has been provided by the Alfred P. Sloan Foundation, the Participating Institutions, the National Science Foundation, and the U.S. Department of Energy Office of Science. 
The SDSS-III web site is http://www.sdss3.org/.
SDSS-III is managed by the Astrophysical Research Consortium for the Participating Institutions of the SDSS-III 
Collaboration including the University of Arizona, the Brazilian Participation Group, Brookhaven National Laboratory, Carnegie Mellon University, 
University of Florida, the French Participation Group, the German Participation Group, Harvard University, the Instituto de Astrofisica de Canarias, the 
Michigan State/Notre Dame/JINA Participation Group, Johns Hopkins University, Lawrence Berkeley National Laboratory, Max Planck Institute for Astrophysics, 
Max Planck Institute for Extraterrestrial Physics, New Mexico State University, New York University, Ohio State University, Pennsylvania State University, 
University of Portsmouth, Princeton University, the Spanish Participation Group, University of Tokyo, University of Utah, Vanderbilt University, University of Virginia, 
University of Washington, and Yale University. 
The Digitized Sky Surveys were produced at the Space Telescope Science Institute under U.S. Government grant NAG W-2166. 
The images of these surveys are based on photographic data obtained using the Oschin Schmidt Telescope on Palomar Mountain and the UK Schmidt Telescope. 
The plates were processed into the present compressed digital form with the permission of these institutions. 
The National Geographic Society - Palomar Observatory Sky Atlas (POSS-I) was made by the California Institute of Technology with grants from the National Geographic Society. 
The Second Palomar Observatory Sky Survey (POSS-II) was made by the California Institute of Technology with funds from the National Science Foundation, 
the National Geographic Society, the Sloan Foundation, the Samuel Oschin Foundation, and the Eastman Kodak Corporation. 
The Oschin Schmidt Telescope is operated by the California Institute of Technology and Palomar Observatory. 
The UK Schmidt Telescope was operated by the Royal Observatory Edinburgh, with funding from the UK Science and 
Engineering Research Council (later the UK Particle Physics and Astronomy Research Council), until 1988 June, and thereafter 
by the Anglo-Australian Observatory. The blue plates of the southern Sky Atlas and its Equatorial Extension (together known as the SERC-J), as well as the 
Equatorial Red (ER), and the Second Epoch [red] Survey (SES) were all taken with the UK Schmidt. 
This research has made use of NASA's Astrophysics Data System Bibliographic Service. 

\appendix
\section{IR Images of Newly Identified Sources}
%FIG 19
\begin{figure}[th] 
\begin{center}$
\begin{array}{c}
\includegraphics[scale=0.30]{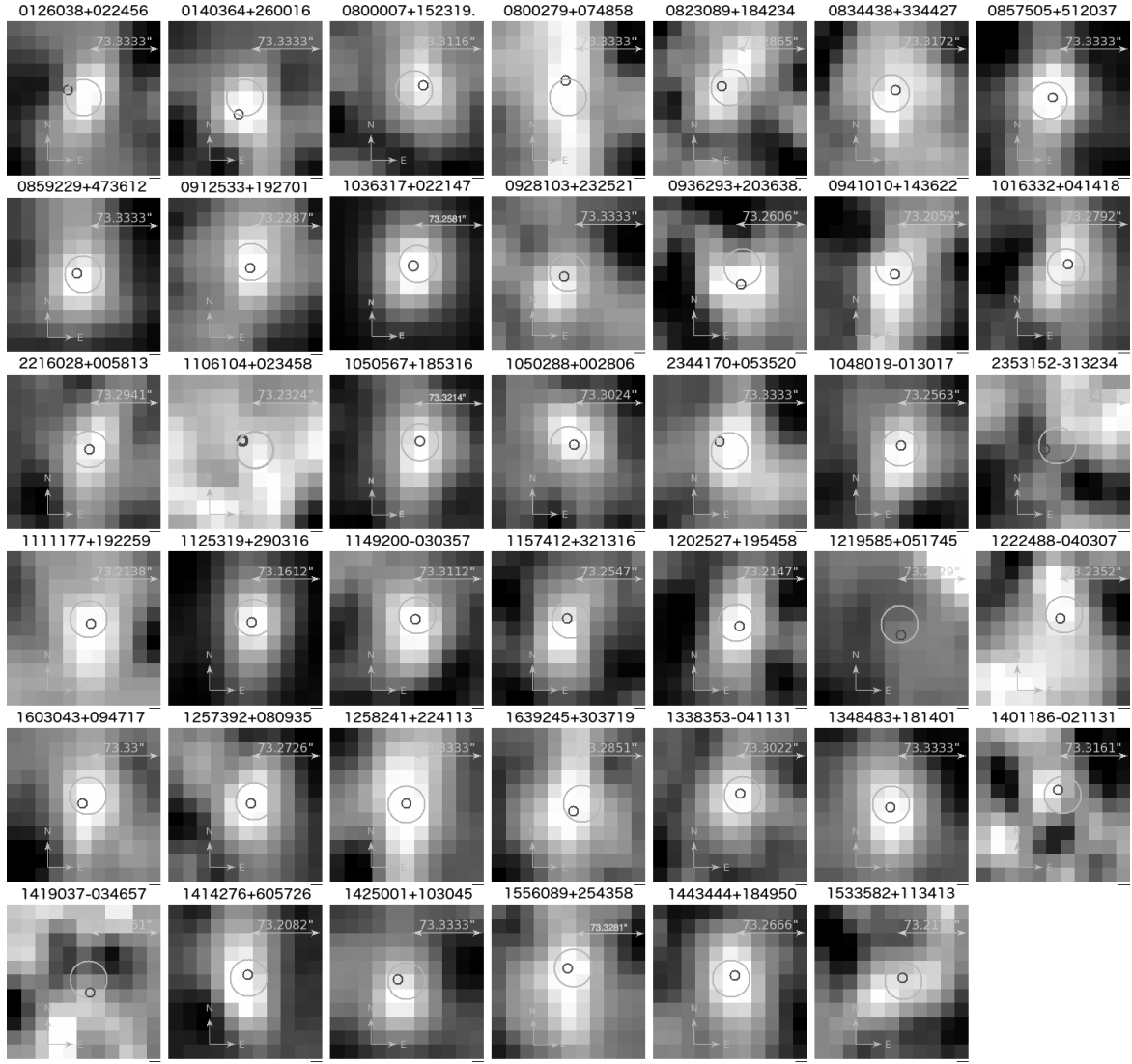}\\ 
\end{array}$
\end{center}
\caption{The \textit{AKARI} 90\micron\ images of the all newly identified ULIRGs and one HLIRG. 
The scale of the images are 165\arcsec$\times$165\arcsec. 
The 5\arcsec\ radius small circles represent the optical counterpart 
and the 20\arcsec\ radius big circle shows the IR source.}
\label{fig:figB1}
\end{figure}
\clearpage

%****************************************************************************
%***********BIBLIOGRAPHY STARTS HERE*****************************************
%****************************************************************************

\clearpage
%****************************************************************************
%***********TABLES START HERE************************************************
%****************************************************************************
% TABLE1
\clearpage
\onecolumngrid
\LongTables
\begin{landscape}
\begin{deluxetable}{llccccccccccccc} 
%\rotate
\tablecolumns{15}
\tabletypesize{\scriptsize}
\setlength{\tabcolsep}{0.01in}
\tablewidth{0pt}
\tablecaption{New ULIRGs Sample}
\tablehead{
\colhead{Name} &
\colhead{AKARI RA} &
\colhead{AKARI DEC}&
\colhead{Other Name}&
\colhead{z\tablenotemark{a}}&
\colhead{$\log(L_{IR}/L_{\odot})$\tablenotemark{b}} &
\colhead{F(65\micron)\tablenotemark{c}} &
\colhead{F(90\micron)\tablenotemark{c}} &
\colhead{F(140\micron)\tablenotemark{c}} &
\colhead{F(160\micron)\tablenotemark{c}} &
\colhead{$r$\tablenotemark{d}} &
\colhead{IC\tablenotemark{e}} &
\colhead{IC\tablenotemark{f}} &
\colhead{Note\tablenotemark{g}}&
\colhead{Spectral\tablenotemark{h}} \\
\colhead{AKARI-FIS-V1} &
\colhead{(J2000)} &
\colhead{(J2000)} &
\colhead{} &
\colhead{} &
\colhead{} &
\colhead{(Jy)} &
\colhead{(Jy)} &
\colhead{(Jy)} &
\colhead{(Jy)} &
\colhead{(mag)} &
\colhead{} &
\colhead{Ref.} &
\colhead{} &
\colhead{Class} \\
\colhead{(1)} &
\colhead{(2)} &
\colhead{(3)} &
\colhead{(4)} &
\colhead{(5)} &
\colhead{(6)} &
\colhead{(7)} &
\colhead{(8)} &
\colhead{(9)} &
\colhead{(10)} &
\colhead{(11)} &
\colhead{(12)} &
\colhead{(13)} &
\colhead{(14)}&
\colhead{(15)}}
\startdata
 \hline
J2216028+005813& 22 16 02.82& +00 58 13.5&SDSS J221602.70+005811.0&0.21$^{\ast}$&$12.83^{+ 0.15}_{- 0.14}$&  0.54$\pm$  0.13&  0.54$\pm$  0.05&  0.96$\pm$  0.63&  1.14$\pm$  0.28&14.30&IIIa&3&\nodata&\nodata\\ 
J0859229+473612& 08 59 22.93& +47 36 11.7&SDSS J085923.61+473610.5&0.180&$12.20^{+ 0.01}_{- 0.09}$&\nodata&  0.48$\pm$  0.06&  1.95$\pm$  0.29&\nodata&14.44&IIIa&3&\nodata& Star Forming\\
J1443444+184950& 14 43 44.44& +18 49 49.7&SDSS J144344.64+184945.7&0.177&$12.21^{+ 0.01}_{- 0.31}$&  0.43$\pm$  0.11&  0.55$\pm$  0.04&  1.69$\pm$  0.53&\nodata&16.03&V&3&\nodata& LINER\\
J0857505+512037& 08 57 50.48& +51 20 37.2&SDSS J085750.79+512032.6&0.366&$12.89^{+ 0.07}_{- 0.02}$&  1.12$\pm$  0.28&  0.68$\pm$  0.06&  1.58$\pm$  0.08&\nodata&14.40&IIIb&3&\nodata& LINER\\ 
J1106104+023458& 11 06 10.37& +02 34 57.8&SDSS J110611.44+023502.2&0.283&$12.23^{+ 0.06}_{- 0.06}$&\nodata&  0.42$\pm$  0.01&  1.42$\pm$  0.36&  0.56$\pm$  0.14&17.17&NI&3&B&Seyfert\\ 
J1157412+321316& 11 57 41.21& +32 13 16.4&SDSS J115741.47+321316.4&0.160&$12.14^{+ 0.01}_{- 0.12}$&  0.66$\pm$  0.16&  0.60$\pm$  0.03&  2.22$\pm$  1.21&  2.23$\pm$  0.40&16.37&V&3&\nodata& Star Forming\\ 
J1149200-030357& 11 49 20.03& -03 03 57.3&SDSS J114920.04-030402.1&0.162&$12.02^{+ 0.02}_{- 0.04}$&  0.19$\pm$  0.05&  0.42$\pm$  0.02&  1.30$\pm$  0.33&  2.58$\pm$  0.21&13.97&V,G&3&\nodata& Star Forming\\ 
J0126038+022456& 01 26 03.80& +02 24 55.9&SDSS J012604.62+022509.9&0.242&$12.22^{+ 0.04}_{- 0.06}$&\nodata&  0.63$\pm$  0.01&  0.70$\pm$  0.18&\nodata&14.73&Tp1,G&3&\nodata& LINER\\
J1556089+254358& 15 56 08.92& +25 43 57.8&SDSS J155609.36+254355.9&0.154&$12.03^{+ 0.01}_{- 0.26}$&  0.56$\pm$  0.14&  0.51$\pm$  0.01&  2.26$\pm$  0.57&\nodata&16.76&IIIa,G&3&\nodata&Composite\\ 
J0140364+260016& 01 40 36.40& +26 00 15.9&SDSS J014037.36+260001.5&0.321&$12.77^{+ 0.06}_{- 0.02}$&  0.57$\pm$  0.14&  0.56$\pm$  0.03&  0.70$\pm$  0.17&  2.51$\pm$  0.15&16.01&IIIa,G&3&\nodata&Seyfert\\ 
J1257392+080935& 12 57 39.15& +08 09 35.1&SDSS J125739.33+080931.7&0.272&$12.24^{+ 0.04}_{- 0.02}$&  0.46$\pm$  0.12&  0.51$\pm$  0.01&\nodata&  0.22$\pm$  0.06&17.72&NI&3&\nodata&QSO\\ 
J0800007+152319& 08 00 00.68& +15 23 18.7&SDSS J080000.05+152326.0&0.274&$12.14^{+ 0.00}_{- 0.00}$&\nodata&  0.43$\pm$  0.03&\nodata&\nodata&14.50&V&3&A&LINER$^{\star}$\\
 J0800279+074858& 08 00 27.92& +07 48 57.6&SDSS J080028.37+074915.5&0.173&$12.12^{+ 0.00}_{- 0.09}$&  0.34$\pm$  0.09&  0.35$\pm$  0.09&  2.29$\pm$  0.57&  2.04$\pm$  0.51&13.10&V,G&3&\nodata& LINER\\ 
J0834438+334427& 08 34 43.82& +33 44 27.2&SDSS J083443.56+334432.5&0.166&$12.13^{+ 0.03}_{- 0.20}$&  0.59$\pm$  0.15&  0.65$\pm$  0.04&\nodata&  2.07$\pm$  0.52&15.90&IIIb&3&A&Composite\\ 
J0823089+184234& 08 23 08.91& +18 42 33.9&SDSS J082309.51+184233.4&0.425&$12.57^{+ 0.43}_{- 0.03}$&  0.45$\pm$  0.11&  0.41$\pm$  0.02&\nodata&\nodata&12.52&V&3&\nodata&\nodata\\ 
J1202527+195458& 12 02 52.69& +19 54 58.4&SDSS J120252.39+195456.7&0.132&$12.05^{+ 0.04}_{- 0.02}$&  0.11$\pm$  0.03&  0.54$\pm$  0.05&  0.69$\pm$  0.10&  3.43$\pm$  0.04&14.50&IIIb&3&\nodata& Star Forming\\
J0912533+192701& 09 12 53.33& +19 27 00.8&SDSS J091253.25+192653.9&0.233&$12.11^{+ 0.09}_{- 0.04}$&  0.17$\pm$  0.04&  0.44$\pm$  0.04&  0.68$\pm$  0.17&  0.93$\pm$  0.23&15.49&V&3&\nodata&Composite\\ 
J0941010+143622& 09 41 01.03& +14 36 22.4&SDSS J094100.81+143614.5&0.384&$12.75^{+ 0.01}_{- 0.05}$&  0.89$\pm$  0.22&  0.77$\pm$  0.07&  0.24$\pm$  0.06&  0.46$\pm$  0.11&17.19&NI&3&B&QSO\\ 
J1016332+041418& 10 16 33.25& +04 14 17.9&SDSS J101633.19+041422.1&0.266&$12.39^{+ 0.03}_{- 0.05}$&  0.51$\pm$  0.13&  0.65$\pm$  0.10&  0.91$\pm$  0.01&  1.34$\pm$  0.33&14.73&IIIb&3&\nodata&Composite\\ 
J1401186-021131& 14 01 18.61& -02 11 30.9&SDSS J140119.02-021126.7&0.172&$12.07^{+ 0.01}_{- 0.08}$&  0.26$\pm$  0.06&  0.30$\pm$  0.08&  1.85$\pm$  0.30&\nodata&15.98&IIIa&3&\nodata&Seyfert\\ 
J1258241+224113& 12 58 24.10& +22 41 13.0&SDSS J125824.16+224113.6&0.208&$12.07^{+ 0.05}_{- 0.04}$&  0.97$\pm$  0.24&  0.60$\pm$  0.05&  1.44$\pm$  1.09&  0.86$\pm$  0.21&16.62&IIIb&3&\nodata&Composite\\
J1036317+022147& 10 36 31.66& +02 21 47.3&SDSS J103631.88+022144.1&0.050&$12.06^{+ 0.03}_{- 0.04}$& 13.32$\pm$  1.09& 14.81$\pm$  0.54& 10.83$\pm$  1.15&  8.46$\pm$  0.74&14.75&IIIb&3&\nodata&Composite\\ 
J1050567+185316& 10 50 56.73& +18 53 16.1&SDSS J105056.78+185316.9&0.219&$12.60^{+ 0.03}_{- 0.06}$&  0.67$\pm$  0.17&  0.93$\pm$  0.10&  3.08$\pm$  2.13&  3.32$\pm$  0.45&14.91&Tp1,G&3&\nodata&Seyfert\\ 
J1111177+192259& 11 11 17.72& +19 22 58.9&SDSS J111117.46+192255.0&0.225&$12.60^{+ 0.03}_{- 0.06}$&\nodata&  0.55$\pm$  0.05&  0.05$\pm$  2.25&\nodata&14.56&IIIb&3&\nodata& Star Forming\\
J1219585+051745& 12 19 58.50& +05 17 44.6&SDSS J121958.11+051735.1&0.487&$12.87^{+ 0.02}_{- 0.04}$&  0.29$\pm$  0.07&  0.82$\pm$  0.09&  0.43$\pm$  0.11&\nodata&15.16&NI&3&A&\nodata\\  
J1414276+605726& 14 14 27.55& +60 57 25.8&SDSS J141427.98+605727.0&0.151&$12.11^{+ 0.00}_{- 0.12}$&  0.37$\pm$  0.09&  0.63$\pm$  0.04&  1.94$\pm$  0.36&  2.75$\pm$  1.38&14.11&V&3&\nodata&Composite\\ 
J0936293+203638& 09 36 29.33& +20 36 37.6&SDSS J093629.03+203620.0&0.175&$12.01^{+ 0.04}_{- 0.02}$&  0.42$\pm$  0.11&  0.57$\pm$  0.09&  2.05$\pm$  0.23&  0.83$\pm$  0.21&14.12&IIIb&3&\nodata&Composite\\ 
J1533582+113413& 15 33 58.15& +11 34 12.7&SDSS J153358.24+113415.8&0.337&$12.32^{+ 0.08}_{- 0.08}$&  0.25$\pm$  0.06&  0.33$\pm$  0.04&  0.46$\pm$  0.11&\nodata&15.29&V,G&3&A&\nodata\\ 
J1348483+181401& 13 48 48.32& +18 14 00.9&SDSS J134848.32+181357.4&0.179&$12.19^{+ 0.03}_{- 0.03}$&  0.42$\pm$  0.11&  0.66$\pm$  0.04&  1.66$\pm$  0.03&  1.17$\pm$  0.29&14.89&IV&3&\nodata&Composite\\ 
J1125319+290316& 11 25 31.92& +29 03 16.2&SDSS J112531.90+290311.3&0.138&$12.27^{+ 0.01}_{- 0.05}$&  1.98$\pm$  0.14&  1.84$\pm$  0.09&  1.53$\pm$  0.56&\nodata&13.34&IV&3&\nodata&QSO\\
J1603043+094717& 16 03 04.29& +09 47 17.5&SDSS J160304.57+094707.8&0.152&$12.02^{+ 0.03}_{- 0.03}$&  0.79$\pm$  0.20&  0.54$\pm$  0.07&  1.70$\pm$  0.01&  0.36$\pm$  0.09&13.11&V&3&\nodata& Star Forming\\ 
J1639245+303719& 16 39 24.50& +30 37 19.1&SDSS J163925.01+303709.8&0.224&$12.11^{+ 0.01}_{- 0.05}$&  0.51$\pm$  0.13&  0.51$\pm$  0.04&  1.14$\pm$  0.23&  0.12$\pm$  0.03&16.53&IIIb&3&\nodata&\nodata\\
J1050288+002806& 10 50 28.80& +00 28 06.0&SDSS J105028.49+002807.7&0.216&$12.38^{+ 0.13}_{- 0.08}$&  0.53$\pm$  0.13&  0.79$\pm$  0.06&  1.79$\pm$  0.37&\nodata&14.77&IIIb&3&\nodata&Composite\\ 
J0928103+232521& 09 28 10.29& +23 25 21.0&SDSS J092810.52+232515.8&0.197&$12.07^{+ 0.08}_{- 0.04}$&  0.09$\pm$  0.02&  0.44$\pm$  0.01&  1.13$\pm$  0.28&\nodata&15.40&IIIb&3&\nodata& Star Forming\\ 
J2344170+053520& 23 44 17.04& +05 35 19.8&SDSS J234417.43+053533.5&0.267&$12.56^{+ 0.01}_{- 0.19}$&  0.64$\pm$  0.16&  0.42$\pm$  0.06&  1.29$\pm$  0.90&  1.80$\pm$  0.45&15.80&IIIa&3&\nodata&Seyfert\\ 
J2353152-313234& 23 53 15.20& -31 32 34.5&TGS430Z217&0.185&$12.04^{+ 0.00}_{- 0.12}$&  0.37$\pm$  0.09&  0.23$\pm$  0.06&  1.67$\pm$  0.44&  2.78$\pm$  0.70&14.29&\nodata&\nodata&\nodata&\nodata\\ 
J1222488-040307& 12 22 48.78& -04 03 07.1&TGN123Z071&0.181&$12.19^{+ 0.02}_{- 0.04}$&  0.16$\pm$  0.04&  0.59$\pm$  0.10&  1.68$\pm$  0.04&  1.09$\pm$  0.27&15.17&\nodata&\nodata&\nodata&\nodata\\
J1419037-034657& 14 19 03.68& -03 46 56.6&TGN145Z052&0.152&$12.09^{+ 0.01}_{- 0.08}$&\nodata&  0.61$\pm$  0.15&  1.86$\pm$  0.66&  2.49$\pm$  0.62&16.95&\nodata&\nodata&\nodata&\nodata\\
J1048019-013017& 10 48 01.87& -01 30 17.4&TGN296Z061&0.167&$12.09^{+ 0.07}_{- 0.07}$&  0.49$\pm$  0.12&  1.10$\pm$  0.15&  1.13$\pm$  0.40&\nodata&13.53&IV&3&\nodata&\nodata\\ 
J1338353-041131& 13 38 35.29& -04 11 31.4&TGN139Z200&0.175&$12.14^{+ 0.12}_{- 0.22}$&  0.40$\pm$  0.10&  0.68$\pm$  0.29&  1.77$\pm$  0.68&\nodata&15.67&\nodata&\nodata&\nodata&\nodata\\
\enddata
\tablenotetext{a}{Based on SDSS optical spectra. The redshifts marked with asterisk are adopted from \citet{2009MNRAS.398..109W} since there are no available SDSS spectra for these sources.}
\tablenotetext{b}{$L_{IR}$ is the total IR luminosity between 8\micron-1000\micron\ measured from the SEDs fitted to the \textit{AKARI} fluxes at 65, 90, 140 and 160 \micron.}
\tablenotetext{c}{The {\it{AKARI}} flux density and the associated uncertainties are adopted from \citet{Yamamura2010}. For the cases for which the flux uncertainties are not available we adopt 25\%\ of the measured flux as the uncertainty.}
\tablenotetext{d}{SDSS Petrosian $r$ magnitude.}
\tablenotetext{e}{Interaction Classes (IC) are described in \S \ref{S:morphprop}. The IC listed in this Table are based on combined $gri$ SDSS images; `\nodata' entries in the Table indicate lack of SDSS image.}
\tablenotetext{f}{References: (1) \citet{Veilleux2002}, (2) \citet{Hwang2007}, (3) This work.}
\tablenotetext{g}{Superposition of other sources in the field: (A) Star; (B) Galaxy.}
\tablenotetext{h}{Classifications based on optical spectra. 
For the sources that are in the SDSS DR10 \textit{emissionlinesPort}\footnote[7]{http://www.sdss3.org/dr10/} catalogue \citep{Thomas2013} we adopt the given BPT classifications: `Star Forming', `Composite', `LINER' and `Seyfert'. 
 In the case of quasars (QSO) we adopt the `spectrotype' classification from the \textit{galSpecInfo} catalogue \citep[see][for the details of SDSS spectroscopic target selection]{Richards2002}. 
The classifications marked with a star are based on our BPT classification based on the emission-line fluxes adopted from SDSS catalogue \textit{galSpecLine}$^{7}$ \citep{Tremonti2004,Brinchmann2004}.}
\label{tab:newULIRGs}
\end{deluxetable}
\clearpage
\end{landscape}
\twocolumngrid

% TABLE2
\clearpage
\onecolumngrid
\LongTables
\begin{landscape}
\begin{deluxetable}{llccccccccccccc} 
%\rotate
\tablecolumns{15}
\tabletypesize{\scriptsize}
\setlength{\tabcolsep}{0.02in}
\tablewidth{0pt}
\tablecaption{Known ULIRGs Sample}
\tablehead{
\colhead{Name} &
\colhead{AKARI RA} &
\colhead{AKARI DEC}&
\colhead{Other Name}&
\colhead{z\tablenotemark{a}}&
\colhead{$\log(L_{IR}/L_{\odot})$\tablenotemark{b}} &
\colhead{F(65\micron)\tablenotemark{c}} &
\colhead{F(90\micron)\tablenotemark{c}} &
\colhead{F(140\micron)\tablenotemark{c}} &
\colhead{F(160\micron)\tablenotemark{c}} &
\colhead{$r$\tablenotemark{d}} &
\colhead{IC\tablenotemark{e}} &
\colhead{IC\tablenotemark{f}} &
\colhead{Note\tablenotemark{g}}&
\colhead{Spectral\tablenotemark{h}} \\
\colhead{AKARI-FIS-V1} &
\colhead{(J2000)} &
\colhead{(J2000)} &
\colhead{} &
\colhead{} &
\colhead{} &
\colhead{(Jy)} &
\colhead{(Jy)} &
\colhead{(Jy)} &
\colhead{(Jy)} &
\colhead{(mag)} &
\colhead{} &
\colhead{Ref.} &
\colhead{} &
\colhead{Class} \\
\colhead{(1)} &
\colhead{(2)} &
\colhead{(3)} &
\colhead{(4)} &
\colhead{(5)} &
\colhead{(6)} &
\colhead{(7)} &
\colhead{(8)} &
\colhead{(9)} &
\colhead{(10)} &
\colhead{(11)} &
\colhead{(12)} &
\colhead{(13)} &
\colhead{(14)}&
\colhead{(15)}}
\startdata
 \hline
 J0857064+190855& 08 57 06.37& +19 08 55.4&SDSS J085706.35+190853.5&0.331&$12.92^{+ 0.01}_{- 0.05}$&\nodata&  0.48$\pm$  0.06&\nodata&  2.56$\pm$  0.10&15.79&NI&3&\nodata&QSO\\ 
 J1022125+241208& 10 22 12.46& +24 12 07.8&SDSS J102212.64+241202.4&0.188&$12.01^{+ 0.22}_{- 0.08}$&\nodata&  0.59$\pm$  0.03&  0.90$\pm$  2.81&\nodata&15.80&IV&3&\nodata& LINER\\ 
 J1422313+260205& 14 22 31.27& +26 02 05.2&SDSS J142231.37+260205.1&0.159&$12.20^{+ 0.11}_{- 0.05}$&  0.96$\pm$  0.24&  1.36$\pm$  0.04&  2.29$\pm$  0.43&\nodata&13.59&IIIa&1&\nodata& Star Forming\\
J1231216+275524& 12 31 21.57& +27 55 24.4&SDSS J123121.37+275524.0&0.212&$12.33^{+ 0.01}_{- 0.09}$&  0.52$\pm$  0.13&  0.40$\pm$  0.10&  1.41$\pm$  0.35&  1.93$\pm$  0.48&17.35&IV&3&\nodata&Composite\\  
J1251200+021900& 12 51 20.03& +02 19 00.2&SDSS J125120.04+021902.4&0.253&$12.48^{+ 0.04}_{- 0.02}$&  1.00$\pm$  0.25&  0.73$\pm$  0.04&  1.86$\pm$  0.09&  0.81$\pm$  0.20&15.00&V&2&\nodata&Composite\\
J0030089-002743& 00 30 08.95& -00 27 43.5&SDSS J003009.08-002744.2&0.242$^{\ast}$&$12.46^{+ 0.03}_{- 0.03}$&  0.37$\pm$  0.09&  0.62$\pm$  0.06&  1.34$\pm$  0.30&  0.40$\pm$  0.10&14.56&IIIb&3&\nodata&\nodata\\ 
J0914140+032200& 09 14 14.01& +03 22 00.4&SDSS J091413.79+032201.4&0.145&$12.07^{+ 0.06}_{- 0.03}$&  1.11$\pm$  0.28&  1.39$\pm$  0.11&  2.29$\pm$  0.51&  0.86$\pm$  0.21&14.48&IIIa&1&\nodata& LINER\\ 
J1105377+311432& 11 05 37.71& +31 14 32.3&SDSS J110537.54+311432.2&0.199&$12.20^{+ 0.05}_{- 0.02}$&  0.71$\pm$  0.18&  0.98$\pm$  0.06&  1.31$\pm$  0.33&  0.59$\pm$  0.15&16.03&IV&1&\nodata&Composite\\ 
J0323227-075612& 03 23 22.75& -07 56 12.1&SDSS J032322.87-075615.3&0.166&$12.14^{+ 0.11}_{- 0.06}$&  0.50$\pm$  0.12&  0.91$\pm$  0.00&  1.80$\pm$  0.25&\nodata&12.99&IV&1&\nodata&Composite\\ 
J1632212+155145& 16 32 21.24& +15 51 44.8&SDSS J163221.38+155145.5&0.242&$12.67^{+ 0.02}_{- 0.04}$&  1.50$\pm$  0.13&  1.46$\pm$  0.03&  2.42$\pm$  0.48&  2.75$\pm$  0.06&14.27&V&1&\nodata&Composite\\
J0148531+002857& 01 48 53.10& +00 28 57.1&SDSS J014852.57+002859.8&0.280&$12.30^{+ 0.26}_{- 0.04}$&  0.70$\pm$  0.17&  0.48$\pm$  0.05&  0.94$\pm$  0.39&\nodata&15.49&IIIa&2&\nodata&Composite\\ 
J0159503+002340& 01 59 50.28& +00 23 39.9&SDSS J015950.25+002340.9&0.163&$12.43^{+ 0.01}_{- 0.04}$&  2.01$\pm$  0.03&  1.82$\pm$  0.17&  2.94$\pm$  0.18&  0.16$\pm$  0.04&15.63&IV&1&\nodata&QSO\\
J1353317+042809& 13 53 31.72& +04 28 08.8&SDSS J135331.57+042805.3&0.136$^{\ast}$&$12.44^{+ 0.01}_{- 0.05}$&  1.26$\pm$  0.31&  1.62$\pm$  0.05&  0.23$\pm$  0.06&  1.61$\pm$  0.40&12.67&IV&1&B&\nodata\\ 
J0244173-003040& 02 44 17.35& -00 30 40.3&SDSS J024417.44-003041.1&0.200&$12.07^{+ 0.01}_{- 0.05}$&  0.36$\pm$  0.09&  0.65$\pm$  0.19&  0.77$\pm$  0.19&  0.86$\pm$  0.22&15.08&IIIa,G&2&\nodata&QSO\\
J1202268-012918& 12 02 26.81& -01 29 18.0&SDSS J120226.76-012915.3&0.150&$12.36^{+ 0.05}_{- 0.01}$&  1.94$\pm$  0.39&  2.54$\pm$  0.20&  3.09$\pm$  0.67&  1.06$\pm$  1.26&15.68&IV,G&3&\nodata&Star Forming$^{\star}$\\ 
J1013477+465402& 10 13 47.75& +46 54 02.1&SDSS J101348.09+465359.6&0.206&$12.24^{+ 0.09}_{- 0.05}$&  0.13$\pm$  0.03&  0.72$\pm$  0.05&\nodata&\nodata&16.78&IV&3&\nodata&Seyfert\\ 
J0858418+104124& 08 58 41.77& +10 41 24.3&SDSS J085841.77+104122.1&0.148&$12.17^{+ 0.05}_{- 0.03}$&  1.00$\pm$  0.25&  1.55$\pm$  0.13&  2.09$\pm$  0.38&  2.62$\pm$  0.94&16.17&IV,G&1&\nodata&Seyfert\\ 
J1347336+121727& 13 47 33.58& +12 17 27.4&SDSS J134733.36+121724.3&0.120&$12.18^{+ 0.03}_{- 0.03}$&  1.90$\pm$  0.47&  1.75$\pm$  0.08&  0.96$\pm$  0.24&  0.97$\pm$  0.24&14.13&IIIb&1&\nodata&Seyfert$^{\star}$\\
J0853252+252646& 08 53 25.21& +25 26 45.6&SDSS J085325.07+252656.0&0.256&$12.37^{+ 0.09}_{- 0.08}$&  0.56$\pm$  0.14&  0.68$\pm$  0.06&  0.68$\pm$  4.39&  1.28$\pm$  0.32&11.72&V&3&B&QSO\\ 
J0825215+383306& 08 25 21.47& +38 33 05.7&SDSS J082521.65+383258.5&0.206&$12.28^{+ 0.10}_{- 0.08}$&  0.26$\pm$  0.06&  0.59$\pm$  0.01&  1.49$\pm$  0.37&\nodata&16.03&IIIb&2&\nodata&Composite\\ 
J0829512+384528& 08 29 51.18& +38 45 27.8&SDSS J082951.39+384523.7&0.195&$12.02^{+ 0.11}_{- 0.07}$&  0.33$\pm$  0.08&  0.53$\pm$  0.09&  0.89$\pm$  0.22&\nodata&13.34&IIIb&2&\nodata&Composite\\ 
J1142035+005135& 11 42 03.51& +00 51 35.5&SDSS J114203.41+005135.8&0.245&$12.10^{+ 0.04}_{- 0.04}$&  0.37$\pm$  0.09&  0.45$\pm$  0.05&  0.15$\pm$  0.04&  0.67$\pm$  0.17&15.26&IV&2&\nodata&Composite\\ 
J0810595+281354& 08 10 59.51& +28 13 54.1&SDSS J081059.61+281352.2&0.336&$12.65^{+ 0.18}_{- 0.14}$&\nodata&  0.63$\pm$  0.01&  1.10$\pm$  0.28&\nodata&15.40&IIIb&2&\nodata&Composite\\ 
J0900252+390400& 09 00 25.21& +39 03 59.8&SDSS J090025.37+390353.7&0.058&$12.03^{+ 0.01}_{- 0.04}$&  5.93$\pm$  0.70&  5.13$\pm$  0.19&  2.62$\pm$  0.91&  1.43$\pm$  0.36&15.97&IIIb&1&\nodata& Star Forming\\ 
J0830197+192040& 08 30 19.74& +19 20 40.0&SDSS J083019.75+192050.0&0.186&$12.03^{+ 0.12}_{- 0.07}$&  0.49$\pm$  0.12&  0.59$\pm$  0.07&\nodata&  1.30$\pm$  0.33&15.02&IIIb&3&\nodata& Star Forming\\ 
J1121293+112233& 11 21 29.25& +11 22 33.3&SDSS J112129.00+112225.7&0.185&$12.32^{+ 0.04}_{- 0.03}$&  0.83$\pm$  0.21&  1.07$\pm$  0.03&  2.16$\pm$  0.11&\nodata&13.37&IIIa&2&\nodata&Seyfert\\ 
J1006038+411223& 10 06 03.83& +41 12 23.4&SDSS J100603.85+411224.8&0.328&$12.42^{+ 0.03}_{- 0.05}$&  0.28$\pm$  0.07&  0.54$\pm$  0.03&  0.47$\pm$  0.12&\nodata&15.36&V&2&\nodata& Star Forming\\ 
J0838034+505516& 08 38 03.36& +50 55 16.5&SDSS J083803.61+505508.9&0.097&$12.03^{+ 0.01}_{- 0.04}$&  2.31$\pm$  0.04&  2.00$\pm$  0.08&  2.62$\pm$  0.40&\nodata&17.30&IV&2&\nodata&Composite\\ 
J0902489+523623& 09 02 48.87& +52 36 22.6&SDSS J090248.90+523624.7&0.157&$12.05^{+ 0.01}_{- 0.06}$&  1.20$\pm$  0.30&  0.87$\pm$  0.04&  0.25$\pm$  0.06&  1.56$\pm$  0.39&15.72&V&1&\nodata&Composite\\ 
J0847504+232113& 08 47 50.37& +23 21 12.8&SDSS J084750.26+232110.9&0.152&$12.01^{+ 0.09}_{- 0.08}$&\nodata&  0.75$\pm$  0.06&  1.66$\pm$  0.39&\nodata&13.62&IIIb&3&\nodata& Star Forming\\ 
J1559301+380843& 15 59 30.13& +38 08 42.7&SDSS J155930.40+380838.8&0.218&$12.19^{+ 0.02}_{- 0.04}$&  0.07$\pm$  0.02&  0.58$\pm$  0.06&  1.36$\pm$  0.14&  2.17$\pm$  0.38&13.85&IV&2&\nodata&Seyfert\\  
J1324197+053705& 13 24 19.74& +05 37 05.4&SDSS J132419.89+053704.7&0.203&$12.66^{+ 0.02}_{- 0.04}$&  1.22$\pm$  0.30&  0.89$\pm$  0.08&\nodata&\nodata&14.68&V&1&\nodata&QSO\\ 
J1102140+380240& 11 02 14.02& +38 02 40.0&SDSS J110214.00+380234.6&0.158&$12.15^{+ 0.06}_{- 0.01}$&  1.12$\pm$  0.28&  1.32$\pm$  0.03&  2.36$\pm$  0.70&  0.46$\pm$  0.20&15.27&IIIb&1&\nodata&Composite\\ 
J1204244+192509& 12 04 24.41& +19 25 08.9&SDSS J120424.54+192509.8&0.168&$12.15^{+ 0.02}_{- 0.04}$&  1.62$\pm$  0.41&  1.24$\pm$  0.03&  1.19$\pm$  0.04&  0.79$\pm$  0.20&15.38&IV&1&\nodata&Composite\\ 
J1108513+065915& 11 08 51.31& +06 59 15.1&SDSS J110851.03+065901.5&0.182&$12.07^{+ 0.09}_{- 0.11}$&  0.30$\pm$  0.07&  0.52$\pm$  0.05&  0.29$\pm$  4.55&  1.68$\pm$  0.42&15.64&IIIa&3&B&QSO\\ 
J1040290+105325& 10 40 29.05& +10 53 25.3&SDSS J104029.17+105318.3&0.136&$12.26^{+ 0.06}_{- 0.03}$&  2.02$\pm$  0.50&  2.13$\pm$  0.13&  1.46$\pm$  2.03&  0.75$\pm$  0.19&14.73&IV&1&\nodata& LINER\\
J1207210+021702& 12 07 21.03& +02 17 01.7&SDSS J120721.45+021657.8&0.222&$12.09^{+ 0.04}_{- 0.04}$&  1.32$\pm$  0.33&  0.58$\pm$  0.02&  0.54$\pm$  0.13&  0.81$\pm$  0.20&14.36&V&2&\nodata&Composite\\ 
J1255482-033908& 12 55 48.19& -03 39 08.2&SDSS J125547.83-033909.6&0.169&$12.06^{+ 0.13}_{- 0.08}$&\nodata&  0.73$\pm$  0.02&\nodata&  1.66$\pm$  0.41&15.53&IV&2&A&QSO\\ 
J0906339+045136& 09 06 33.93& +04 51 35.5&SDSS J090634.03+045127.6&0.125&$12.02^{+ 0.06}_{- 0.04}$&  1.62$\pm$  0.40&  1.64$\pm$  0.07&  2.94$\pm$  1.55&  2.43$\pm$  0.61&13.68&IV&1&\nodata&Composite\\ 
J1153144+131432& 11 53 14.39& +13 14 32.1&SDSS J115314.23+131427.9&0.127&$12.26^{+ 0.05}_{- 0.02}$&  2.37$\pm$  0.16&  2.57$\pm$  0.14&  2.76$\pm$  0.55&  1.03$\pm$  1.03&13.50&IV&1&\nodata&Composite\\ 
J1202054+112813& 12 02 05.41& +11 28 13.2&SDSS J120205.59+112812.2&0.194&$12.19^{+ 0.05}_{- 0.03}$&  0.07$\pm$  0.02&  1.03$\pm$  0.03&\nodata&  0.87$\pm$  0.22&14.85&IIIa&2&A& Star Forming\\
J1006432+091726& 10 06 43.16& +09 17 26.3&SDSS J100643.50+091727.5&0.171&$12.10^{+ 0.07}_{- 0.01}$&  0.79$\pm$  0.20&  1.02$\pm$  0.08&  1.22$\pm$  0.30&  0.42$\pm$  0.10&13.72&IIIb&2&\nodata&Composite\\  
J1052232+440849& 10 52 23.24& +44 08 48.6&SDSS J105223.52+440847.6&0.092&$12.06^{+ 0.03}_{- 0.03}$&  3.39$\pm$  0.27&  3.53$\pm$  0.12&  4.11$\pm$  0.37&  2.96$\pm$  1.01&15.33&IV&1&\nodata&Composite\\  
J1254008+101115& 12 54 00.82& +10 11 14.6&SDSS J125400.80+101112.4&0.319&$12.58^{+ 0.03}_{- 0.05}$&  0.69$\pm$  0.17&  0.79$\pm$  0.05&\nodata&  0.99$\pm$  0.25&12.99&V&3&\nodata&QSO\\ 
J1348397+581854& 13 48 39.66& +58 18 54.1&SDSS J134840.08+581852.0&0.158&$12.12^{+ 0.03}_{- 0.04}$&  0.53$\pm$  0.13&  1.36$\pm$  0.03&  0.93$\pm$  0.23&  1.26$\pm$  0.63&15.08&IIIb&1&\nodata&Composite\\
J1015153+272717& 10 15 15.33& +27 27 17.1&SDSS J101515.35+272717.1&0.210&$12.09^{+ 0.03}_{- 0.04}$&  0.84$\pm$  0.21&  0.64$\pm$  0.02&  0.86$\pm$  0.21&  0.74$\pm$  0.18&13.88&V&3&\nodata&Composite\\
J1356100+290538& 13 56 09.98& +29 05 38.0&SDSS J135609.99+290535.1&0.109&$12.04^{+ 0.02}_{- 0.05}$&  1.70$\pm$  0.43&  1.78$\pm$  0.05&  2.72$\pm$  0.56&  0.26$\pm$  0.07&14.21&IIIb&1&\nodata&Composite\\
  J1502320+142132& 15 02 31.95& +14 21 32.4&SDSS J150231.96+142135.3&0.162&$12.12^{+ 0.04}_{- 0.03}$&  0.26$\pm$  0.06&  1.69$\pm$  0.08&  2.37$\pm$  0.28&  0.77$\pm$  0.19&15.15&Tp1&1&\nodata&Composite\\
J1336237+391733& 13 36 23.74& +39 17 32.5&SDSS J133624.06+391731.1&0.179&$12.37^{+ 0.06}_{- 0.06}$&  1.38$\pm$  0.34&  1.03$\pm$  0.06&  2.58$\pm$  0.36&  2.74$\pm$  1.36&16.01&IV,G&1&\nodata& QSO\\ 
J1141215+405951& 11 41 21.52& +40 59 51.3&SDSS J114122.03+405950.3&0.149&$12.01^{+ 0.18}_{- 0.04}$&  0.58$\pm$  0.14&  1.08$\pm$  0.05&  1.57$\pm$  0.61&\nodata&15.66&V&1&\nodata&Composite\\ 
J1433271+281157& 14 33 27.14& +28 11 57.0&SDSS J143327.52+281159.9&0.175&$12.12^{+ 0.16}_{- 0.07}$&  0.46$\pm$  0.11&  0.83$\pm$  0.05&  1.64$\pm$  0.41&\nodata&13.94&IIIb&3&\nodata&Seyfert\\
J1450544+350835& 14 50 54.40& +35 08 34.7&SDSS J145054.16+350837.9&0.206&$12.34^{+ 0.11}_{- 0.14}$&  0.34$\pm$  0.08&  0.72$\pm$  0.07&  1.81$\pm$  0.90&\nodata&14.35&V&3&\nodata&Composite\\ 
J1406380+010258& 14 06 37.97& +01 02 58.1&SDSS J140638.20+010254.6&0.236&$12.35^{+ 0.01}_{- 0.05}$&  0.99$\pm$  0.25&  0.74$\pm$  0.09&  0.18$\pm$  0.05&\nodata&16.03&IV&2&\nodata&Composite\\ 
J1522382+333135& 15 22 38.17& +33 31 35.4&SDSS J152238.10+333135.9&0.125&$12.03^{+ 0.01}_{- 0.05}$&  0.91$\pm$  0.23&  1.26$\pm$  0.04&  1.22$\pm$  0.25&\nodata&17.23&IV&1&\nodata&Composite$^{\star}$\\ 
 J1505390+574305& 15 05 39.04& +57 43 04.6&SDSS J150539.55+574307.1&0.151&$12.02^{+ 0.02}_{- 0.08}$&  1.14$\pm$  0.28&  0.87$\pm$  0.05&  1.61$\pm$  0.97&  0.73$\pm$  2.51&14.70&IIIb&1&\nodata& Star Forming\\ 
J1441041+532011& 14 41 04.11& +53 20 10.8&SDSS J144104.38+532008.7&0.105&$12.03^{+ 0.01}_{- 0.04}$&  2.03$\pm$  0.45&  1.78$\pm$  0.13&  1.77$\pm$  0.41&  0.91$\pm$  2.13&16.49&Tp1&1&\nodata& LINER$^{\star}$\\
 J1706529+382010& 17 06 52.87& +38 20 09.9&SDSS J170653.27+382007.1&0.168&$12.15^{+ 0.02}_{- 0.04}$&  0.79$\pm$  0.20&  0.96$\pm$  0.02&  1.68$\pm$  0.75&  0.19$\pm$  0.05&15.16&IIIa,G&2&A& LINER\\ 
J1649140+342510& 16 49 14.01& +34 25 09.8&SDSS J164914.09+342513.2&0.113&$12.07^{+ 0.08}_{- 0.01}$&  2.28$\pm$  0.36&  2.24$\pm$  0.08&  2.69$\pm$  0.95&  1.96$\pm$  3.66&15.55&IIIb&1&\nodata& Star Forming\\
J0823127+275140& 08 23 12.66& +27 51 39.6&SDSS J082312.61+275139.8&0.168&$12.07^{+ 0.03}_{- 0.03}$&  0.53$\pm$  0.13&  1.04$\pm$  0.02&  0.96$\pm$  0.24&  1.26$\pm$  0.32&18.30&IV&1&\nodata& Star Forming\\
J1213460+024844& 12 13 45.99& +02 48 43.8&SDSS J121346.11+024841.5&0.073&$12.25^{+ 0.03}_{- 0.03}$&  7.10$\pm$  0.51&  8.69$\pm$  0.31&  6.17$\pm$  0.36&  3.90$\pm$  1.32&15.42&IIIb&1&\nodata&Composite\\
J1346511+074720& 13 46 51.09& +07 47 20.0&SDSS J134651.09+074719.0&0.135&$12.09^{+ 0.06}_{- 0.02}$&  1.71$\pm$  0.43&  1.49$\pm$  0.10&  2.25$\pm$  0.86&  0.61$\pm$  0.15&16.68&Tp1,G&1&\nodata&Composite\\ 
J2257246-262120& 22 57 24.65& -26 21 20.5&TGS123Z162&0.164&$12.16^{+ 0.05}_{- 0.06}$&  1.27$\pm$  0.32&  1.17$\pm$  0.17&  1.53$\pm$  1.17&  1.68$\pm$  0.11&15.22&IIIa&2&\nodata&\nodata\\ 
J2223286-270006& 22 23 28.57& -27 00 05.7&TGS178Z172&0.131&$12.19^{+ 0.05}_{- 0.02}$&  1.87$\pm$  0.08&  1.90$\pm$  0.09&  2.90$\pm$  0.87&  1.06$\pm$  0.26&12.70&IIIb&1&\nodata&\nodata\\ 
J1132417-053940& 11 32 41.68& -05 39 40.2&TGN111Z322&0.230&$12.18^{+ 0.07}_{- 0.05}$&  0.44$\pm$  0.11&  0.64$\pm$  0.02&  0.81$\pm$  0.20&\nodata&14.46&IV&2&\nodata&\nodata\\ 
J0238167-322036& 02 38 16.66& -32 20 36.3&TGS465Z105&0.198&$12.31^{+ 0.02}_{- 0.03}$&  1.08$\pm$  0.27&  0.95$\pm$  0.11&  1.66$\pm$  0.10&  2.33$\pm$  0.58&13.90&V&2&\nodata&\nodata\\ 
J0238126-473813& 02 38 12.64& -47 38 12.9&TGS875Z072&0.098&$12.07^{+ 0.02}_{- 0.03}$&  2.76$\pm$  0.32&  3.24$\pm$  0.09&  3.99$\pm$  0.42&  2.34$\pm$  0.66&14.53&V&2&\nodata&\nodata\\ 
J0237297-461544& 02 37 29.66& -46 15 44.4&TGS875Z471&0.206&$12.37^{+ 0.06}_{- 0.06}$&  0.90$\pm$  0.56&  1.13$\pm$  0.07&  1.04$\pm$  1.04&  1.99$\pm$  0.50&15.94&V&2&\nodata&\nodata\\ 
J0048064-284820& 00 48 06.37& -28 48 20.0&TGS288Z046&0.110&$12.12^{+ 0.04}_{- 0.02}$&  2.56$\pm$  0.64&  2.48$\pm$  0.16&  4.08$\pm$  0.36&\nodata&16.85&IIIa&1&\nodata&\nodata\\ 
J0112165-273819& 01 12 16.49& -27 38 18.8&TGS213Z002&0.222&$12.37^{+ 0.01}_{- 0.24}$&  0.43$\pm$  0.11&  0.45$\pm$  0.04&  0.83$\pm$  0.99&  1.79$\pm$  0.45&15.05&IIIa&2&\nodata&\nodata\\  
J0138061-324519& 01 38 06.14& -32 45 18.6&TGS509Z038&0.198&$12.12^{+ 0.08}_{- 0.02}$&  1.23$\pm$  0.31&  0.75$\pm$  0.07&  0.89$\pm$  0.22&  0.40$\pm$  0.10&16.99&V&2&\nodata&\nodata\\ 
J0302108-270725& 03 02 10.82& -27 07 24.7&TGS238Z241&0.221&$12.42^{+ 0.06}_{- 0.03}$&  0.98$\pm$  0.24&  1.22$\pm$  0.15&  1.07$\pm$  0.27&  0.42$\pm$  0.11&15.98&IV&2&\nodata&\nodata\\ 
J0118266-253607& 01 18 26.61& -25 36 06.8&TGS147Z020&0.237&$12.12^{+ 0.04}_{- 0.03}$&  0.67$\pm$  0.17&  0.54$\pm$  0.03&  0.50$\pm$  0.13&  0.59$\pm$  0.15&16.32&V&2&\nodata&\nodata\\ 
J0152042-285116& 01 52 04.20& -28 51 16.5&TGS302Z057&0.184&$12.06^{+ 0.06}_{- 0.04}$&  0.46$\pm$  0.11&  0.73$\pm$  0.06&  2.14$\pm$  0.60&  1.02$\pm$  0.26&15.63&IIIa&2&\nodata&\nodata\\ 
J0159138-292436& 01 59 13.82& -29 24 36.1&TGS304Z128&0.140&$12.06^{+ 0.08}_{- 0.02}$&  2.16$\pm$  0.40&  1.36$\pm$  0.05&  1.61$\pm$  0.36&  0.63$\pm$  0.16&16.72&IV&1&\nodata&\nodata\\ 
J1329391-034654& 13 29 39.11& -03 46 53.7&TGN137Z043&0.222&$12.19^{+ 0.33}_{- 0.00}$&  0.58$\pm$  0.14&  0.75$\pm$  0.06&\nodata&  2.49$\pm$  3.85&16.39&IV&2&\nodata&\nodata\\  
J1112034-025414& 11 12 03.36& -02 54 13.8&TGN232Z018&0.106&$12.14^{+ 0.07}_{- 0.02}$&  2.55$\pm$  0.39&  2.93$\pm$  0.26&  1.47$\pm$  0.28&\nodata&14.31&IV&1&\nodata&\nodata\\  
J2307212-343838& 23 07 21.24& -34 38 38.4&TGS538Z137&0.208&$12.22^{+ 0.04}_{- 0.02}$&  0.76$\pm$  0.19&  0.89$\pm$  0.10&  1.43$\pm$  0.36&  0.34$\pm$  0.59&18.43&V&2&\nodata&\nodata\\ 
J2208493-344627& 22 08 49.25& -34 46 27.4&TGS528Z076&0.174&$12.21^{+ 0.05}_{- 0.14}$&  0.80$\pm$  0.20&  0.67$\pm$  0.02&  1.88$\pm$  0.45&\nodata&14.93&V&2&\nodata&\nodata\\  
\enddata 
\tablenotetext{e}{See \S \ref{S:morphprop} for the details of the interaction classes. The IC listed in this Table are mostly adopted from the references given in column (13).} 
\tablecomments{For a,  b, c, d, f, g, h see notes in Table \ref{tab:knownULIRGs}.} 
\label{tab:knownULIRGs}
\end{deluxetable}
\clearpage
\end{landscape}
\twocolumngrid

% TABLE3 
\clearpage
\onecolumngrid
\LongTables
\begin{landscape}
\begin{deluxetable}{llccccccccccccc} 
%\rotate
\tablecolumns{15}
\tabletypesize{\scriptsize}
\setlength{\tabcolsep}{0.02in}
\tablewidth{0pt}
\tablecaption{New HLIRG}
\tablehead{
\colhead{Name} &
\colhead{AKARI RA} &
\colhead{AKARI DEC}&
\colhead{Other Name}&
\colhead{z\tablenotemark{a}}&
\colhead{$\log(L_{IR}/L_{\odot})$\tablenotemark{b}} &
\colhead{F(65\micron)\tablenotemark{c}} &
\colhead{F(90\micron)\tablenotemark{c}} &
\colhead{F(140\micron)\tablenotemark{c}} &
\colhead{F(160\micron)\tablenotemark{c}} &
\colhead{$r$\tablenotemark{d}} &
\colhead{IC\tablenotemark{e}} &
\colhead{IC\tablenotemark{f}} &
\colhead{Note\tablenotemark{g}}&
\colhead{Spectral\tablenotemark{h}} \\
\colhead{AKARI-FIS-V1} &
\colhead{(J2000)} &
\colhead{(J2000)} &
\colhead{} &
\colhead{} &
\colhead{} &
\colhead{(Jy)} &
\colhead{(Jy)} &
\colhead{(Jy)} &
\colhead{(Jy)} &
\colhead{(mag)} &
\colhead{} &
\colhead{Ref.} &
\colhead{} &
\colhead{Class} \\
\colhead{(1)} &
\colhead{(2)} &
\colhead{(3)} &
\colhead{(4)} &
\colhead{(5)} &
\colhead{(6)} &
\colhead{(7)} &
\colhead{(8)} &
\colhead{(9)} &
\colhead{(10)} &
\colhead{(11)} &
\colhead{(12)} &
\colhead{(13)} &
\colhead{(14)}&
\colhead{(15)}}
\startdata
 \hline
J1425001+103045& 14 25 00.06& +10 30 44.7&SDSS J142500.73+103043.6&0.480&$13.34^{+ 0.01}_{- 0.16}$&  0.95$\pm$  0.24&  0.61$\pm$  0.06&  1.97$\pm$  0.48&  1.97$\pm$  0.49&18.51&V&3&\nodata&Composite\\
\enddata
\tablecomments{For a,  b, c, d, e, f, g, h see notes in Table \ref{tab:knownULIRGs}.}
\label{tab:newHLIRG}
\end{deluxetable}
\clearpage
\end{landscape}
\twocolumngrid

% TABLE4
\clearpage
\onecolumngrid
\LongTables
\begin{landscape}
\begin{deluxetable}{llccccccccccccc} 
%\rotate
\tablecolumns{15}
\tabletypesize{\scriptsize}
\setlength{\tabcolsep}{0.02in}
\tablewidth{0pt}
\tablecaption{Unconfirmed ULIRG Candidates}
\tablehead{
\colhead{Name} &
\colhead{AKARI RA} &
\colhead{AKARI DEC}&
\colhead{Other Name}&
\colhead{z\tablenotemark{a}}&
\colhead{$\log(L_{IR}/L_{\odot})$\tablenotemark{b}} &
\colhead{F(65\micron)\tablenotemark{c}} &
\colhead{F(90\micron)\tablenotemark{c}} &
\colhead{F(140\micron)\tablenotemark{c}} &
\colhead{F(160\micron)\tablenotemark{c}} &
\colhead{$r$\tablenotemark{d}} &
\colhead{IC\tablenotemark{e}} &
\colhead{IC\tablenotemark{f}} &
\colhead{Note\tablenotemark{g}}&
\colhead{Spectral\tablenotemark{h}} \\
\colhead{AKARI-FIS-V1} &
\colhead{(J2000)} &
\colhead{(J2000)} &
\colhead{} &
\colhead{} &
\colhead{} &
\colhead{(Jy)} &
\colhead{(Jy)} &
\colhead{(Jy)} &
\colhead{(Jy)} &
\colhead{(mag)} &
\colhead{} &
\colhead{Ref.} &
\colhead{} &
\colhead{Class} \\
\colhead{(1)} &
\colhead{(2)} &
\colhead{(3)} &
\colhead{(4)} &
\colhead{(5)} &
\colhead{(6)} &
\colhead{(7)} &
\colhead{(8)} &
\colhead{(9)} &
\colhead{(10)} &
\colhead{(11)} &
\colhead{(12)} &
\colhead{(13)} &
\colhead{(14)}&
\colhead{(15)}}
\startdata
 \hline
J0058034-025406& 00 58 03.44& -02 54 05.8&SDSS J005802.70-025401.8&0.267&$12.27^{+ 0.03}_{- 0.43}$&\nodata&  0.21$\pm$  0.05&  5.13$\pm$  1.49&\nodata&14.31&IIIb&3&B&\nodata\\  
J1444024+000415& 14 44 02.37& +00 04 15.0&SDSS J144402.86+000411.9&0.330&$12.66^{+ 0.02}_{- 0.06}$&  0.09$\pm$  0.02&  0.23$\pm$  0.06&  0.96$\pm$  0.24&  3.74$\pm$  0.32&13.44&IIIa&3&\nodata&\nodata\\ 
J0545295-011324& 05 45 29.50& -01 13 24.5&SDSS J054529.53-011318.0&0.884&$12.82^{+ 0.02}_{- 0.07}$&  0.41$\pm$  0.10&  0.57$\pm$  0.20&\nodata&  0.24$\pm$  0.06&12.76&NI&3&\nodata&\nodata\\  
J1352265+020815& 13 52 26.48& +02 08 14.9&SDSS J135227.76+020816.5&2.853&$13.01^{+ 0.12}_{- 0.06}$&\nodata&  0.39$\pm$  0.07&  0.76$\pm$  0.18&  3.28$\pm$  0.82&15.70&IIIa&3&\nodata&\nodata\\  
J1716527+271405& 17 16 52.73& +27 14 04.6&SDSS J171653.48+271406.2&0.539&$12.21^{+ 0.09}_{- 0.11}$&\nodata&  0.23$\pm$  0.06&  1.97$\pm$  0.36&  0.17$\pm$  0.04&11.74&V&3&A&\nodata\\ 
\enddata
\tablecomments{For a,  b, c, d, e, f, g, h see notes in Table \ref{tab:knownULIRGs}.}
\label{tab:susULIRGs}
\end{deluxetable}
\clearpage
\end{landscape}
\twocolumngrid

% TABLE5
\LongTables
\begin{deluxetable*}{llcllccc} 
%\rotate
\tablecolumns{8}
\tabletypesize{\footnotesize}
\setlength{\tabcolsep}{0.03in}
\tablewidth{0pt}
\tablecaption{Stellar Masses, Star Formation Rates, Oxygen Abundances, Optical Colors and Absolute Magnitudes}
\tablehead{
\colhead{AKARI-FIS-V1 Name} &
\colhead{IRAS Name} &
\colhead{$\log(M_{star} (M_{\odot}))$}&
\colhead{SFR(IR)}&
\colhead{SFR(\Halpha)}&
\colhead{12+$\log$(O/H)} &
\colhead{u$^{0.1}-$r$^{0.1}$} &
\colhead{$M_{r}^{0.1}$}  \\
\colhead{} &
\colhead{} &
\colhead{}&
\colhead{(M$_{\odot}$ year$^{-1}$)}&
\colhead{(M$_{\odot}$ year$^{-1}$)}&
\colhead{} &
\colhead{(mag)} &
\colhead{(mag)}  \\
\colhead{(1)} &
\colhead{(2)} &
\colhead{(3)} &
\colhead{(4)} &
\colhead{(5)} &
\colhead{(6)} &
\colhead{(7)} &
\colhead{(8)}}
\startdata
 \hline
J2216028+005813& F22134+0043 & \nodata                         & $1153_{320}^{463}$ & \nodata                            & \nodata           & \nodata          \\
J0859229+473612& F08559+4748 & $10.74_{0.06}^{0.10}$ & $275_{49}^{2}$ & $37_{10}^{10}$   &8.94$\pm$0.08&2.92$\pm$0.12&-21.41$\pm$0.01\\
J1443444+184950& F14414+1902 & $10.34_{0.02}^{0.08}$ & $281_{141}^{10}$ & $14_{19}^{19}$ & \nodata           &2.47$\pm$0.12&-21.07$\pm$0.01\\
J0857505+512037& F08542+5132 & $10.96_{0.04}^{0.03}$ & $1325_{61}^{247}$ & $77_{87}^{87}$& \nodata           &2.23$\pm$0.43&-21.35$\pm$0.01\\
J1106104+023458& \nodata            & $11.14_{0.03}^{0.03}$ & $292_{39}^{46}$ & $7_{86}^{86}$    & \nodata           &2.59$\pm$0.29&-21.77$\pm$0.01\\
J1157412+321316& F11550+3233  &$10.58_{0.08}^{0.03}$ & $238_{57}^{1}$ & $62_{22}^{22}$   &8.79$\pm$0.10&1.45$\pm$0.02&-21.70$\pm$0.01\\
J1149200-030357&  F11467-0247  & \nodata                        & $178_{16}^{10}$ & \nodata                               & \nodata           &1.72$\pm$0.03&-21.29$\pm$0.01\\
J0126038+022456& F01234+0209 &$10.88_{0.08}^{0.40}$ & $288_{34}^{31}$ &$0_{0}^{0}$           & \nodata           &3.18$\pm$0.48&-21.06$\pm$0.01\\
J1556089+254358& F15540+2552 &$10.47_{0.13}^{0.06}$ & $183_{83}^{5}$ & $35_{14}^{14}$      &8.89$\pm$0.11&2.19$\pm$0.06&-21.08$\pm$0.01\\
J0140364+260016& F01378+2545 &$11.03_{0.32}^{0.18}$ & $1021_{52}^{152}$ & \nodata                             & \nodata           &1.61$\pm$0.67&-21.40$\pm$0.02\\
J1257392+080935& F12551+0825 & \nodata                        & $297_{11}^{32}$ & \nodata                              & \nodata           &0.43$\pm$0.01&-22.52$\pm$0.01\\
J0800007+152319& \nodata            & \nodata                       & $237_{0}^{3}$ & \nodata                                  & \nodata           &1.34$\pm$0.06&-21.40$\pm$0.01\\
J0800279+074858& \nodata            &$11.26_{0.34}^{0.01}$ & $228_{42}^{3}$ & $0_{4}^{4}$          & \nodata           &3.06$\pm$0.17&-21.11$\pm$0.01\\
J0834438+334427&  F08315+3354 &$10.42_{0.02}^{0.17}$ & $231_{86}^{17}$ & $69_{42}^{42}$ & \nodata           &2.27$\pm$0.08&-21.30$\pm$0.01\\
J0823089+184234&  \nodata            &$10.63_{0.12}^{0.01}$ & $634_{44}^{1068}$ & \nodata                        & \nodata           &1.77$\pm$0.15&-21.45$\pm$0.01\\
J1202527+195458&  F12002+2011 &$10.58_{0.30}^{0.02}$ & $193_{6}^{17}$ & $51_{11}^{11}$    &8.92$\pm$0.08&1.89$\pm$0.03&-21.06$\pm$0.01\\
J0912533+192701&  \nodata            &$10.71_{0.04}^{0.17}$ & $222_{20}^{49}$ & $19_{14}^{14}$ &8.63$\pm$0.28&2.87$\pm$0.19&-22.04$\pm$0.01\\
J0941010+143622& F09382+1449 & \nodata                         & $968_{101}^{32}$ & \nodata                           & \nodata           &1.37$\pm$0.06&-22.29$\pm$0.01\\
J1016332+041418& F10139+0429 &$10.40_{0.01}^{0.32}$ & $423_{44}^{31}$ & $38_{21}^{21}$ &8.40$\pm$0.22&1.93$\pm$0.09&-21.37$\pm$0.01\\
J1401186-021131&  \nodata            &$10.80_{0.25}^{0.01}$ & $201_{35}^{4}$ & $6_{3}^{3}$         & \nodata           &2.01$\pm$0.06&-21.98$\pm$0.01\\
J1258241+224113& \nodata            &$10.09_{0.04}^{0.06}$ & $201_{16}^{25}$ & $17_{14}^{14}$&8.59$\pm$0.23&2.77$\pm$0.65&-19.93$\pm$0.02\\
J1036317+022147& \nodata            &$9.98_{0.23}^{0.15}$ & $196_{14}^{12}$ & $7_{2}^{2}$         &8.69$\pm$0.12&1.68$\pm$0.01&-20.43$\pm$0.01\\
J1050567+185316& F10482+1909 &$10.40_{0.07}^{0.14}$ & $678_{80}^{56}$ & $154_{83}^{83}$& \nodata           &2.49$\pm$0.11&-21.23$\pm$0.01\\
J1111177+192259& \nodata            &$10.27_{0.14}^{0.12}$ & $243_{36}^{2}$ & $42_{8}^{8}$          &8.83$\pm$0.15&0.98$\pm$0.03&-21.62$\pm$0.01\\
J1219585+051745& \nodata            &$10.88_{0.09}^{0.45}$ & $1277_{94}^{74}$ & \nodata                             & \nodata           &2.52$\pm$1.19&-21.66$\pm$0.01\\
J1414276+605726& F14129+6111  &$10.16_{0.06}^{0.06}$ & $220_{51}^{0}$ & $28_{9}^{9}$&8.69$\pm$0.10&1.07$\pm$0.02&-20.69$\pm$0.01\\
J0936293+203638& F09336+2049 &$10.60_{0.15}^{0.01}$ & $177_{8}^{16}$ & $29_{11}^{11}$&9.00$\pm$0.23&2.44$\pm$0.09&-21.34$\pm$0.01\\
J1533582+113413& \nodata            &$11.34_{0.32}^{0.04}$ & $357_{67}^{78}$ & \nodata                        & \nodata           &2.11$\pm$0.41&-22.18$\pm$0.01\\
J1348483+181401& F13464+1828 &$10.16_{0.07}^{0.08}$ & $264_{17}^{18}$ & $39_{18}^{18}$&8.78$\pm$0.84&2.75$\pm$0.30&-19.73$\pm$0.02\\
J1125319+290316& \nodata            & \nodata                         & $320_{32}^{10}$ & \nodata                        & \nodata           &1.96$\pm$0.03&-21.24$\pm$0.01\\
J1603043+094717&  F16006+0955 &$10.51_{0.02}^{0.18}$ & $181_{10}^{14}$ & $57_{21}^{21}$& \nodata           &2.37$\pm$0.06&-21.54$\pm$0.01\\
J1639245+303719& F16374+3043 & \nodata                         & $223_{25}^{5}$ & \nodata                        & \nodata           &1.69$\pm$0.08&-21.64$\pm$0.01\\
J1050288+002806& \nodata            & $10.26_{0.22}^{0.01}$ & $416_{66}^{148}$ & $63_{26}^{26}$&8.74$\pm$0.13&1.31$\pm$0.04&-21.58$\pm$0.01\\
J0928103+232521& \nodata            &$10.01_{0.03}^{0.32}$ & $202_{18}^{44}$ & $10_{5}^{5}$&9.06$\pm$0.09&2.18$\pm$0.17&-20.18$\pm$0.02\\
J2344170+053520&  F23417+0518 &$11.61_{0.11}^{0.09}$ & $625_{218}^{14}$ & $43_{275}^{275}$& \nodata           &4.00$\pm$1.64&-22.39$\pm$0.01\\
J2353152-313234& \nodata            & \nodata                         & $186_{43}^{0}$ & \nodata                        & \nodata           & \nodata  \\            
J1222488-040307& \nodata            & \nodata                         & $269_{16}^{20}$ & \nodata                        &  \nodata           & \nodata  \\
J1419037-034657& \nodata            & \nodata                         & $213_{4}^{36}$ & \nodata                        & \nodata           & \nodata  \\
J1048019-013017& \nodata            & \nodata                         & $210_{40}^{29}$ & \nodata                        &  \nodata           & \nodata  \\
J1338353-041131& \nodata            & \nodata                         & $239_{95}^{76}$ & \nodata                        &  \nodata           & \nodata \\ 
J0857064+190855& F08542+1920 & \nodata                        & $1425_{139}^{34}$ & \nodata                        & \nodata           &0.16$\pm$0.01&-23.15$\pm$0.01\\
 J1022125+241208&  F10194+2427 &$10.78_{0.09}^{0.03}$ & $176_{27}^{113}$ & \nodata                        & \nodata           &2.39$\pm$0.07&-21.91$\pm$0.01\\
 J1422313+260205& F14202+2615 &$10.31_{0.06}^{0.36}$ & $274_{30}^{83}$ & $29_{8}^{8}$&8.68$\pm$0.09&1.15$\pm$0.02&-21.47$\pm$0.01\\
J1231216+275524&  F12288+2811 &$10.58_{0.04}^{0.15}$ & $367_{69}^{6}$ & $7_{5}^{5}$&8.54$\pm$0.24&2.44$\pm$0.09&-21.92$\pm$0.01\\
J1251200+021900& F12487+0235 &$10.60_{0.01}^{0.24}$ & $521_{18}^{56}$ & $67_{40}^{67}$&8.40$\pm$0.32&1.67$\pm$0.04&-22.24$\pm$0.01\\
J0030089-002743& F00275-0044 & \nodata                        & $496_{28}^{34}$ & \nodata                        & \nodata           & \nodata \\
J0914140+032200& F09116+0334 &$10.56_{0.01}^{0.06}$ & $200_{11}^{28}$ & $374_{303}^{303}$& \nodata           &2.21$\pm$0.03&-21.92$\pm$0.01\\
J1105377+311432&  F11028+3130 &$9.97_{0.04}^{0.06}$ & $272_{13}^{33}$ & $3_{1}^{1}$& \nodata           &1.90$\pm$0.10&-20.72$\pm$0.01\\
J0323227-075612& F03209-0806 &$10.41_{0.03}^{0.12}$ & $235_{27}^{66}$ & $37_{10}^{10}$&8.75$\pm$1.21&1.37$\pm$0.03&-21.51$\pm$0.01\\
J1632212+155145& F16300+1558 &$10.65_{0.17}^{0.01}$ & $802_{60}^{38}$ & $43_{36}^{36}$&8.69$\pm$0.21&1.75$\pm$0.06&-22.07$\pm$0.01\\
J0148531+002857& F01462+0014 &$10.47_{0.24}^{0.03}$ & $340_{29}^{284}$ & $74_{37}^{37}$& \nodata           &1.33$\pm$0.04&-21.72$\pm$0.01\\
J0159503+002340& F01572+0009 & \nodata                        & $467_{41}^{15}$ & \nodata                        & \nodata           & \nodata \\
J1353317+042809&  F13509+0442 & \nodata                        & $473_{9}^{47}$ & \nodata                        & \nodata           & \nodata \\
J0244173-003040& F02417-0043 & \nodata                        & $202_{20}^{6}$ & \nodata                        & \nodata           & \nodata \\
J1202268-012918&  F11598-0112 & \nodata                        & $394_{6}^{51}$ & \nodata                        & \nodata           & \nodata \\
J1013477+465402& F10107+4708 &$10.17_{0.05}^{0.30}$ & $302_{31}^{68}$ & \nodata                        &  \nodata           &1.98$\pm$0.13&-20.91$\pm$0.01\\
J0858418+104124& F08559+1053 &$10.67_{0.04}^{0.06}$ & $254_{14}^{29}$ & \nodata                        & \nodata           &2.18$\pm$0.03&-21.91$\pm$0.01 \\
J1347336+121727& F13451+1232 & \nodata                        & $259_{15}^{18}$ & \nodata                        & \nodata           & \nodata \\
J0853252+252646& F08504+2538 & \nodata                        & $405_{68}^{91}$ & \nodata                        & \nodata           & \nodata \\
J0825215+383306& F08220+3842 &$10.65_{0.01}^{0.06}$ & $329_{57}^{87}$ &$26_{18}^{18}$&8.64$\pm$0.20&2.68$\pm$0.11&-21.70$\pm$0.01\\
J0829512+384528& F08266+3855 &$9.66_{0.05}^{0.02}$ & $180_{27}^{51}$ &$26_{19}^{19}$&8.40$\pm$0.27&2.11$\pm$0.25&-19.45$\pm$0.02\\
J1142035+005135&  F11394+0108 &$10.16_{0.07}^{0.21}$ & $214_{19}^{22}$ &$75_{43}^{43}$& \nodata           &1.08$\pm$0.03&-21.63$\pm$0.01\\
J0810595+281354& F08079+2822 &$10.28_{0.01}^{0.36}$ & $775_{212}^{415}$ &$36_{19}^{19}$&8.06$\pm$0.26&1.23$\pm$0.06&-21.43$\pm$0.01\\
J0900252+390400&  F08572+3915 &$9.42_{0.02}^{0.04}$ & $186_{16}^{6}$ &$5_{2}^{2}$&8.63$\pm$0.14&1.84$\pm$0.03&-19.40$\pm$0.01\\
J0830197+192040& F08274+1930 &$10.24_{0.07}^{0.16}$ & $186_{28}^{60}$ &$18_{5}^{5}$&8.89$\pm$0.07&1.00$\pm$0.03&-21.03$\pm$0.01\\
J1121293+112233& F11188+1138 &$10.46_{0.01}^{0.01}$ & $361_{20}^{39}$ &$81_{27}^{27}$& \nodata           &1.67$\pm$0.03&-21.97$\pm$0.01\\
J1006038+411223& F10030+4126 &$10.18_{0.01}^{0.19}$ & $455_{50}^{35}$ &$66_{40}^{40}$&8.56$\pm$0.27&1.38$\pm$0.07&-21.46$\pm$0.01\\
J0838034+505516& F08344+5105 &$9.96_{0.01}^{0.08}$ & $186_{16}^{6}$ &$18_{5}^{5}$&8.63$\pm$0.19&1.94$\pm$0.03&-20.48$\pm$0.01\\
J0902489+523623& F08591+5248 &$10.57_{0.11}^{0.01}$ & $195_{25}^{6}$ &$35_{12}^{12}$& \nodata           &2.15$\pm$0.04&-21.68$\pm$0.01\\
J0847504+232113& F08449+2332 &$10.23_{0.06}^{0.17}$ & $176_{29}^{42}$ &$36_{8}^{8}$&8.81$\pm$0.07&1.06$\pm$0.02&-20.88$\pm$0.01\\
J1559301+380843&  F15577+3816 &$10.70_{0.05}^{0.01}$ & $268_{23}^{16}$ &$23_{10}^{10}$& \nodata           &1.89$\pm$0.08&-21.56$\pm$0.01\\
J1324197+053705&  F13218+0552 & \nodata                        & $795_{60}^{42}$ & \nodata                        & \nodata           & \nodata \\
J1102140+380240&  F10594+3818 &$10.47_{0.04}^{0.12}$ & $245_{6}^{40}$ &$45_{13}^{13}$&8.89$\pm$0.30&1.45$\pm$0.02&-21.50$\pm$0.01\\
J1204244+192509&  F12018+1941 &$10.15_{0.04}^{0.02}$ & $242_{18}^{11}$ &$65_{35}^{35}$&7.62$\pm$0.37&1.90$\pm$0.05&-21.30$\pm$0.01\\
J1108513+065915&  F11062+0715 & \nodata                        & $200_{43}^{45}$ & \nodata                        & \nodata           & \nodata \\
J1040290+105325& F10378+1108 &$10.35_{0.01}^{0.18}$ & $311_{17}^{43}$ &$43_{14}^{14}$& \nodata           &2.21$\pm$0.05&-21.35$\pm$0.01\\
J1207210+021702& F12047+0233 &$10.20_{0.11}^{0.24}$ & $211_{16}^{20}$ &$48_{18}^{18}$&8.89$\pm$0.10&1.00$\pm$0.03&-21.50$\pm$0.01\\
J1255482-033908&  F12532-0322 & \nodata                        & $195_{33}^{69}$ & \nodata                        & \nodata           & \nodata \\
J0906339+045136&  F09039+0503 &$10.45_{0.33}^{0.01}$ & $181_{14}^{29}$ &$36_{14}^{14}$&8.22$\pm$0.20&1.82$\pm$0.04&-20.82$\pm$0.01\\
J1153144+131432&  F11506+1331 &$10.08_{0.01}^{0.05}$ & $311_{11}^{37}$ &$47_{28}^{28.}$&8.40$\pm$0.28&2.17$\pm$0.07&-20.51$\pm$0.01\\
J1202054+112813&  F11595+1144 &$10.17_{0.01}^{0.50}$ & $268_{14}^{30}$ &$71_{25}^{25}$&8.75$\pm$0.17&1.22$\pm$0.02&-21.61$\pm$0.01\\
J1006432+091726&  F10040+0932 &$10.34_{0.03}^{0.12}$ & $218_{11}^{36}$ &$26_{15}^{15}$&8.75$\pm$0.15&1.55$\pm$0.04&-21.06$\pm$0.01\\
J1052232+440849&  F10494+4424 &$10.22_{0.03}^{0.04}$ & $196_{10}^{14}$ &$20_{5}^{5}$&8.37$\pm$0.16&2.52$\pm$0.06&-20.45$\pm$0.01\\
J1254008+101115& F12514+1027 & \nodata                        & $659_{75}^{50}$ & \nodata                        & \nodata           & \nodata \\
J1348397+581854& F13469+5833 &$10.46_{0.13}^{0.06}$ & $224_{16}^{16}$ &$13_{6}^{6}$&8.77$\pm$0.40&2.23$\pm$0.07&-21.02$\pm$0.01\\
J1015153+272717& F10124+2742 &$10.36_{0.01}^{0.32}$ & $213_{19}^{17}$ &$49_{19}^{19}$&8.45$\pm$0.16&1.91$\pm$0.06&-21.26$\pm$0.01\\
J1356100+290538& F13539+2920 &$11.00_{0.10}^{0.06}$ & $190_{18}^{7}$ &$39_{14}^{14}$& \nodata           &3.09$\pm$0.11&-20.41$\pm$0.01\\
J1502320+142132& F15001+1433 &$10.44_{0.01}^{0.07}$ & $225_{15}^{20}$ &$71_{28}^{28}$&8.38$\pm$0.21&2.07$\pm$0.04&-21.58$\pm$0.01 \\
J1336237+391733&  F13342+3932 & \nodata                        & $405_{47}^{66}$ & \nodata                        & \nodata           & \nodata \\
J1141215+405951& F11387+4116 &$10.45_{0.02}^{0.06}$ & $174_{13}^{93}$ &$70_{55}^{55}$&8.14$\pm$1.22&2.47$\pm$0.06&-21.08$\pm$0.01\\
J1433271+281157& F14312+2825 &$10.72_{0.06}^{0.01}$ & $227_{31}^{102}$ &$83_{32}^{32}$& \nodata           &2.23$\pm$0.05&-21.79$\pm$0.01\\
J1450544+350835& F14488+3521 &$10.86_{0.02}^{0.07}$ & $379_{103}^{115}$ &$183_{57}^{57}$&8.79$\pm$0.15&1.55$\pm$0.02&-22.33$\pm$0.01 \\
J1406380+010258& F14041+0117 &$10.11_{0.01}^{0.18}$ & $387_{37}^{12}$ &$11_{8}^{22}$&7.79$\pm$0.67&1.28$\pm$0.04&-21.40$\pm$0.01 \\
J1522382+333135& F15206+3342 & \nodata                        & $184_{18}^{6}$ &  \nodata                        & \nodata           & \nodata \\
J1505390+574305& F15043+5754 &$10.45_{0.08}^{0.06}$ & $181_{30}^{9}$ &$19_{11}^{11}$&8.85$\pm$0.13&1.29$\pm$0.03&-20.69$\pm$0.01 \\
J1441041+532011& F14394+5332 & \nodata                        & $186_{16}^{5}$ &  \nodata                        & \nodata           & \nodata \\
J1706529+382010& F17051+3824 &$10.26_{0.01}^{0.01}$ & $241_{20}^{10}$ &$14_{9}^{9}$& \nodata           &2.17$\pm$0.06&-20.94$\pm$0.01 \\
J1649140+342510& F16474+3430 &$9.63_{0.15}^{0.32}$ & $201_{4}^{43}$ &$6_{2}^{2}$&8.87$\pm$0.09&1.73$\pm$0.06&-19.69$\pm$0.01 \\
J0823127+275140& F08201+2801 &$10.38_{0.20}^{0.11}$ & $201_{12}^{16}$ &$14_{4}^{4}$&8.70$\pm$0.10&1.58$\pm$0.04&-21.19$\pm$0.01 \\
J1213460+024844& F12112+0305 &$10.28_{0.01}^{0.01}$ & $306_{21}^{23}$ &$9_{3}^{3}$&8.39$\pm$0.33&1.78$\pm$0.03&-20.14$\pm$0.01 \\
J1346511+074720& F13443+0802 &$10.57_{0.01}^{0.05}$ & $211_{10}^{33}$ &$20_{3}^{3}$&8.97$\pm$0.04&2.08$\pm$0.03&-21.49$\pm$0.01 \\
J2257246-262120& F22546-2637 & \nodata                        & $249_{29}^{32}$ &  \nodata                        & \nodata           & \nodata \\
J2223286-270006& F22206-2715 & \nodata                        & $266_{13}^{30}$ &  \nodata                        & \nodata           & \nodata \\
J1132417-053940&  F11300-0522 & \nodata                        & $161_{27}^{43}$ &  \nodata                        & \nodata           & \nodata \\
J0238167-322036& F02361-3233 & \nodata                        & $350_{25}^{19}$ &  \nodata                        & \nodata           & \nodata \\
J0238126-473813& F02364-4751 & \nodata                        & $201_{14}^{11}$ &  \nodata                        & \nodata           & \nodata \\
J0237297-461544& F02356-4628 & \nodata                        & $400_{53}^{59}$ &  \nodata                        & \nodata           & \nodata \\
J0048064-284820& F00456-2904 &  \nodata                        & $228_{11}^{23}$ &  \nodata                        & \nodata           & \nodata \\
J0112165-273819&  F01098-2754 & \nodata                        & $404_{170}^{7}$ &  \nodata                        & \nodata           & \nodata \\
J0138061-324519&  F01358-3300 & \nodata                        & $224_{8}^{43}$ &  \nodata                        & \nodata           & \nodata \\
J0302108-270725& F03000-2719 & \nodata                        & $457_{70}^{28}$ &  \nodata                        & \nodata           & \nodata \\
J0118266-253607& F01160-2551 &  \nodata                        & $227_{13}^{23}$ &  \nodata                        & \nodata           & \nodata \\
J0152042-285116& F01497-2906 &  \nodata                        & $197_{18}^{28}$ &  \nodata                        & \nodata           & \nodata \\
J0159138-292436& F01569-2939 &   \nodata                        & $195_{6}^{37}$ &   \nodata                        & \nodata           & \nodata \\
J1329391-034654&  F13270-0331 &  \nodata                        & $268_{0}^{305}$ &  \nodata                        &  \nodata           & \nodata \\
J1112034-025414&  F11095-0238 &   \nodata                        & $236_{11}^{44}$ &  \nodata                        & \nodata           & \nodata \\
J2307212-343838& F23046-3454 &   \nodata                        & $285_{11}^{30}$  \nodata                        & \nodata           & \nodata \\
J2208493-344627& F22058-3501 &   \nodata                        & $280_{75}^{33}$   \nodata                        & \nodata           & \nodata \\
J1425001+103045& F14225+1044 &$10.77_{0.05}^{0.04}$ & $3766_{1151}^{101}$ &$139_{80}^{80}$&8.64$\pm$0.17&1.19$\pm$0.06&-22.62$\pm$0.00\\
\enddata
\tablecomments{ (1) \textit{AKARI} FIS-V1 Catalog Name. (2) \textit{IRAS} Faint Source Catalog Name. (3) Stellar mass adopted from \citet{Maraston2012}. (4) SFRs derived from IR luminosity. 
(5) SFRs derived from \Halpha\ luminosity. (6) Oxygen abundances derived in this work. (7) u$^{0.1}-$r$^{0.1}$ color. (8)Absolute magnitude in $r$ band.}
\label{tab:sparameters}
\end{deluxetable*}
\clearpage

\end{document}